\documentclass[a4paper,11pt]{article}
\pdfoutput=1
\usepackage{color}
\usepackage{epsfig}
\usepackage{amsmath}
\usepackage{amssymb}
\usepackage{graphicx}
\usepackage{jheppub} % for details on the use of the package, please
\newcommand{\beq}{\begin{equation}}
\newcommand{\eeq}{\end{equation}}

\def \w {u}

\title{\boldmath Structure constants at wrapping order}

\author[\dagger]{Benjamin Basso,}
\author[\mathcal{x}]{Vasco Gon\c calves}
\author[\mathcal{z}]{and Shota Komatsu}

\affiliation[\dagger]{Laboratoire de Physique Th\'eorique de l'\'Ecole Normale Sup\'erieure, CNRS,\\ Universit\'e de Recherche PSL, Sorbonne Universit\'es, Universit\'e Pierre et Marie Curie,\\
24 rue Lhomond, Paris 75005, France}
\affiliation[\mathcal{x}]{ICTP South American Institute for Fundamental Research, IFT-UNESP,\\
S\~ao Paulo, SP Brazil 01440-070}
\affiliation[\mathcal{z}]{Perimeter Institute for Theoretical Physics,\\
Waterloo, Ontario N2L 2Y5, Canada}

\abstract{We consider structure constants of single trace operators in planar $\mathcal{N}=4$ Super-Yang-Mills theory within the hexagon framework. The standard procedure for forming a three point function out of two hexagons develops divergences when the effects of virtual particles wrapping around the operators are taken into account. In this paper, we explain how to renormalize these divergences away and obtain definite predictions, at the leading wrapping order, for some of the structure constants that parameterize the OPE of two chiral primaries. We test our method at weak coupling against the four loop planar correction to the BPS-BPS-Konishi OPE coefficient, derived recently in the field theory. At strong coupling, we compare our expressions with the structure constants obtained in string theory for three semiclassical strings.}

\begin{document}

\maketitle
\flushbottom

\renewcommand{\thefootnote}{\arabic{footnote}}

\section{Introduction}\label{Intro}

Integrability has inspired and fueled new strategies for computing observables in planar $\mathcal{N}=4$ SYM and in its string theory dual \cite{Maldacena:1997re}. This has been illustrated in many ways in recent years, with examples ranging from scattering amplitudes \cite{Alday:2009dv,Alday:2010vh,Alday:2010ku,Basso:2013vsa,Basso:2013aha}, Wilson loops \cite{Correa:2012hh,Drukker:2012de,Bajnok:2013sya,Toledo:2014koa,Gromov:2015dfa}, form factors of local operators~\cite{Sasha}, one- \cite{deLeeuw:2015hxa,Buhl-Mortensen:2015gfd,Buhl-Mortensen:2016pxs,deLeeuw:2016ofj}, two- \cite{Gromov:2013pga,Cavaglia:2014exa,Borsato:2016xns}, three- \cite{Escobedo:2010xs,Bajnok:2015hla,short,Bajnok:2015ftj, Shota}, four-point functions \cite{Caetano:2011eb,Caetano:2012ac,Eden:2016xvg,Thiago&Shota,Basso:2017khq} to, stunningly, all the higher planar correlators of the theory \cite{Thiago&Shota}. It appears also possible, in some circumstances, to descend all the way down to individual Feynman integrals, by cleverly twisting the theory \cite{Gurdogan:2015csr,Caetano:2016ydc}.

Despite the ongoing progress, we still lack a systematic way to exploit integrability. (Even just to ascertain that it is correctly implemented in the many integrability stamped computations eludes us.) In lack of a universal top-down approach, one instead relies on bottom-up searches for ``coarser" building blocks. The latter must be small enough to be amenable to and solved by integrable bootstrap methods but big enough to capture some interesting ``physical bits'' of the observable under study. The combination of gauge and string theoretical ideas has been a leitmotif in all these atomic studies, hinting both at how to geometrically pull the elements apart and at how to algebraically bring them back together. One might hope that in the long run a more comprehensive framework will emerge from all of these investigations.

In this paper, we will be concerned with a particular bottom-up strategy, the hexagon method, developed for the structure constants of single trace operators \cite{short}. The idea here is to cut open the pair of pants representing a three point function into two hexagonal patches. These hexagons have the benefit of concentrating the physics of the structure constants in their cores, revealing new asymptotically flat boundaries where standard scattering and form factor techniques can be applied straightforwardly. They are the atoms of the structure constant, resumming an infinite class of planar $\sim 1/\sqrt{N}$ subdiagrams or, equivalently, describing an open string version of the string splitting, in the spirit of \cite{Witten:1985cc} though for strings in $AdS$. They come with a lot of symmetries which nicely combine with the worldsheet bulk integrability and allow the hexagon form factors to be boostrapped at any value of the 't Hooft coupling \cite{short}.

Ultimately, one would like to wrap the hexagons around the pair of pants and obtain a closed string amplitude $\sim 1/N$. Firstly, one must find a way to glue the patches together along their common edges. The common attitude is to include an infinite series of so called mirror corrections which take into account the finite size effects of the closed geometry. Physically, these are sums over virtual particles flowing through the various channels which are cut open by the decompactification procedure and sewed back together by the wrapping procedure -- an improved Feynman diagrammatic expansion of sort, although not controlled by the strength of the coupling but by the size of the geometry. Lastly, one must figure out how to efficiently generate and sum up all of these effects.

To date, the wrapping itinerary has only been completed for the determination of the spectrum of planar scaling dimensions. In this case, the mirror corrections are due to particles wrapping around the cylinder \cite{Ambjorn:2005wa,Bajnok:2008bm} and when seen from the right angle appear to be thermal in nature \cite{Zamolodchikov:1989cf}. This observation allows one to take several steps at a time and immediately resum all the finite size effects into a familiar system of TBA equations \cite{Gromov:2009tv,Bombardelli:2009ns,Gromov:2009bc,Arutyunov:2009ur}. As a cherry on the cake, these equations can be dramatically simplified and cast into a less familiar but arguably more fruitful covariant framework known as the Quantum Spectral Curve \cite{Gromov:2013pga} (see~\cite{Marboe:2014gma,Marboe:2014sya,Gromov:2015wca,Gromov:2015vua,Gromov:2016rrp,Hegedus:2016eop} for a sample of recent applications).

For the structure constants, the resummation and its covariantization are still out of reach as there are further difficulties standing in the way. Indeed, the naive hexagon gluing procedure suffers from divergences at wrapping order \cite{3loops}. They come from mirror particles that wind closely around the punctures of the planar surface or equivalently escape through the half infinite tubes that stick out of the pair of pants. Taming this issue will be our main concern in this paper. We shall argue that it can be resolved by renormalizing the operator insertion. More concretely, we shall view the problem as stemming from an overconfident use of the naive measures of integration and address it, in a somewhat standard manner, by regularizing the Hilbert space of the mirror particles and absorbing the regulator dependence in the wave function used to describe the inserted operator. This subtraction procedure should in principle remove all of the divergences affecting the hexagonal construction, at least in the planar limit and for non extremal correlators. In this paper, we shall analyze it in detail on the particular case where only one of the operators renormalizes. This set-up is realized by the structure constants involving two BPS and one non-BPS operators. We will furthermore restrict ourselves to non-BPS operators falling in a closed rank one subsector of the full theory. For them we shall obtain an improved hexagon formula which is explicitly finite and valid at the leading wrapping order. It can be seen as a first step towards a complete ``thermalization" of the hexagon gluing procedure.

Besides making the ``sixfold way'' safer for the higher point functions, the motivation for our study is to catch up with the most advanced field theory techniques for computing higher loop correlators \cite{Grisha,Drummond:2013nda,Chicherin:2015edu,Eden,Vasco,EdenPaul}, which culminate in the structure constant $C^{\bullet\circ\circ}$ between the Konishi operator $\mathcal{O}_{\mathcal{K}}$ and two length two chiral primaries $\mathcal{O}_{\textbf{20}}$ (or, in dual words, the first ``stringy" correction in the OPE $\mathcal{O}_{\textbf{20}}(1)\times \mathcal{O}_{\textbf{20}}(0) = 1+C^{\circ\circ\circ}\mathcal{O}_{\textbf{20}}(0) + C^{\bullet\circ\circ}\mathcal{O}_{\mathcal{K}}(0) + \ldots$). This OPE coefficient is currently known through four loops \cite{Vasco,EdenPaul} (see \cite{DO,Eden} for the earlier loop orders)
\beq\label{QFT}
\begin{aligned}
\left(\frac{C^{\bullet\circ\circ}}{C^{\circ\circ\circ}}\right)^2 =\,\,& \frac{1}{6}-2g^2+(28+12\zeta_{3})g^4-(384+64\zeta_{3}+200\zeta_{5})g^6 \\
&+ (4976+656\zeta_{3}+1960\zeta_{5}+144\zeta_{3}^2+2940\zeta_{7})g^8 + O(g^{10})\, ,
\end{aligned}
\eeq
where $g^2 = \lambda/(4\pi)^2$ is the 't Hooft coupling and $\zeta_{z} = \zeta(z)$ is the Riemann zeta function. This robust field theory result is an ideal testing ground for our method, since four loops is the onset of the divergences in the hexagon formalism for the Konishi operator. In this paper we will explain how to reproduce it using the renormalized hexagon formula.

Finally, in the opposite limit, at strong coupling, we will see that our improvement partially resolves a tension that exists between the bare hexagon approach and the string theory prediction~\cite{Shota} for the semiclassical decay of the GKP string into two BMN vacua. 

The paper is structured as follows. In Section \ref{Sect2} we review the recipe for gluing two hexagons together and explain what goes wrong at wrapping order. We proceed in Section \ref{Sect3} with the regularization of the divergences and show that they can be absorbed in the two point function normalization of the excited operator. Section \ref{Sect4} contains the derivation of our wrapping formula for the structure constants with two BPS and one non-BPS operators. We extract its predictions at weak and strong coupling in Section \ref{test} and compare them with the gauge and string theory results. We conclude in Section \ref{Sect6}. Several appendices provide details on the computations.

\section{Hexagons and structure constants}\label{Sect2}

\begin{figure}
\begin{center}
\includegraphics[scale=0.40]{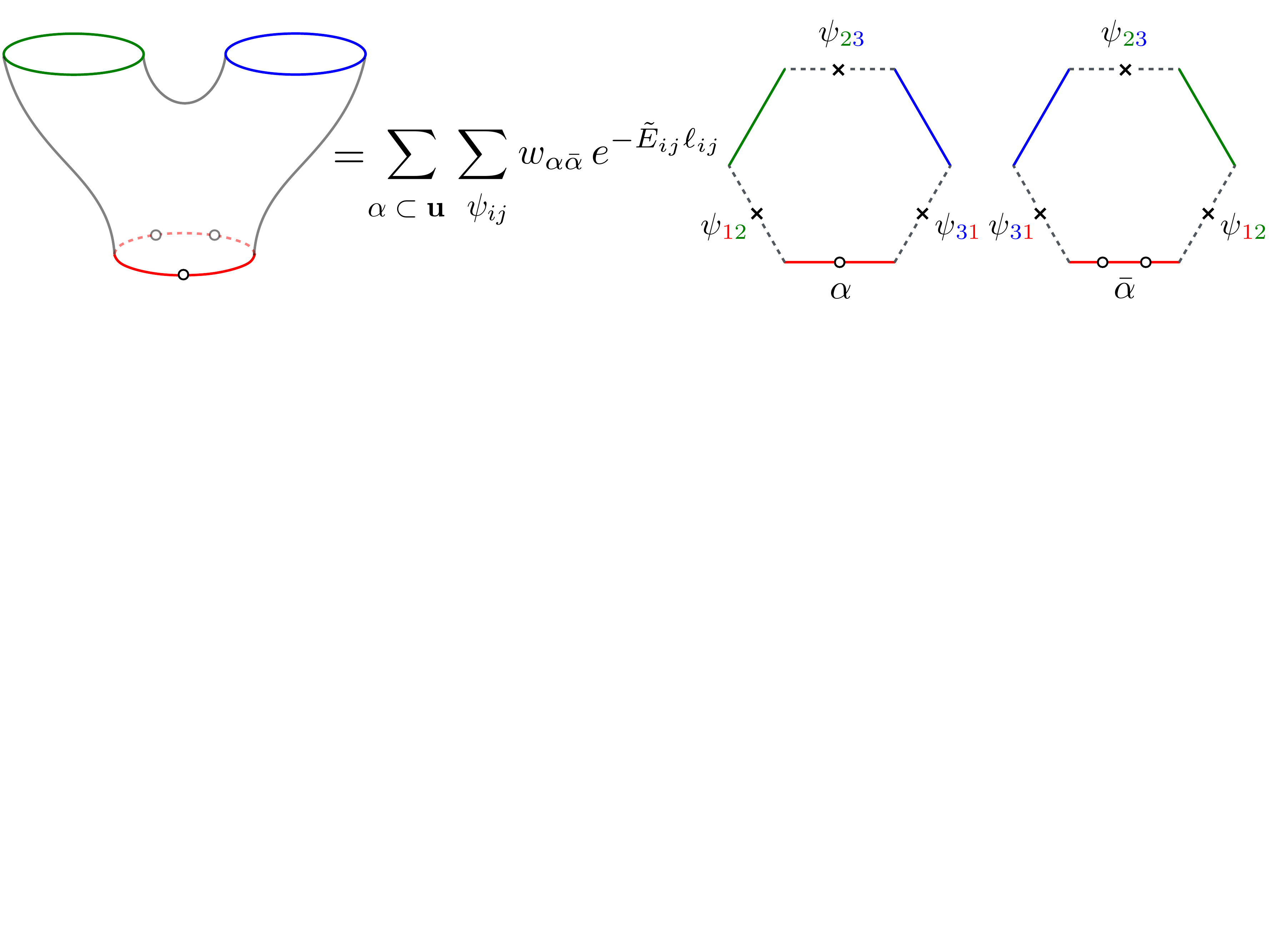}
\end{center}
\vspace{-7cm}
\caption{A pair of pants is made of two hexagons that are stitched together along their three common edges (dashed lines). Each one of the three spin chains lying at its boundary, represented here by the three circles in the left panel, is cut open into two segments, each becoming an edge of one or the other of the two hexagons. A Bethe state is a linear superposition of magnons (seen as dots in the picture), propagating on the spin chain and interacting by means of a pairwise factorized scattering. It is uniquely characterized by the set of magnons' rapidities $\textbf{u}$. The hexagon cutting procedure splits the state into two substates, $|\textbf{u}\mathcal{i} = \sum_{\alpha \cup \bar{\alpha} = \textbf{u}}w_{\alpha\bar{\alpha}} |\alpha\mathcal{i}\otimes |\bar{\alpha}\mathcal{i}$, with each half living on one hexagon and with the two halves being entangled with an appropriate Bethe wave function $w_{\alpha\bar{\alpha}}$, see \cite{short,Escobedo:2010xs}. Gluing the hexagons back together along the dashed line $ij$, stretching between the spin chains $i$ and $j$, amounts to summing over a basis of so called mirror states $\psi_{ij}$ of the Hilbert space connecting these two boundaries. The contribution of a mirror state $\psi_{ij}$ is damped by the factor $\exp{(-\tilde{E}_{ij}\ell_{ij})}$ where $\tilde{E}_{ij}$ is its energy and where $\ell_{ij}$ is the distance separating the (centers of the) two hexagons along the direction perpendicular to the dashed line $ij$.}\label{hexagons}
\end{figure}

In the hexagon approach, the structure constant is viewed as a finite volume correlator of two hexagon operators, as shown in figure \ref{hexagons}. When all the distances are large, the hexagons decouple and can be studied separately. The quantities of interest are then the amplitudes for creation - annihilation of fundamental excitations (magnons) on the six boundaries of an hexagon. They are the so called hexagon form factors.

All the amplitudes can be obtained starting from a configuration where the $M$ magnons all sit together on the same spin chain segment on the hexagon. These form factors were bootstrapped in \cite{short} and argued to take the form
\beq\label{HFF}
\mathcal{h}\mathfrak{h}|\chi_{A_{1}\dot{A}_{1}}(u_{1})\ldots \chi_{A_{M}\dot{A}_{M}}(u_{M})\mathcal{i} = \prod_{i<j} h(u_{i}, u_{j}) \times \mathcal{M}_{A_{1}\dot{A}_{1}\, ,\ldots ,\, A_{M}\dot{A}_{M}}\, ,
\eeq
where $\chi_{A\dot{A}}(u)$ stands for a magnon carrying a rapidity $u$, corresponding to a momentum $p(u)$ and an energy $E(u)$, and a pair $(A, \dot{A})$ of $SU(2|2)_{L}\times SU(2|2)_{R}$ fundamental indices. The dynamical part $h(u, v)$ of the two-particle process is easy to write \cite{short}
\beq\label{huv}
h(u, v) = \frac{x(u^{-})-x(v^{-})}{x(u^{-})-x(v^{+})}\frac{1-1/x(u^{-})x(v^{+})}{1-1/x(u^{+})x(v^{+})} \frac{1}{\sigma(u, v)} = \frac{u-v}{u-v-i} + O(g^2)\, ,
\eeq
in terms of the Zhukowski map $x(u) = (u+\sqrt{u^2-4g^2})/2g = u/g + O(g)$, the shifted rapidities $u^{\pm} = u\pm i/2$ and the BES dressing phase $\sigma(u, v)$ \cite{BES}. It obeys the Watson equation $h(u, v)/h(v, u) = S(u, v)$, where $S(u,v)$ is the dynamical factor of the two-body S-matrix. The matrix part $\mathcal{M}$ preserves a diagonal $PSU(2|2)_{D}$ subgroup inside $SU(2|2)^2$ and was conjectured in \cite{short} to be given by the matrix element of the factorized $SU(2|2)$ S-matrix \cite{Beisert06} with the left and right components of the magnons entering as incoming and outgoing states.

\begin{figure}
\begin{center}
\includegraphics[scale=0.40]{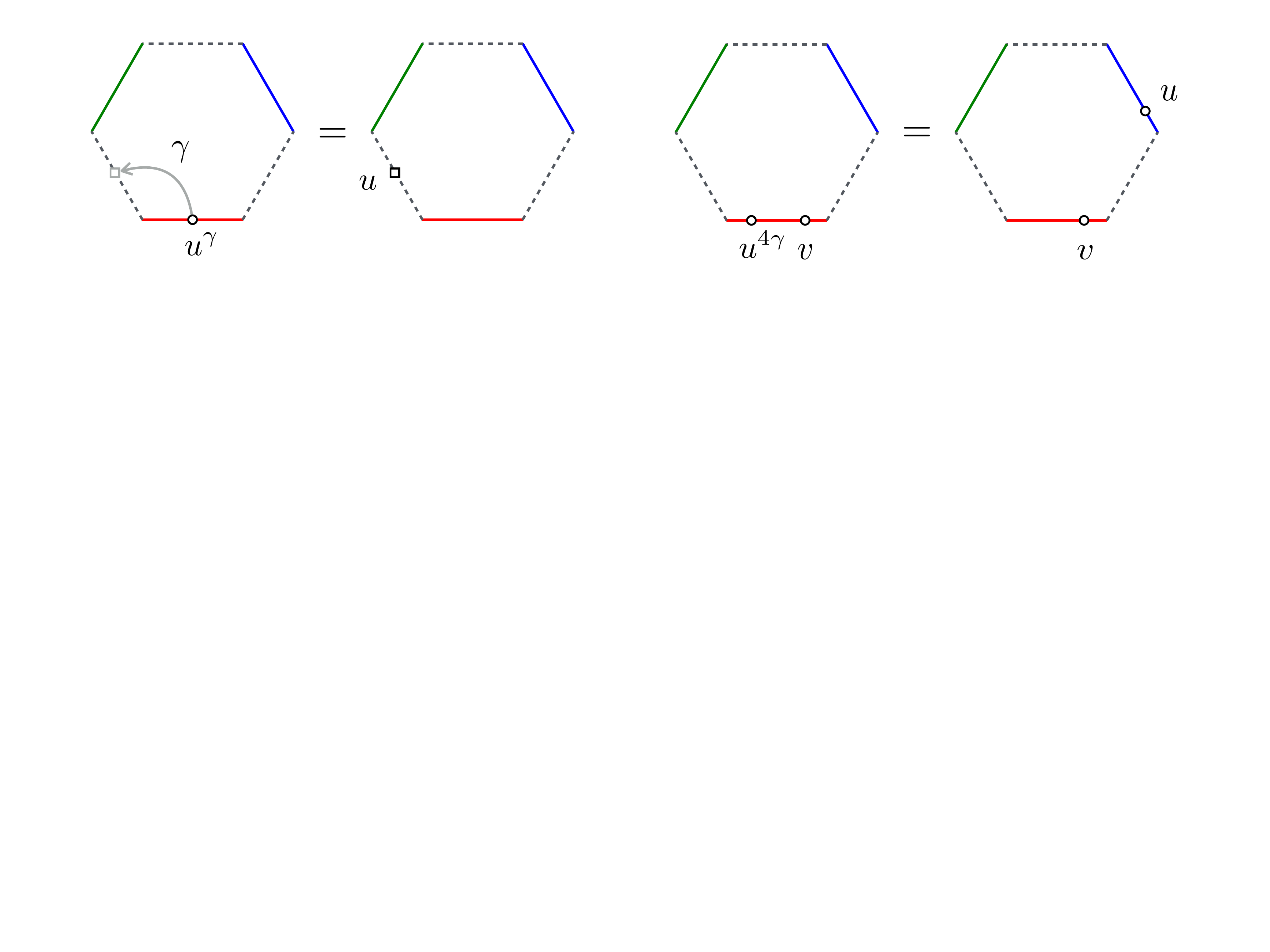}
\end{center}
\vspace{-7.7cm}
\caption{We can move an excitation around by using the mirror rotation $\gamma$. The latter maps rapidities, $\gamma : u\rightarrow \gamma(u) \equiv u^{\gamma}$, from the real to the mirror kinematics. In a relativistic theory, the mirror rotation coincides with an euclidean boost by $90$ degrees, $\gamma : \theta \rightarrow \theta^{\gamma} = \theta + i\pi/2$. For the $\mathcal{N}=4$ spin chain, it consists in analytically continuing the rapidity $u$ to the second sheet of the Zhukowski rapidity $x^{+} = x(u+\frac{i}{2})$, with $x= (u+\sqrt{u^2-4g^2})/2g$. One can move the excitation farther away on the hexagon, through further analytical continuations, as illustrated in the right panel for a $4\gamma = 360^{\circ}$ rotation.}\label{mirror}
\end{figure}

Starting from the canonical configuration (\ref{HFF}), one can reach a generic configuration by transporting some magnons to the other edges of the hexagon. The basic move is the mirror transformation $\gamma$ depicted in figure \ref{mirror}. This is a transformation that swaps the roles of space and time,
\beq
E(u^{\gamma}) = i\tilde{p}(u)\ ,  \qquad p(u^{\gamma}) = i\tilde{E}(u)\, ,
\eeq
{and is such that a magnon in the mirror kinematics has positive energy $\tilde{E}$ and real momentum $\tilde{p}$.}%
\footnote{{The former eigenvalue is conjugate to the spin chain coordinate $\ell$, which is a discrete euclidean time for the mirror magnon, while the latter is an imaginary scaling dimension, conjugate to the continuous coordinate $\sigma$ parameterizing the dilatations. Equivalently, the magnon has wave function  $e^{E\sigma+ip\ell} = e^{-\tilde{E}\ell+i\tilde{p}\sigma}$.}}
These are the states one should sum over when sewing the hexagons together along the mirror edges, as done in figure \ref{hexagons}. {One can also move an excitation farther away by combining several mirror transformations. In particular, one obtains the amplitude for the transition of a magnon from a spin chain edge to another by applying a $4\gamma = 360^{\circ}$ rotation to the rapidity $u$ of the creation amplitude (\ref{huv}), as depicted in the right panel of figure \ref{mirror}. We quote here the result of this analytical continuation,
\beq\label{htrans}
h(u^{4\gamma}, v) = \frac{1}{h(v, u)}\, ,
\eeq
referring the reader to \cite{short} for its derivation and for a thorough discussion of the mirror algebra in general.}

The form factors (\ref{HFF}) completely characterize the hexagon operator. The next step is to join two hexagons together into a single pair of pants. In the following, we recall the net result of the standard gluing procedure. We then comment on the trouble appearing when trying to evaluate this result at wrapping order.

\subsection{The hexagon prescription}

The form factor expansion of the pair of pants amplitude naturally takes the form of an infinite series, with a sum over a basis of states for each one of the three mirror Fock spaces along which the two hexagons overlap. (These are the cuts in figure \ref{hexagons}, labelled by their endpoints, $12, 23, 31$, and referred to as mirror channels in the following). We write it as
\beq\label{bare-exp}
\mathcal{A} = \sum_{n_{ij} \geqslant 0} \mathcal{A}_{(n_{12}, n_{23}, n_{31})}\, ,
\eeq
where the amplitude $\mathcal{A}_{(n_{ij})}$ represents the contribution with $n_{ij}$ mirror magnons in the channel $ij$. The latter amplitude is exponentially suppressed $\sim \exp{(-\tilde{E}_{ij}\ell_{ij})}$, at the integrand level, with $\tilde{E}_{ij}$ the energy of the state and with $\ell_{ij}$ the distance, in spin chain unit, separating the two hexagons along the channel $ij$. Denoting by $L_{1,2,3}$ the lengths of the three operators $\mathcal{O}_{1,2,3}$ at the boundary, we have $\ell_{12} =  (L_{1}+L_{2}-L_{3})/2$ and similarly for $\ell_{23}, \ell_{31}$, by cyclic permutations.

We shall consider the situation where one of the operators, say $\mathcal{O}_{1}$, is non-BPS and carries a single type of excitations (all ``polarized" along the same direction). For definiteness, we draw this operator from the $\mathfrak{sl}(2)$ subsector: $\mathcal{O}_{1}\sim \textrm{tr}\, \mathcal{D}^{M} Z^{L_{1}}$, where $\mathcal{D}$ is a lightcone covariant derivative and $Z$ a complex scalar field. It has Lorentz spin $M$, for the number of magnons $\mathcal{D}$, and R-charge $L_{1}$, for the spin chain length. Its fine structure is encoded in the set of magnons' momenta, or, equivalently, the set of Bethe roots $\textbf{u} = \{u_{1}, \ldots , u_{M}\}$.%
\footnote{We assume that $\mathcal{O}_{1}$ is a conformal primary, implying that all the rapidities are finite, $u_{i}\neq \infty$ for $u_{i}\in \textbf{u}$.}
The periodicity of the chain forces the roots $\textbf{u}$ to be quantized according to the Bethe ansatz equations,
\beq\label{ABA}
e^{ip_{i} L_{1}}\prod_{j\neq i}S_{ij} = 1\, ,
\eeq
where $p_{i} = p(u_{i})$, with $p(u) = -i\log{(x(u^+)/x(u^-))}$ the spin chain momentum, and where $S_{ij} = S(u_{i}, u_{j})$ is the S-matrix in the diagonal $\mathfrak{sl}(2)$ subsector. The roots are also subject to the zero momentum condition, $\prod_{i}e^{ip_{i}} = 1$, arising from the cyclic property of single trace operators. The other two operators, $\mathcal{O}_{2}$ and $\mathcal{O}_{3}$, are BPS and solely characterized by their lengths.

The asymptotic amplitude $\mathcal{A}_{\textrm{asympt}} = \mathcal{A}_{(0, 0, 0)}$ projects to the vacuum state in each mirror channel and thus dominates the sum (\ref{bare-exp}) at large lengths, $\ell_{ij} \gg 1$. It is set to $1$ for three empty spin chains at the boundary, that is when $M=0$ in our set-up. The more general formula is given as a sum over the $2^M$ partitions of the Bethe state $\textbf{u}$. It was first derived in \cite{Vieira:2013wya} at weak coupling using the spin chain tailoring method \cite{Escobedo:2010xs}. Its hexagonal generalization is expected to be valid to all loops for asymptotically large lengths, $\ell_{ij} \gg 1$. It reads
\beq\label{Aasympt}
\mathcal{A}_{\textrm{asympt}} = \sum_{\alpha \cup \bar{\alpha} = \textbf{u}}a_{\alpha\bar{\alpha}} = \sum_{\alpha \cup \bar{\alpha} =  \textbf{u}} (-1)^{|\bar{\alpha}|} \frac{e^{ip_{\bar{\alpha}}\ell_{13}}}{h_{\alpha\bar{\alpha}}}\, ,
\eeq
with $|\bar{\alpha}|$ the number of roots in the subset $\bar{\alpha} \subset \textbf{u}$,
\beq
p_{\bar{\alpha}} = \sum_{u_i\in \bar{\alpha}} p(u_{i})\, , \qquad h_{\alpha\bar{\alpha}} = \prod_{u_{i}\in \alpha, u_{j}\in \bar{\alpha}} h(u_{i}, u_{j})\, ,
\eeq
and where $h(u, v)$ is the hexagon form factor (\ref{huv}). As pictured in figure \ref{hexagons}, this sum directly results from the cutting into two of the Bethe state at the boundary, with each subset of roots falling on one of the two hexagons.

Notice that in our normalization the contribution with $\bar{\alpha} = \varnothing$ in (\ref{Aasympt}) has been set to $1$. This is so because we pulled out the overall factor $\prod_{i<j} h_{ij}$ and absorbed it in the normalization of the wave function, see equation (\ref{asyC}) below. As a result of this operation, the latter wave function is invariant under permutation of any two magnons. The other contributions in (\ref{Aasympt}) are obtained by shifting some magnons from one hexagon to the other, through the bridge of Wick contractions connecting the operators $1$ and $3$. The price to pay for this move is the phase $e^{ip_{\bar{\alpha}}\ell_{31}}$ appearing in (\ref{Aasympt}). The sign $(-1)^{|\bar{\alpha}|}$ has a kinematical origin \cite{Thiago&Shota,Eden:2016xvg}.

Next come the contributions involving mirror magnons, which correct the asymptotic answer when $\ell_{ij}\sim 1$. Their general expression, for the structure constants of interest, was worked out in \cite{3loops}
\beq\label{general}
\begin{aligned}
\mathcal{A}_{(n_{ij})} = \sum_{\alpha}a_{\alpha\bar{\alpha}}\int \frac{d\mu(\textbf{\w}_{123})}{n_{12}!n_{23}!n_{31}!}\frac{e^{i\sum_{ij}p(\textbf{\w}^{\gamma}_{ij})\ell_{ij}}h(\textbf{\w}_{12}^{\gamma}, \alpha)h(\textbf{\w}^{\gamma}_{31}, \bar{\alpha})T(\textbf{\w}_{12}^{\gamma})T(\textbf{\w}_{23}^{-\gamma})T(\textbf{\w}_{31}^{\gamma})}{h(\bar{\alpha}, \textbf{\w}_{12}^{\gamma})h(\alpha, \textbf{\w}_{31}^{\gamma})h(\textbf{u}, \textbf{\w}_{23}^{-\gamma})h(\textbf{\w}^{\gamma}_{12}, \textbf{\w}^{\gamma}_{31})h(\textbf{\w}^{\gamma}_{31}, \textbf{\w}^{\gamma}_{12})}\, .
\end{aligned}
\eeq
{It involves integration over the elements in $\textbf{u}_{123} = \cup_{ij}\textbf{u}_{ij}$, where $\textbf{\w}_{ij} = \{\w_{ij,1}, \ldots , \w_{ij, n_{ij}}\}$ is the set of the rapidities of the $n_{ij}$ magnons in the mirror channel $ij$, and an implicit summation over the bound state label $a \in \mathbb{N}^*$ is understood for each mirror magnon $\left|u^{\gamma}, a\right>$ in the set $\textbf{u}_{123}$. (Magnons in the Bethe state $\textbf{u}  = \alpha\cup \bar{\alpha}$ are fundamental, i.e., $a=1$ for all of them.)} The multi-magnon measure is given by
\beq\label{mu123}
d\mu(\textbf{\w}_{123}) = \prod_{ij}\frac{d\textbf{u}_{ij}}{(2\pi)^{n_{ij}}}\mu(\textbf{u}^{\gamma}_{ij})h(\textbf{u}^{\gamma}_{ij}, \textbf{u}^{\gamma}_{ij})\, ,
\eeq
with the individual measure $\mu_{a}(\w^{\gamma})$ of the magnon $\left|u^{\gamma}, a\right>$ defined by
\beq\label{mu}
\textrm{res}_{u=v} \frac{i\mu_{a}(u^{\gamma})}{h_{ab}(u^{\gamma}, v^{\gamma})} = \delta_{ab}\, .
\eeq
We also adopt the convention that functions of sets are defined as products over the sets' elements, $f(\textbf{u}, \ldots) = \prod_{u_{i}\in \textbf{u}} f_{a_{i}\ldots}(u_{i}, \ldots)$, with the diagonal elements being removed whenever two identical sets appear in the arguments. E.g.,
\beq
\int \frac{d\textbf{\w}_{ij}}{(2\pi)^{n_{ij}}}\mu(\textbf{u}^{\gamma}_{ij})h(\textbf{u}^{\gamma}_{ij}, \textbf{u}^{\gamma}_{ij}) = \prod_{k=1}^{n_{ij}}\sum_{a_{k}=1}^{\infty}\int \frac{du_{ij, k}}{2\pi} \prod_{l=1}^{n_{ij}}\mu_{a_l}(u^{\gamma}_{ij, l}) \prod_{\substack{m, n = 1\\ m\neq n}}^{n_{ij}}h_{a_{m}a_{n}}(u^{\gamma}_{ij, m}, u^{\gamma}_{ij, n})\, . 
\eeq
As a departure from the conventions of \cite{short,3loops} we work here with the forward transfer matrix,
\beq\label{Tbegin}
T_{a}(u) = \textrm{tr}_{a}\, \mathcal{S}_{a1}(u, \textbf{u})\, ,
\eeq
with $\mathcal{S}_{a1}(u, \textbf{u})$ the left factor of the scattering matrix $\mathbb{S}_{a1} = S_{a1} \mathcal{S}_{a1}\dot{\mathcal{S}}_{a1}$ between a bound state $|u, a\mathcal{i}$ and the Bethe state $\textbf{u}$ and with the trace running over the $4a$ left degrees of freedom of the bound state.%
\footnote{One could equivalently work with the right components, since only left-right symmetric Bethe states appear in the OPE of two chiral primaries \cite{Basso:2017khq}.} This transfer matrix can be related to the backward transfer matrix used in \cite{short,3loops} by a crossing transformation or by complex conjugation, see Appendix \ref{AFF}. 

Finally, the normalized structure constant $C^{\bullet\circ\circ}_{123}/C^{\circ\circ\circ}_{123}$ is obtained by dividing (\ref{Aasympt}) by the norm of the (asymptotic) Bethe wave function of the state $\textbf{u}$. Namely,
\beq\label{asyC}
\left(C^{\bullet\circ\circ}_{123}/C^{\circ\circ\circ}_{123}\right)^2 = \frac{d\mu(\textbf{u})}{d\textbf{m}} \times \mathcal{A}^2 = \frac{\mu(\textbf{u})h(\textbf{u}, \textbf{u})}{G_{1}} \times \mathcal{A}^2\, ,
\eeq
where $\textbf{m}$ are the mode numbers of the on-shell state $\textbf{u}$ and where $d\mu(\textbf{u})/d\textbf{m}$ is the ratio of the real counterpart of the multi-magnon measure~(\ref{mu123}),
\beq
(2\pi)^{M}\frac{d\mu(\textbf{u})}{d\textbf{u}} = \prod_{i=1}^{M}\mu(u_{i}) \prod_{\substack{j, k=1\\ j\neq k}}^{M} h(u_{j}, u_{k})\, ,
\eeq
and of the square of the Gaudin norm,
\beq\label{G1}
G_{1} = (2\pi)^M \frac{d\textbf{m}}{d\textbf{u}} = \textrm{det}\, \{L_{1}\partial_{u_{i}}p_{j} -i\partial_{u_{i}}\sum_{k\neq j}\log{S(u_{j}, u_{k})}\}\, .
\eeq

\subsection{Divergences at wrapping order}

The first virtual corrections, $\mathcal{A}_{(0, 1, 0)}, \mathcal{A}_{(1, 0, 0)}$ and $\mathcal{A}_{(0, 0, 1)}$, were analyzed in \cite{short,Eden:2015ija,3loops}. They correspond to processes where an excitation is created on one side of the three point function, transported to the other side, through one of the three mirror channels, and immediately destroyed after that. The excitation has no time to wind around the operators lying at the boundary. For that to happen, one necessarily needs at least two of the three mirror channels to be excited simultaneously. The simplest configurations in this class correspond to $\mathcal{A}_{(1, 0, 1)}, \mathcal{A}_{(1, 1, 0)}$ and $\mathcal{A}_{(0, 1, 1)}$ and they are the ones we will be mostly considering in this paper.

In fact, the amplitude that is the most relevant to our study is the one for which the magnon is giving a chance to wrap around the excited operator. We reproduce it here, for the reader's convenience,
\beq\label{101}
\begin{aligned}
\mathcal{A}_{(1, 0, 1)} = \sum_{a, b\geqslant 1}\int \frac{du dv}{(2\pi)^2} &\frac{\mu_{a}(u^{\gamma})\mu_{b}(v^{\gamma})T_{a}(u^{\gamma})T_{b}(v^{\gamma})}{h_{ab}(u^{\gamma}, v^{\gamma})h_{ba}(v^{\gamma}, u^{\gamma})}e^{ip_{a}(u^{\gamma})\ell_{12}+ip_{b}(v^{\gamma})\ell_{31}} \\
&\qquad \times \sum_{\alpha \cup \bar{\alpha} = \textbf{u}} a_{\alpha\bar{\alpha}}\frac{h_{a1}(u^{\gamma}, \alpha)h_{b1}(v^{\gamma}, \bar{\alpha})}{h_{1a}(\bar{\alpha}, u^{\gamma})h_{1b}(\alpha, v^{\gamma})}\, .
\end{aligned}
\eeq
The two remaining amplitudes, in which the two mirror magnons are surrounding one of the two BPS operators, do not really bring anything new. Indeed, as one can see from (\ref{general}), their integrands factorize into those for the constitutive individual channels. This factorization is a result of a supersymmetric cancellation and it is indicating that one cannot really wrap anything around a BPS operator. Furthermore, these two amplitudes appear to be subleading at weak coupling; they were estimated in \cite{short} to start contributing at five loops, at the earliest. So, in principle, truncating the series of mirror corrections at (\ref{101}) should be enough to obtain (\ref{QFT}) at four loops. The problem is that (\ref{101}) does not really make sense, because the mirror hexagon transition $1/h_{ab}(u^{\gamma}, v^{\gamma})$ has a pole at $u=v$ when $b=a$. Note that it is that same pole which gave us, early on, the integration measure for a single magnon, see equation (\ref{mu}). {(The pole is manifest in the transition (\ref{htrans}), since $h(u, v)$ vanishes at $u=v$, see equation (\ref{huv}). Here, we are facing it, as well as its consequences, in the mirror kinematics, $\{u, v\} \rightarrow \{u^{\gamma}, v^{\gamma}\}$.)}

It is tempting to try to solve the problem directly by using some $i0$ prescription. However, that alone turns out to be insufficient; the $i0$ version of the integral does not pass all the checks and e.g.~does not yield to an agreement with the field theory result (\ref{QFT}).
Though the $i0$ regularized amplitude arguably captures a big chunk of the answer (most of the two-magnon phase space is covered by its two-fold integral), it is not the full answer. The singularity that we are seeing here really is the tip of an iceberg of wrapping corrections, which call for a more careful treatment of the hexagon gluing procedure.

To find a more sensible way of addressing the problem, it helps remembering why we got a pole in the first place. The latter is a built-in kinematical singularity and is an essential part of the bootstrap axioms for the hexagon form factors~\cite{short}; it is a common axiom for the class of (non-local) twist operators the hexagon belongs to, see \cite{Cardy:2007mb} and \cite{Basso:2013vsa,Basso:2013aha,Basso:2014jfa} for other examples. It expresses the fact that a ``magnon-antimagnon" pair, with no overall charges, decouples from the bulk of the hexagon geometry. We shall review this link in more detail in the next section. What we want to stress here and illustrate on (\ref{101}) are its main consequences for the divergences of $\mathcal{A}$.

\begin{figure}
\begin{center}
\includegraphics[scale=0.40]{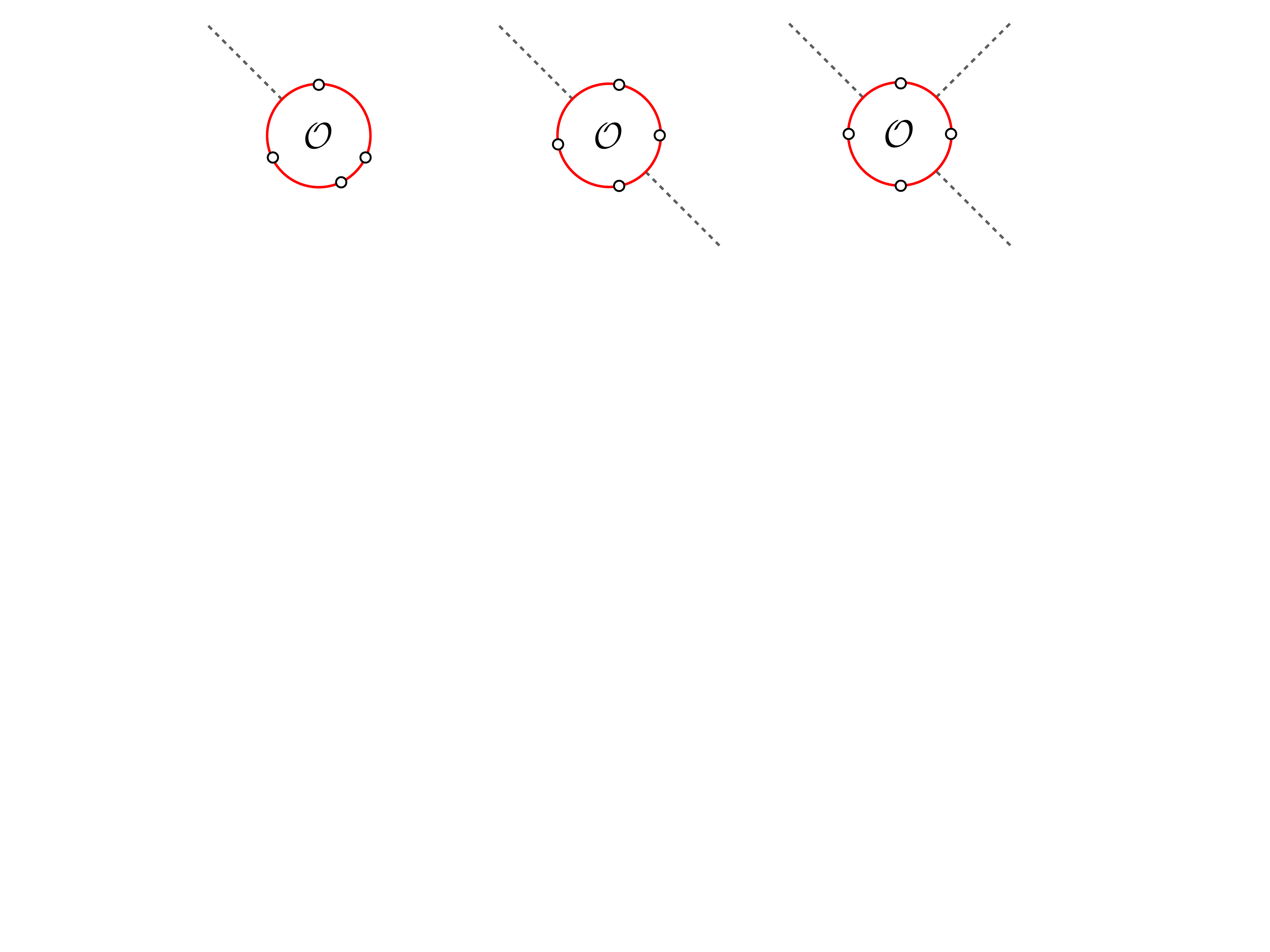}
\end{center}
\vspace{-8cm}
\caption{The divergences triggered by the wrapping effects are local: they depend on the operator $\mathcal{O}$ but not on the details of the geometry it is inserted in, like e.g.~the number of times we cut the operator or the way we distribute the magnons on the various subchains. For instance, the three inequivalent cuttings of the operator shown here, corresponding, respectively, to the insertion of the operator in a 2-, 3- and 4-pt function, should develop the same divergences at wrapping order. If we canonically normalize the 2-pt function, then the divergences should cancel out in all the higher point functions.}\label{cutting}
\end{figure}

Namely, the decoupling condition endows the divergences with some universal factorization properties. Setting $v=u$ in (\ref{101}) and stripping out the pole,
\beq\label{singular}
\frac{\mu_{a}(u^{\gamma})\mu_{b}(v^{\gamma})}{h_{ab}(u^{\gamma}, v^{\gamma})h_{ba}(v^{\gamma}, u^{\gamma})} \sim \frac{\delta_{ab}}{(u-v)^2}\, ,
\eeq
reveals, indeed, that the residue factorizes into two parts,
\beq\label{residue}
e^{ip_{a}(u^{\gamma})L_{1}}T_{a}(u^{\gamma})^2\frac{h_{a1}(u^{\gamma}, \textbf{u})}{h_{1a}(\textbf{u}, u^{\gamma})}\sum_{\alpha \cup \bar{\alpha} = \textbf{u}} a_{\alpha\bar{\alpha}} = Y_{a}(u^{\gamma})\times\mathcal{A}_{\textrm{asympt}} \, ,
\eeq
with all the dependence on the partition of the Bethe state and on the details of the geometry (e.g. the way the length is distributed among the mirror channels or the dynamical information about of the hexagons) being captured by the asymptotic amplitude. The remaining factor
\beq\label{Yfirst}
Y_{a}(u^{\gamma}) := e^{-\tilde{E}_{a}(u)L_{1}}S_{a1}(u^{\gamma}, \textbf{u})T_{a}(u^{\gamma})^2\, ,
\eeq
which we simplified using the Watson equation $h_{a1}(u^{\gamma}, \textbf{u}) = S_{a1}(u^{\gamma}, \textbf{u})h_{1a}(\textbf{u}, u^{\gamma})$, is, on the contrary, only sensitive to the ``local" properties of the operator which is being probed, see figure \ref{cutting}. This factorization is universal in the sense that it applies regardless of the shape of the geometry. For more complicated backgrounds or processes, only the prefactor $\mathcal{A}_{\textrm{asympt}}$ in (\ref{residue}) will differ and change accordingly.%
\footnote{This is so for non-extremal and planar geometries. Additionnal divergences are expected at the non-planar level or for extremal configurations. In the latter case, they should indicate the need to take into account the mixing with double traces.}
This is readily verified at the level of the general formula (\ref{general}).

These features clearly point towards the resolution of the problem: the divergences are property of the state $\textbf{u}$, not of the pair of pants itself, and they factor out of the amplitudes. Schematically,
\beq
\mathcal{A} = \sqrt{Z(\textbf{u})}\times (\mathcal{A}_{\textrm{asympt}} + \ldots )\, ,
\eeq
where the terms in brackets is whatever finite piece remains after the divergent (state-dependent) factor
\beq\label{Zu}
Z(\textbf{u}) = 1+\infty\times \sum_{a}\int Y_{a} +\ldots \, ,
\eeq
has been stripped out. The main goal of this paper is to make these formulae precise. Before getting into that, let us point out yet another important corollary of the decoupling condition (and of the Watson equation) which relates to the interpretation of the ``properly normalized" residue (\ref{Yfirst}). Namely, the latter ought to be same as the amplitude associated to a magnon wrapping the Bethe state on the cylinder (that is, on the two-point function geometry). One easily verifies that, indeed, the residue (\ref{Yfirst}) is the ``$Y$ function" that controls the wrapping correction to the scaling dimension \cite{Bajnok:2008bm,Gromov:2009tv}. It is here merely truncated to its leading wrapping form, since the mirror magnon is winding only once around the operator. This is in line with the claimed property that the divergences (\ref{Zu}) only care about what is happening far down the leg of the pair of pants.

These few basic facts about the divergences plaguing the hexagon series are all we need to know to make sense of them and renormalize them away. They also show that one cannot easily amend the hexagon form factors, such as e.g.~remove their poles, without taking the risk of polluting the geometry with some artificial features. This is why we shall follow a different route: we shall excise the boundary part of the geometry that is problematic and absorb it into the definition of the operator insertion.

\section{Regularized wrapping procedure}\label{Sect3}

In this section, we explain how to regularize the divergences of the hexagonal formula for the three-point function and give a proper meaning to (\ref{Zu}). We shall also review and rederive some important facts about the wrapping corrections on the cylinder, which we will need later on.

\subsection{The regularized octagon approach}\label{Gluing}

\begin{figure}
\begin{center}
\includegraphics[scale=0.40]{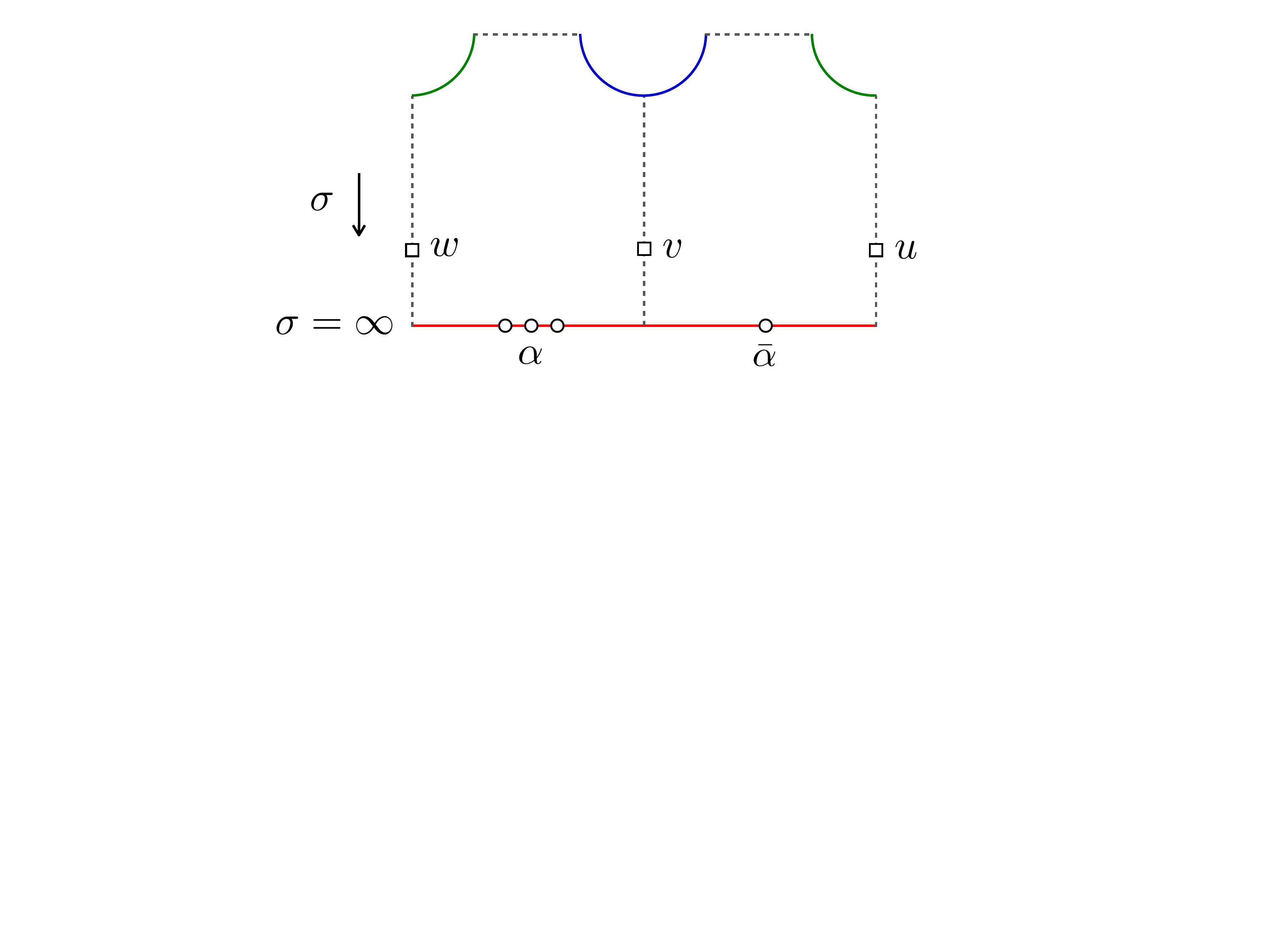}
\end{center}
\vspace{-6.7cm}
\caption{We obtain an octagon by stitching two hexagons together along one mirror edge. We then consider the octagon transition between an incoming mirror magnon $w$ in the far left and an outgoing mirror magnon $u$ in the far right. This transition develops a simple pole at $w = u$, as soon as there is a mirror magnon $v$ in the intermediate mirror channel to mediate the interaction from one hexagon to the other. This pole results from two basic facts: 1) an excitation traveling close to the bottom boundary decouples from the bulk of the geometry and moves freely, up to the diffractionless scattering with the roots $\textbf{u} = \alpha\cup \bar{\alpha}$, 2) the distance to the boundary along the mirror direction $\sigma$ is infinite.
}\label{oct1}
\end{figure}

To isolate the problem with the stitching recipe, we shall glue the patches together, slowly but surely, by going through a two-step procedure. In the first step, we consider the much simpler problem of gluing two hexagons along a single mirror channel, which we choose to be the channel $31$. It can be thought of as a partially decompactified structure constant in which one bridge remains finite, $\ell_{12} = \ell_{23} = \infty$ and $\ell_{31} = O(1)$. The resulting object is depicted in figure \ref{oct1} and it has the shape of an octagon. A state $\textbf{u}$ is placed on the spin chain at the bottom of the picture and split into two subsets of roots, $\alpha \cup \bar{\alpha} = \textbf{u}$, like for the structure constant. For convenience, we view this state as being part of the definition of the first and second hexagons, $\mathcal{H}_{\alpha}$ and $\mathcal{H}_{\bar{\alpha}}$. The octagon under study is then represented by the octagon operator
\beq\label{octagon}
\mathcal{O}_{\alpha\bar{\alpha}} = \mathcal{H}_{\bar{\alpha}}(\ell_{31})\mathcal{H}_{\alpha}(0)\, ,
\eeq
where we use the Heisenberg picture, with the first operator being centered around the origin $0$ and the second one being shifted to the euclidean time $\ell_{31}$,
\beq
\mathcal{H}_{\bar{\alpha}}(\ell_{31}) = e^{\tilde{H}\ell_{31}}\mathcal{H}_{\bar{\alpha}}\, e^{-\tilde{H}\ell_{31}}\, ,
\eeq
with $\tilde{H}$ the Hamiltonian of the mirror theory. In the second step, we wrap the octagon on the pair of pants by putting the mirror theory on a circle of length $L_{1}$ and tracing over its Hilbert space. It leads to
\beq\label{thermal}
\sum_{\alpha}a_{\alpha\bar{\alpha}}\, \textrm{tr}\, \{e^{-\tilde{H}L_{1}} \mathcal{O}_{\alpha\bar{\alpha}}\}\, ,
\eeq
after reinstating the sum over the partitions. This second step gives us back the infinite series of contributions to the un-normalized structure constant (\ref{bare-exp}). More precisely, the trace only captures the contributions with $n_{23} = 0$, since we are not gluing the two mirror edges at the top of the octagon in figure \ref{oct1}. We will not need to include the missing terms here, since, as alluded to before, magnons in the channel $23$ do not source wrapping divergences when the operators $2$ and $3$ are BPS. As we will now argue, the divergences entirely come from the trace in (\ref{thermal}), if the product of the two hexagons defining the octagon (\ref{octagon}) is taken appropriately. Said differently, we can push the divergences to the last step in the process, that is the identification of the outer boundary of the octagon. More is actually true and the general statement is that there is no real issue in gluing hexagons together in a linear sequence.

For the sake of clarity, we shall consider the situation where a single magnon is flowing through the hexagons. So we add one magnon with rapidity $w$ in the far left of the octagon, one with rapidity $u$ in the far right and we restrict the sum over intermediate states inbetween the two hexagons in (\ref{octagon}) to the one-magnon subsector, $1 \rightarrow \int d\mu(v^{\gamma}) |v^{\gamma}\mathcal{i}\mathcal{h}v^{\gamma}|$; see figure \ref{oct1}. The latter one-magnon states are the lightest excited states which can mediate the singularity across the octagon and, as such, are the first ones to consider. They are also the only states that we need to cover the case of the amplitude (\ref{101}). Finally, to simplify the exposition, we suppress all the dependence on the flavor degrees of freedom and ignore the summation over the bound states. In sum, the octagon amplitude that we are describing is given by
\beq\label{hhoct}
\int \frac{dv}{2\pi}\frac{\mu(v^{\gamma})}{h(w^{\gamma}, v^{\gamma})h(v^{\gamma}, u^{\gamma})}\times e^{-\ell_{31}(\tilde{E}(v)-\tilde{E}(u))}\sum_{\alpha}a_{\alpha\bar{\alpha}}\, \frac{h(w^{\gamma}, \alpha)h(v^{\gamma}, \bar{\alpha})}{h(\alpha, v^{\gamma})h(\bar{\alpha}, u^{\gamma})} \, .
\eeq
As written, the integral is of course ill defined, due to the poles in the hexagon transitions connecting $w$ to $u$ and $v$ to $u$. The important difference with the divergence encountered in (\ref{101}) is that, here, we are entitled to use the $i0$ prescription to get rid of the problem. Indeed, the procedure we are discussing should be identical to the one designed to glue pentagons together in the OPE approach to null polygonal Wilson loops \cite{Basso:2013aha}, which comes equipped with such an $i0$ prescription.

Let us review the main argument here. It traces back to the kinematical origin of the pole, which we sketched before. Namely, the pole in the transition $1/h \sim \mathcal{h}u^{\gamma}|\mathcal{H}(0)|v^{\gamma}\mathcal{i}$ comes from the fact that the hexagon is defined in infinite volume. It is therefore possible for a particle to propagate arbitrarily far away from the core of the hexagon. When it happens, the particle breaks free and goes straight from the asymptotic past to the asymptotic future. The sole remaining effect of the geometry on the decoupled excitation is to set a bound to the domain the excitation can freely propagate through. Put differently, there is a point along the mirror space direction $\sigma$, which we choose to be $\sigma=0$, where the curvature can no longer be neglected. Hence, we get the free contribution to the hexagon form factor $\mathcal{h}u^{\gamma}|\mathcal{H}(0)|v^{\gamma}\mathcal{i}$ by overlapping an incoming and an outgoing free wave over the half infinite interval $\sigma \in [0, \infty]$,
\beq\label{poles}
\int_{0}^{\infty}d\sigma \, e^{i\sigma (\tilde{p}(v)-\tilde{p}(u)+i0)} = \frac{i}{\tilde{p}(v)-\tilde{p}(u)+i0}\, .
\eeq
This contribution dominates all the others in the diagonal limit $v\rightarrow u$, since nowhere else is the overlap expected to produce large effects. Note that this is a manifestation of the non-local nature of the hexagon operator. For a local operator, we would also get a contribution from $\sigma \in [-\infty, 0]$, which would be the complex conjugate of (\ref{poles}). The overlap would be complete and we would find the familiar disconnected contribution $\mathcal{h}u^{\gamma}|v^{\gamma}\mathcal{i} \sim 2\pi\delta(u-v)$ to the form factor. On the hexagon, the two particles $u$ and $v$ can only freely face each other in the region integrated in (\ref{poles}) and nowhere else. This is the reason why we are getting a pole, with a residue that relates to the infinite volume normalization of the one-particle states, as prescribed in (\ref{mu}).

The representation (\ref{poles}) also makes clear why the integral in (\ref{hhoct}) should be handled with an $i0$ prescription. This choice simply results from the Fourier transform of the step function and it guarantees that the kinematical singularity is the same for the hexagon and the octagon (or for any polygon), since the shifted polar part $i/(v-u+i0)$ is invariant under convolution,
\beq\label{convol}
-\int \frac{dv}{2\pi} \frac{1}{(w-v+i0)(v-u+i0)} = \frac{i}{w-u+i0}\, .
\eeq
Hence, the nature of the singularity does not depend on how many times we cut the spin chain at the boundary, as shown in figure \ref{cutting}.

The normalized residue, which we met in the previous section, is found by looking at what happens at the very bottom of the octagon. It arises from the interaction of the mirror magnon $u=v=w$ with the state $\textbf{u}$ at $\sigma = \infty$. (Though the mirror magnon can decouple from the bulk of the geometry by travelling close to the boundary, there is no way it cannot feel the presence of the state $\textbf{u}$ that is sitting there.) Thanks to integrability, the mirror excitation merely passes through the state and picks up the scattering phases. In the non-diagonal case, we get a monodromy matrix describing the scattering between the free magnon $u=v=w$ and the Bethe state $\textbf{u}$. After taking the trace, it leads to the $Y$ function which we met in the residue of (\ref{101}) -- up to the factor $\exp{(-\tilde{E}(u)L_{1})}$ which we have stripped out in (\ref{thermal}).

Now comes the true problem. Namely, since the octagon transition has a pole at $w=u$,
\beq\label{Odiv}
\mathcal{h}u^{\gamma}|e^{-\tilde{H}L_{1}}\mathcal{O}_{\alpha\bar{\alpha}}|w^{\gamma}\mathcal{i} \sim \frac{iY(u^{\gamma})}{\mu(u^{\gamma})(w-u+i0)} \times \mathcal{h}0|\mathcal{O}_{\alpha\bar{\alpha}}|0\mathcal{i}\, ,
\eeq
it is not immediately clear what it means to wrap the octagon around the pair of pants. Put differently, we cannot take the trace in (\ref{thermal}) without first clarifying what we mean by the diagonal limit $w\rightarrow u$ of the octagon transition.%
\footnote{Note that the $i0$'s in (\ref{convol}) does not help when $w=u$, since the poles at $v = w+i0$ and the one at $v=u-i0$ pinch the contour of integration, in the diagonal limit.} The same difficulty is encountered in the low temperature expansion of the vevs of local operators, see~\cite{Leclair:1999ys,Saleur:1999hq,Pozsgay:2007gx,Pozsgay:2009pv,Pozsgay:2010cr}. The only difference is that, in the latter case, one tries to make sense of the disconnected part $\sim \delta(u-v)$ while, here, we are dealing with a pole (and there is no disconnected component sensu stricto).%
\footnote{Note that there are also ambiguities at defining the diagonal limit of the \textit{connected} form factors of a local operator, when more than one particles are decoupling at the same type. See \cite{Pozsgay:2009pv,Pozsgay:2010cr} and references therein. This problem is relevant for the higher wrapping corrections.} Except for that, the two problems are the same and they can be regularized in the same way, namely, by putting the system in finite volume : $\sigma \leqslant \sigma_{\textrm{cut-off}} = R/2 \gg 1$. In finite volume, there is not enough room for the overlap of the two plane waves in (\ref{poles}) to develop a divergence. The half delta function is regularized and replaced by the cut off $R$,
\beq\label{regularization}
\int_{0}^{R/2} d\sigma \, e^{i\sigma(\tilde{p}(w)-\tilde{p}(u))} = \frac{e^{i(\tilde{p}(w)-\tilde{p}(u))R/2}-1}{i(\tilde{p}(w)-\tilde{p}(u))} \rightarrow R/2\, ,
\eeq
where the last expression is obtained for the diagonal case, $w=u$. We can view this equation as an effective way of disposing of the pole (\ref{singular}) in the amplitude (\ref{101}) or equivalently in (\ref{Odiv}). Namely, after taking into account the Jacobian for converting between rapidity and momentum, it gives us the regularized version of the amplitude (\ref{101}) as
\beq\label{101r}
\mathcal{A}_{(1, 0, 1)} (R) = \frac{R}{2}\mathcal{A}_{\textrm{asympt}}\sum_{a\geqslant 1}\int \frac{d\tilde{p}_{a}(u)}{2\pi} Y_{a}(u^{\gamma}) + O(1)\, .
\eeq
As we will now explain, this expression matches precisely with the expected normalization factor correcting the norm of the state.

\subsection{Wave function renormalization}\label{Luescher}

\begin{figure}
\begin{center}
\includegraphics[scale=0.40]{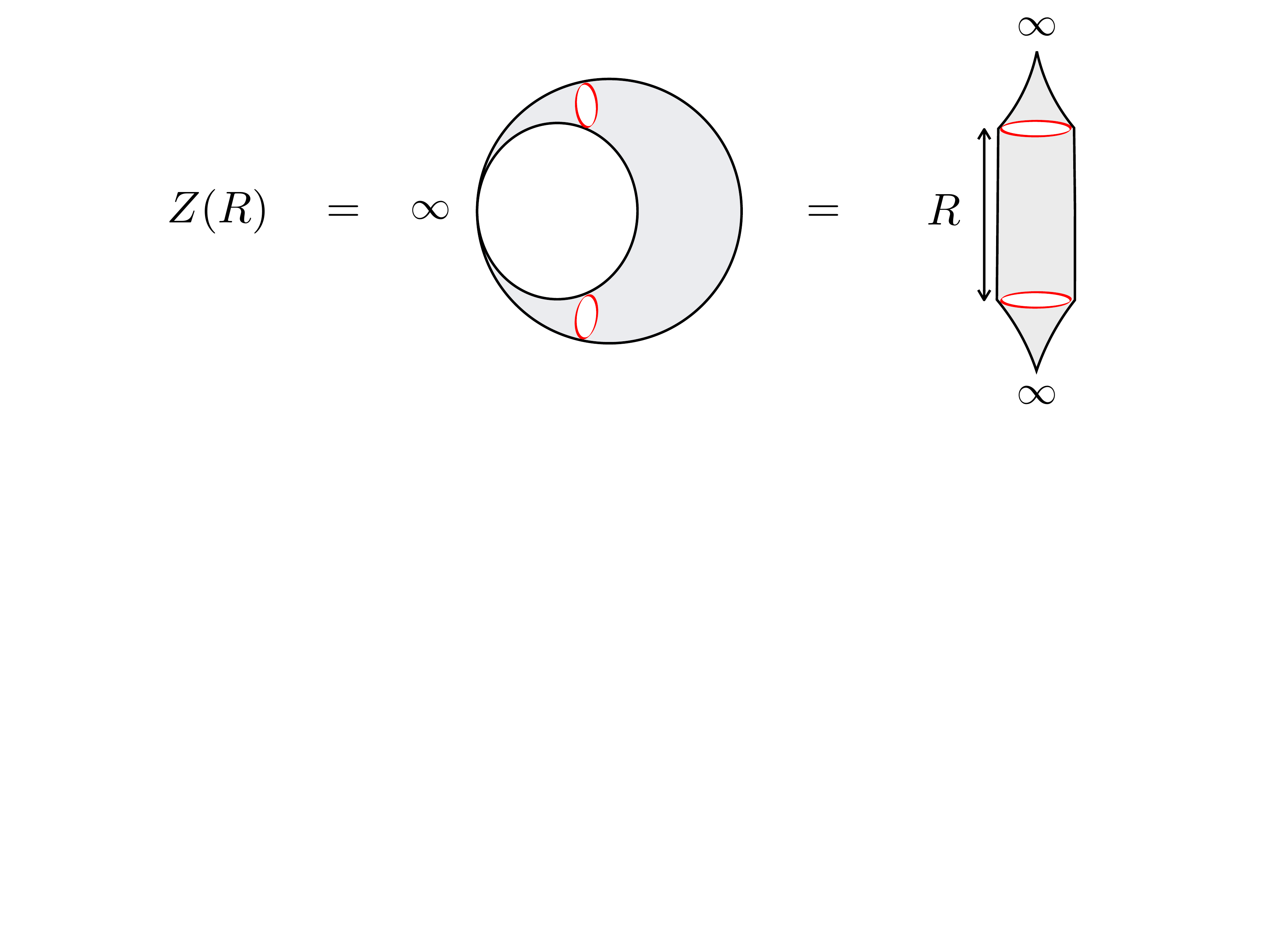}
\end{center}
\vspace{-6cm}
\caption{We identify the wave function normalization factor $Z$ with the excited state partition function. The latter is computed for a pinched torus of length $R\gg 1$ and period $L_{1}$, with the pinching being such that only the Bethe state of interest can propagate.}\label{torus}
\end{figure}

The second main ingredient in the hexagon formula (\ref{asyC}) accounts for the normalization of the wave function. This one too must be corrected due to the wrapping effects. For the consistency of our approach, it must be true that these effects leads to divergences which, after regularization, cancel those coming from the pair of pants amplitude. This is actually granted by the factorization property mentioned earlier, assuming we have the right to identify the overall normalization factor $Z$ of the Bethe wave function with the excited state partition function on a torus of mirror size $R\gg 1$. The latter is the natural regularization of the two-point function amplitude, that is the amplitude associated to a cylinder with the Bethe state $\textbf{u}$ on the past and future boundaries, see figure \ref{torus}.

\begin{figure}
\begin{center}
\includegraphics[scale=0.37]{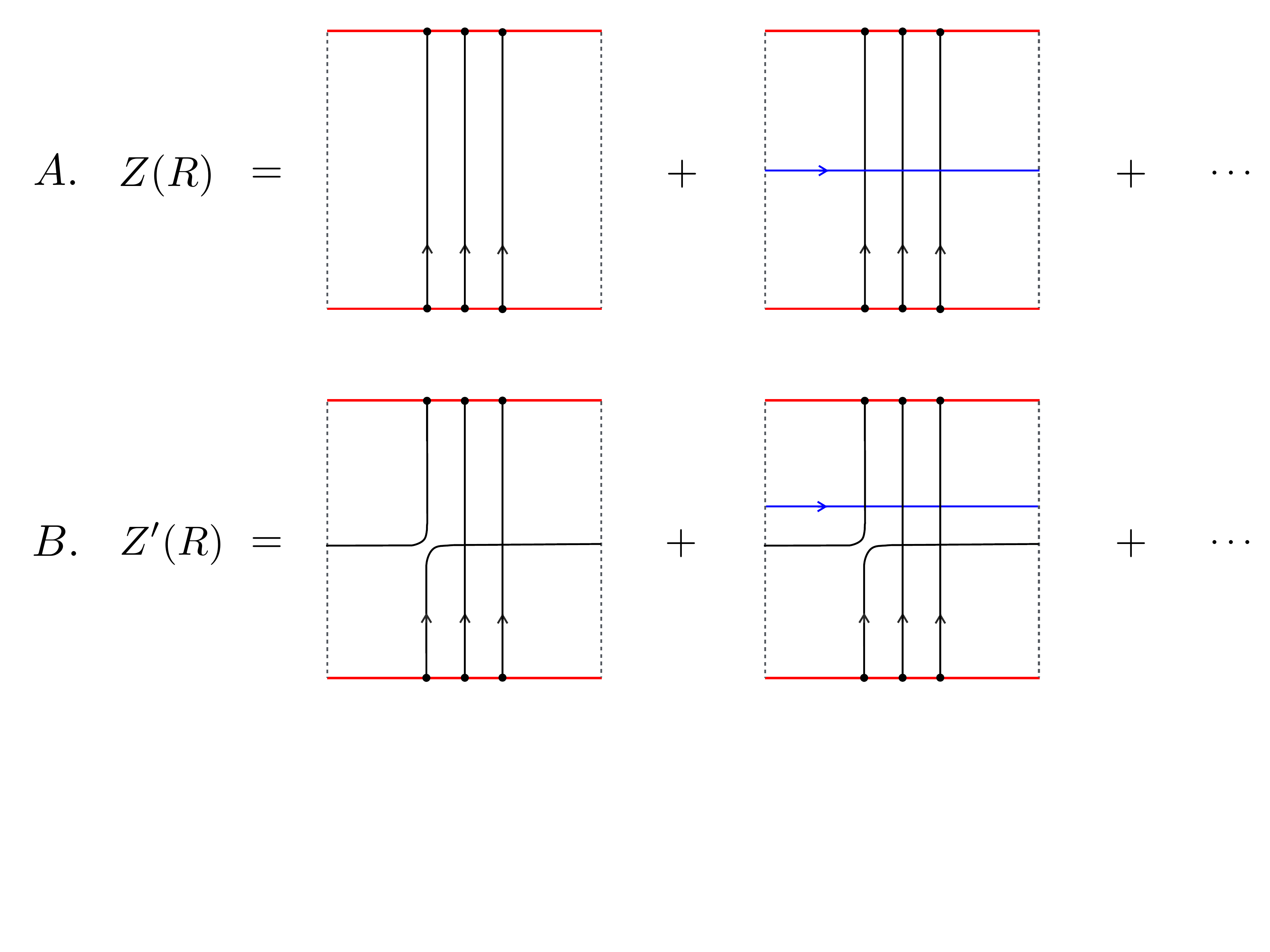}
\end{center}
\vspace{-2.3cm}
\caption{Low temperature $1/L_{1} \ll 1$ expansion of the excited state partition function. The starting point is a square of real size $L_{1}\gg 1$ (solid top and bottom lines) and mirror size $R\gg 1$ (dashed left and right lines). It supports a Bethe state $\textbf{u}$ represented by a bundle of vertical lines stretching from the bottom to the top. We wrap the square around the cylinder by identifying the two parallel mirror edges. It results in a trace over a complete basis of mirror Bethe states (of the cut off mirror Hilbert space). The leading low temperature contribution comes from the mirror vacuum (empty dashed mirror lines). Next comes the first wrapping correction, with one mirror magnon going around the thermal circle. We must sum over the magnon's flavors and integrate over its momentum with a proper $R$ dependent measure. The dots stand for the infinite tail of multi-particle mirror states. In $A.$ the real magnon lines all go straight from the bottom to the top. In $B.$ one of the magnon winds once around the world before reaching the top. The two processes are indistinguishable for periodic Bethe states, resulting in the Bethe ansatz equations $Z(R) = Z'(R)$.}\label{ZR}
\end{figure}

More explicitely, we form a cylinder by gluing the mirror edges of a square amplitude as shown in panel $A.$ of the figure \ref{ZR}. This turns the cylinder into an infinite sum over mirror states or, using the usual $90^\circ$ rotation trick \cite{Zamolodchikov:1989cf}, into a thermal trace over the Hilbert space of the mirror theory. The corresponding partition function $Z$ starts with $1$, in our normalization, for the mirror vacuum, which dominates at low temperature $1/L_{1} \ll 1$. Then comes the contribution of a single mirror magnon, which wraps around the thermal circle. It is controlled by the $Y$ function (\ref{Yfirst}), since we must bring the mirror magnon through the state $\textbf{u}$, sum over the flavors and add the usual damping factor with the energy. Equivalently, the ``$\Box$" amplitude for a single magnon is
\beq
\mathcal{h}\tilde{p}_a (v)|e^{-\tilde{H}\tilde{L}_{1}}\, \Box \, |\tilde{p}_{a}(u)\mathcal{i} = 2\pi Y_{a}(u^{\gamma}) \delta(\tilde{p}(u)-\tilde{p}(v))\, ,
\eeq
after tracing over the flavors. To get the contribution to the partition function for all the one-particle states, we must of course sum over the momentum $\tilde{p}$ of the mirror magnon. As for the pair of pants, this is where the divergences lie, since we cannot really trace over $\delta$ function normalized states. (More generally, one cannot really define the partition function on a non-compact manifold, due to the bulk contributions to the free energy.) The natural exit in thermodynamics is to introduce some discreteness by putting the system in a large but finite volume $R$. So, we assume that the cylinder is actually a long torus of length $R \gg 1$, with periodic boundary conditions at the boundary, and write down the Bethe ansatz equations for the quantized mirror momenta. For a single mirror magnon, there is no interaction and thus $R\tilde{p} = 2\pi n$ where $n\in \mathbb{Z}$. Hence, the first wrapping correction to the partition function is
\beq\label{Rz}
\delta \log{Z(R)} = \sum_{n\in \mathbb{Z}} Y_{a}(u_{n}^{\gamma})  = R\sum_{a\geqslant 1}\int \frac{d\tilde{p}_{a}(u)}{2\pi} Y_{a}(u^{\gamma}) \, ,
\eeq
where, in the last equality, we used that the set of Bethe states is dense at large $R$ and replaced the sum over the mode number $n$ by an integral over the momentum $\tilde{p}$, with help of the Jacobian $2\pi dn = Rd\tilde{p}$. Without surprise, this is the same expression as the one found for the regularized amplitude (\ref{101r}), up to a factor $1/2$.

The analysis of the torus partition function is, of course, textbook material. Here, we have recalled its main lines for completeness and put the emphasis on the pedestrian low temperature approach rather than on the TBA derivation. (See \cite{Bombardelli:2013yka,Ahn:2011xq} for further readings and for discussions of the higher wrapping corrections.) The latter approach is much more efficient at giving us the all order result,
\beq\label{fullZ}
\log{Z(R)} = R\sum_{a\geqslant 1} \int \frac{d\tilde{p}_{a}(u)}{2\pi} \log{(1+Y_{a}(u^{\gamma}))} + o(R^0)\, ,
\eeq
with $Y$ standing now for the full solution to the TBA equations and up to the finite $R$ corrections, which are expected to be suppressed at large $R$ for periodic boundary conditions.%
\footnote{In the linearized approximation (\ref{Rz}) there is no discretization effects $\sim 1/R^k$ coming from the infinite tail of boundary terms in the Euler-Maclaurin summation formula, since the function we are sampling decays at infinity as well as all of its derivatives. Hence, the finite $R$ corrections are likely to be exponentially small in that case at least.}

\subsection{The shift of the Bethe roots}

There is yet one more effect, triggered by the virtual particles, that we must take into account. This is the shift of the Bethe roots \cite{Bajnok:2008bm} or, equivalently, the finite size corrections to the quantization conditions of the Bethe rapidities $\textbf{u}$. We recall below how this effect comes about. The expert reader can directly jump to the next section.

We consider a cylinder with the same boundary conditions as before but with one of the magnon in the Bethe state going once around the world, as in panel $B.$ in figure~\ref{ZR}. Proper Bethe states are those for which there is, in the end, no difference between the two partition functions, $Z(R)$ and $Z'(R)$, for the two cylinders. Equating them asymptotically provides the ABA equations (\ref{ABA}), as well known. For the first wrapping correction, we must take into account the mirror process in the second square of panel $B.$ in \ref{ZR}. We can apply the same strategy as before to regularize the infinite volume divergences inherent to the thermal trace. However, we cannot use the same quantized mirror states as in (\ref{Rz}). The reason that is for $Z'(R)$ a Bethe rapidity $u_{i}$ is crossing the mirror cut and that cannot stay unnoticed by the mirror magnon. A natural assumption is that this rapidity disturbs the mirror Hilbert space like a defect would do. A similar strategy was used in \cite{Bombardelli:2013yka} to derive the NLO L\"uscher formula. Put differently, the real magnon $u_{i}$ acts like a non-dynamical impurity and modifies the Bethe ansatz equations in the mirror channel, which become
\beq\label{mmABA}
e^{i\tilde{p}R}S(u^{\gamma}, u_{i}) = 1\, .
\eeq
The first wrapping correction to $\log{Z'(R)}$ is thus given by the same sum as in (\ref{Rz}) but with a shifted density of states, $2\pi dn = Rd\tilde{p} -idu\,\partial_{u}\log{S(u^{\gamma}, u_{i})}$, resulting in
\beq\label{shiftZ}
\delta \log{(Z'(R)/Z(R))} =  - i\int \frac{du}{2\pi} e^{-L_{1}\tilde{E}(u)} \partial_{u}S(u^{\gamma}, u_{i})\prod_{j\neq i}S(u^{\gamma}, u_{j})\, .
\eeq
The ratio $Z'(R)/Z(R)$ is finite in the infinite volume limit, $R\rightarrow \infty$, as it should be. However, the BAEs, $1 = Z'(R)/Z(R)$, are shifted, because of the non-zero RHS in (\ref{shiftZ}). This is in line with the Bajnok-Janik formula for the exact quantization conditions,
\beq\label{exact}
e^{ip_{\bar{\alpha}}L_{1}+i\Phi_{\bar{\alpha}}}S_{\bar{\alpha}\alpha} = 1\, ,
\eeq
where $S_{\alpha\bar{\alpha}} = \prod_{i, j\in \alpha,\bar{\alpha}}S_{ij}$ and $\Phi_{\bar{\alpha}} = \sum_{i\in \bar{\alpha}}\Phi_{i}$. Were the theory diagonal and consisting of a single type of particle, the phase $\Phi_{i}$ would be literally the integral in (\ref{shiftZ}), up to the imaginary unit, at the leading wrapping order. The scattering theory we are interested in is not diagonal and counts infinitely many bound states. The equations (\ref{exact}) remain the same but one must slightly generalize the discussion to get the correct expression for $\Phi_{i}$.

\begin{figure}
\begin{center}
\includegraphics[scale=0.37]{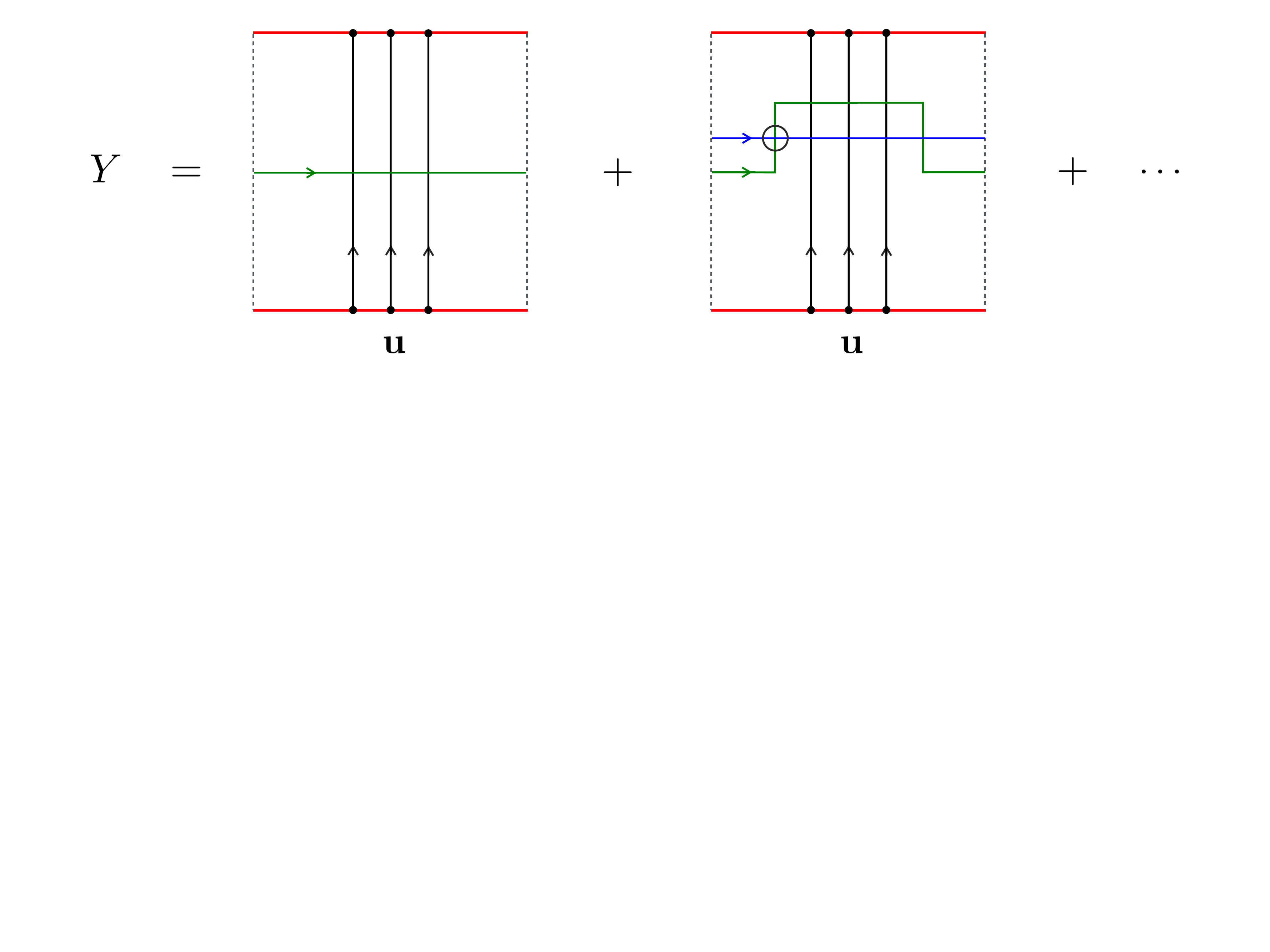}
\end{center}
\vspace{-6.3cm}
\caption{Low temperature expansion of the $Y$ function. We introduce a probe (the green transversal line winding around the cylinder) by pulling a magnon out of the thermal mirror bath. When the probe rapidity matches with one of the Bethe roots $\textbf{u}$, we recover the effect of a real magnon going once around the world. The leading term is controlled by the dressed transfer matrix. At the next order, the probe starts interacting with the excitations in the thermal mirror bath. This interaction is controlled by the mirror scattering kernel, that is, the derivative of the logarithm of the mirror-mirror S-matrix. In the figure, we indicate with a circle the interaction point where a derivative w.r.t.~the rapidity of the thermal mirror line (in blue) is taken.}\label{Yf}
\end{figure}

The matrix degrees of freedom can be encompassed with the help of the $Y$ function. We re-introduce it more formally as the thermal expectation value of the dressed transfer matrix,
\beq\label{reY}
Y_{b}(v^{\gamma}) = Z_{R}^{-1}\times \mathcal{h}\textrm{tr}_{b}\, \{e^{-L_{1}\tilde{H}}\, \mathbb{S}_{b1}(v^{\gamma}, \textbf{u})\}\mathcal{i}_{R, v^{\gamma}}\, ,
\eeq
with $\mathbb{S}_{b1} = S_{b1} \mathcal{S}_{b1} \dot{\mathcal{S}}_{b1}$ the full S-matrix. The subscripts at the foot of the angle brackets are there to remind us that the Hilbert space of mirror states we trace over depends not only on the mirror cut off $R$ but also on the defect $v^{\gamma}$ introduced in the system. As before, the defect is a magnon wrapping around the thermal circle. It comes here equipped with an additional bound state label $b = 1, 2, \ldots\,$ and one must sum over the bound state flavors. (Note that the dependence on $R$  in (\ref{reY}) cancels out in the end.)

The $Y$ function can be viewed as a way of pulling a magnon $|v^{\gamma}, b\mathcal{i}$ out of the mirror thermal ensemble. Its low temperature expansion is shown graphically in figure \ref{Yf}. It reads, up to higher wrapping effects,
\beq\label{Ysecond}
\begin{aligned}
&Y_{b}(v^{\gamma}) = Y^{\circ}_{b}(v^{\gamma}) \\
&\,\,-i\sum_{a\geqslant 1}\int \frac{du}{2\pi} \, e^{i(p_{a}(u^{\gamma})+p_{b}(v^{\gamma}))L_{1}}\, \textrm{tr}_{a\otimes b}\{\mathbb{S}_{ba}(v^{\gamma}, u^{\gamma})\mathbb{S}_{b1}(v^{\gamma}, \textbf{u})\mathbb{S}_{a1}(u^{\gamma}, \textbf{u})\partial_{u}\mathbb{S}_{ab}(u^{\gamma}, v^{\gamma})\}\, ,
\end{aligned}
\eeq
with $Y^{\circ}_{b}(v^{\gamma}) = e^{ip_{b}(v^{\gamma})L_{1}}S_{b1}(v^{\gamma}, \textbf{u})T_{b}(v^{\gamma})\dot{T}_{b}(v^{\gamma})$ being the asymptotic value of $Y$.%
\footnote{The states of interest are diagonal in flavors, $T_{b}(v^{\gamma}) = \dot{T}_{b}(v^{\gamma})$, showing the agreement with the previous expression (\ref{Yfirst}) for $Y^{\circ}$.}
The finite size corrections in the RHS of (\ref{Ysecond}) are obtained by following the same logic as in the abelian set-up. One must first diagonalize the $u^{\gamma}$-$v^{\gamma}$ scattering matrix and then run the earlier analysis for each invariant channel separately, using $v^{\gamma}$ in place of $u_{i}$ in the quantization conditions (\ref{mmABA}). Finally, the sum over the scattering eigenstates must be recognized as being the same as the trace of the scattering kernel $-id\log{\mathbb{S}_{ab}}(u^{\gamma}, v^{\gamma})$ over the $a\otimes b$ module.

The nice thing about the $Y$ function is that it gives us a direct access to the exact Bethe roots $\textbf{u}$. Indeed, according to the excited state TBA analysis \cite{Dorey:1996re}, the roots must be such that $Y_{1}(u_i) = -1$. This property can also be understood from the formula (\ref{reY}), which is directly parameterized in terms of the exact roots $\textbf{u}$, after noticing that $\mathbb{S}_{b1}(v^{\gamma}, u_{i}) = -\mathbb{P}$ when $v^{\gamma} = u_{i}$ and $b =1$, with $\mathbb{P}$ the permutation operator. Pictorially, the latter identity implies that the horizontal line for $v^{\gamma}$ and the vertical line for $u_{i}$ in figure \ref{Yf} merge into a single (self-avoiding) trajectory, hence reproducing the motion of a magnon $u_i$ winding around the cylinder. Once plugged into (\ref{Ysecond}), or used in the second square in figure \ref{Yf}, it leads to the factorization of $Y^{\circ}_{1}(u_{i})$ out of the integral and leaves us with a trace over $a$ and a derivative $\partial_{u}$ acting on $\mathbb{S}_{a1}(u^{\gamma}, u_{i})$. In sum, it yields, for the phase controlling the shift of the roots in (\ref{exact}),
\beq\label{Phis}
\Phi_{\alpha}  = -i\log{(Y_{1}(\alpha)/Y^{\circ}_{1}(\alpha))} = -\sum_{a\geqslant 1}\int \frac{du}{2\pi} e^{ip_{a}(u^{\gamma})L_{1}} \textrm{tr}_{a}\, \big\{\mathbb{S}_{a1}(u^{\gamma}, \bar{\alpha})\partial_{u}\mathbb{S}_{a1}(u^{\gamma}, \alpha)\big\}\, ,
\eeq
up to higher wrapping corrections, in agreement with the formula proposed by Bajnok and Janik in \cite{Bajnok:2008bm}.

As we shall see, the phase shift (\ref{Phis}) induces a leading wrapping correction to the structure constant, contributing at the same loop order at weak coupling as the other wrapping corrections, to be derived shortly. This contrasts with what happens for the energy of the state for which the effect of $\Phi_{i}$ is subleading  \cite{Bajnok:2008bm,Bajnok:2009vm}. (This is specific to this theory and stems from the fact that the anomalous dimension $\sim g^2$ at weak coupling.)

\section{L\"uscher formula for structure constants}\label{Sect4}

\subsection{Renormalized structure constants}

\begin{figure}
\begin{center}
\includegraphics[scale=0.40]{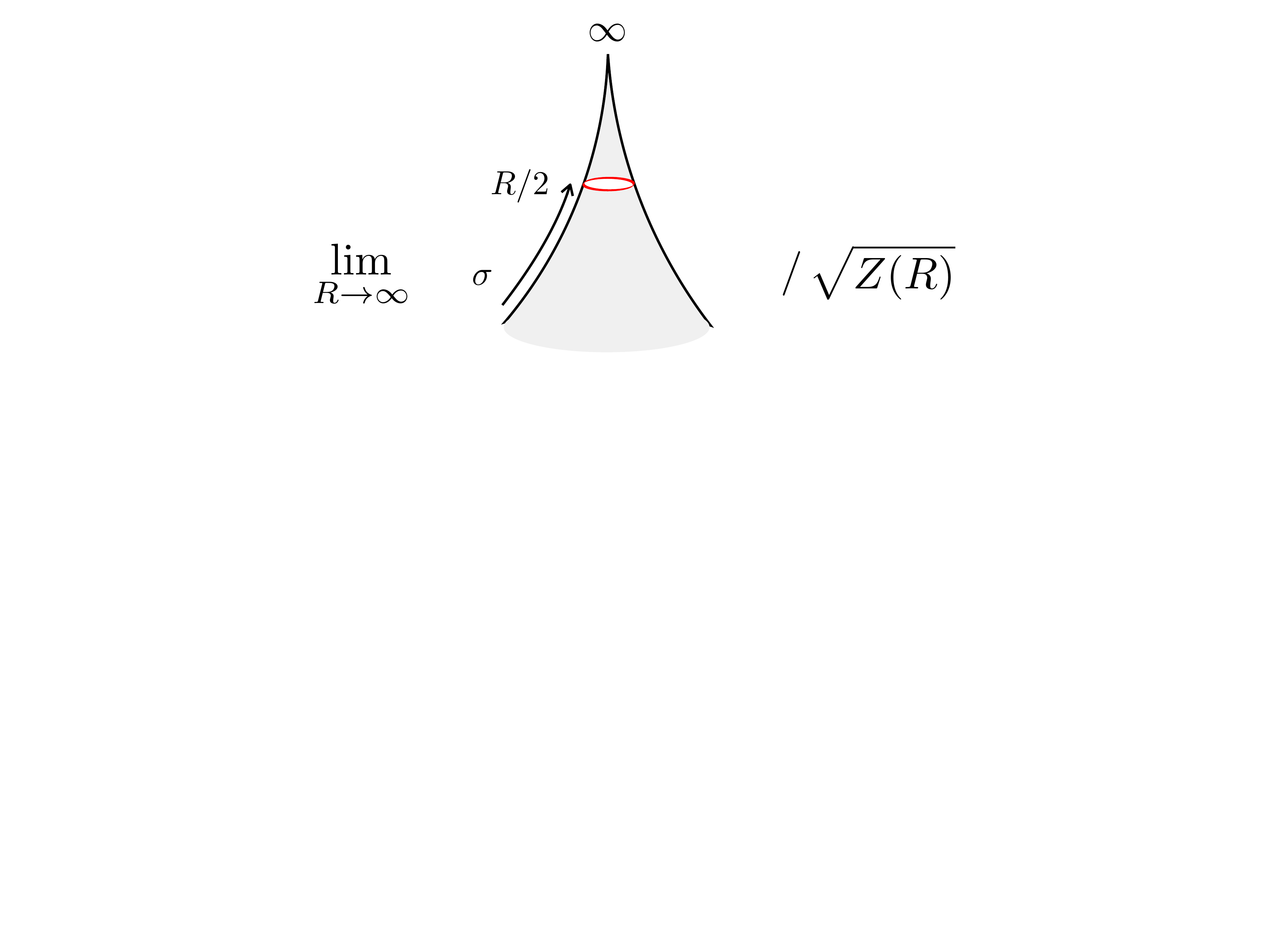}
\end{center}
\vspace{-6.7cm}
\caption{We regularize the operator insertion on a planar surface by cutting off the cone around the puncture or, pictorially, by pulling the state off the tip and placing it at the position $\sigma = R/2\gg 1$ along the mirror spatial direction $\sigma$. The renormalized operator insertion is defined as the quotient of the regularized operator by the square root of the excited state partition function $Z(R)$, in the limit where the cut off $R$ is sent to infinity. This procedure remove the local divergences affecting the insertion of the operator on the three and higher point planar geometries.}\label{puncture}
\end{figure}

According to our previous discussion, we should get a meaningful expression for the structure constant if we both renormalize our operator insertion, as shown in figure \ref{puncture}, and account for the shift of the roots. These two considerations lead us to replace the original ``bare" hexagon formula (\ref{asyC}) by the following, renormalized, one,
\beq\label{RC}
\left(C^{\bullet\circ\circ}_{123}/C^{\circ\circ\circ}_{123}\right)^2 = \frac{\mu(\textbf{u})h(\textbf{u}, \textbf{u})}{G^{\textrm{exact}}_{1}} \times \lim_{R\rightarrow \infty} Z_{R}^{-1} \times \mathcal{A}_{R}^2\, ,
\eeq
where $\mathcal{A}_{R}$ is the finite volume regularization of the sum $\mathcal{A}$ and where $Z_{R}$ is the excited state partition function (\ref{fullZ}). The Bethe roots $\textbf{u}$ in (\ref{RC}) are on shell w.r.t.~to the exact quantization conditions (\ref{exact}) and, accordingly, $G^{\textrm{exact}}_{1}$ is the Gaudin determinant for the exact momenta, that is, $G^{\textrm{exact}}_{1}$ is given by (\ref{G1}) with $\partial_{u_i}\Phi_{j}$ added into the brackets. Each factor in (\ref{RC}) is manifestly finite since all the divergences are cut off in finite volume. The claim is that their product stays finite as one removes the cut off and send $R\rightarrow \infty$. Put differently, the limit defining the renormalized series,
\beq\label{RenA}
\mathcal{A}^2_{\textrm{ren}} = \lim_{R\rightarrow \infty} Z_{R}^{-1} \times \mathcal{A}_{R}^2\, ,
\eeq
should exist. This is clearly so for the leading wrapping contribution $\mathcal{A}^{R}_{(1, 0, 1)}-\tfrac{1}{2}\mathcal{A}_{\textrm{asympt}}\delta Z_{R}$, which is seen to be $R$ independent after using equations (\ref{Rz}) and (\ref{101r}). Testing the finiteness of (\ref{RenA}) in general and tracing the exponentiation of the regulator dependence of $\mathcal{A}_{R}$ all the way back to the kinematical singularities of the multi-magnon hexagon transitions are two interesting problems which we leave for a future investigation. In the rest of this section we shall instead work at obtaining a concise formula, akin to the L\"uscher formula for the energy \cite{Bajnok:2009vm}, for the $(1, 0, 1)$ contribution to the renormalized series $\mathcal{A}_{\textrm{ren}}$.

\subsection{L\"uscher formula}\label{Lues}

Below wrapping order, nothing has changed and $\mathcal{A}_{\textrm{ren}}$ is identical to the bare series $\mathcal{A}$. The two series start differing at the level of the $(1, 0, 1)$ integral. The renormalized form of this integral is obtained by expanding the octagon transition around $w=u$ and dropping the polar part $\sim 1/(w-u)$. Indeed, in the regularized theory, this corresponds to dropping the term linear in $R$ and keeping the $O(R^0)$ part that lies behind. {As in section \ref{Gluing}, here we focus on the fundamental mirror magnon for simplicity and drop the bound state indices from the formulae.}

To begin with, we write the integrand of the octagon transition in the form
\beq
\textrm{int}(w, v, u) = -\frac{\Psi(w, v, u)}{(w-v+i0)(v-u+i0)}\, ,
\eeq
where $\Psi(w, v, u)$ is whatever multiplies the poles in the product of the two hexagon form factors. The latter function is regular in the diagonal limit $w\rightarrow u$, by definition. One then easily strips out the pole at $w=u$ by doing a partial fraction decomposition in $v$,
\beq\label{int}
\textrm{int}(w, v, u) = \frac{i}{w-u+i0}\times \bigg[\frac{i}{w-v+i0}+\frac{i}{v-u+i0}\bigg]\times \Psi(w, v, u)\, .
\eeq
Setting $w = u$ in the two last factors, one recovers that the polar part has support on the set of wrapping configurations $w=v=u$,
\beq\label{polar}
\textrm{polar part} = \frac{i}{w-u+i0}\times 2\pi \delta(u-v) \times \Psi(u, u, u)\, ,
\eeq
as indicated by the delta function $\delta(u-v)$. The renormalized amplitude is the $O(1)$ term in the Taylor expansion of (\ref{int}) around $w = u$,
\beq\label{luescher}
\textrm{finite part} = i\partial_{w}\bigg\{\bigg[\frac{i}{w-v+i0}+\frac{i}{v-u+i0}\bigg]\times \Psi(w, v, u)\bigg\}_{w  = u}\, .
\eeq
The natural next step is to split the formula (\ref{luescher}) into two parts, for the two factors in (\ref{luescher}) the derivative $\partial_{w}$ can act on. The action on the first factor gives what we call the bulk contribution,
\beq\label{bulk}
\textrm{bulk} = \frac{\Psi(u, v, u)}{(u-v+i0)^2}\, .
\eeq
It is identical to the integrand of the bare amplitude, if not for the presence of the $i0$'s. The latter prescription plays an essential role since it makes the singularity at $u=v$ integrable. It does it by simply avoiding the set of wrapping configurations at $v=u$. The second contribution does the exact opposite and, like the polar part, lives on the support of these configurations. It is given as a contact term localized at $u=v$ and is obtained by letting the derivative acts on the second factor in (\ref{luescher}),
\beq\label{contact}
\textrm{contact} = 2\pi \delta(u-v) i\partial_{w} \Psi(w, v, u)\big|_{w=u}\, .
\eeq
It is the essential extra term predicted by our analysis.%
\footnote{The partition into a bulk contribution and a contact term is a bit arbitrary, since it relies on a choice of prescription to integrate the double pole at $u=v$. We would obviously get a slightly different contact term if we switched to a principal value prescription, for instance.}
We think of it as describing the renormalized effect of a magnon wrapping closely the excited operator.

The details of the contact term (\ref{contact}) depend on what is sitting inside $\Psi(w, v, u)$. Putting aside the matrix part, the integrand factorizes, up to the sum over the partitions, and yields
\beq\label{psi}
\Psi = -\frac{\sqrt{\mu(w^{\gamma})}(w-v)\mu(v^{\gamma})(v-u)\sqrt{\mu(u^{\gamma})}}{h(w^{\gamma}, v^{\gamma})h(v^{\gamma}, u^{\gamma})} e^{-\tilde{E}(v)\ell_{31}-\tilde{E}(u)\ell_{12}}\sum_{\alpha} a_{\alpha\bar{\alpha}} \frac{h(w^{\gamma}, \alpha)h(v^{\gamma}, \bar{\alpha})}{h(\alpha, v^{\gamma})h(\bar{\alpha}, u^{\gamma})}\, .
\eeq
The derivative in (\ref{contact}) can act on both the first and the last factors in the RHS of (\ref{psi}). They contain, respectively, the self-interaction of the wrapped mirror magnon and its interaction with the Bethe roots. The analysis of the self-interaction is a bit technical and is deferred to the appendix \ref{Ct}. The outcome is simple to describe. Only the phase of the hexagon amplitude $h(w^{\gamma}, v^{\gamma})$ in (\ref{psi}) contributes to the derivative, when $u=v=w$, and it is controlled by the S-matrix, $S(u^{\gamma}, v^{\gamma}) = h(u^{\gamma}, v^{\gamma})/h(v^{\gamma}, u^{\gamma})$. The Watson equation also allows us, when $w=v=u$, to collapse the last factor in (\ref{psi}) into $S(u^{\gamma}, \textbf{u})$ and thus strip out the sum over the partitions. Taking everything into account yields the first contact term,
\beq\label{first}
\textrm{first} = \mathcal{A}_{\textrm{asympt}}\times \frac{1}{2}\, e^{-\tilde{E}(u)L_{1}}S(u^{\gamma}, \textbf{u}) K(u^{\gamma}, u^{\gamma})\, ,
\eeq
where we used that $\ell_{12}+\ell_{31} = L_{1}$. It is controlled by the (abelian part of the) scattering kernel,
\beq
K(u, v) = \frac{\partial}{i\partial u}\log{S(u, v)}\, ,
\eeq
which is evaluated in (\ref{first}) at coinciding rapidities in the mirror kinematics. The second contact term comes from differentiating the factor $h(w^{\gamma}, \alpha)$ in (\ref{psi}), connecting the mirror rapidity $w$ to the subset of roots that live on the first hexagon. This one is obviously partition dependent and readily given by
\beq\label{second}
\textrm{second} = \sum_{\alpha} a_{\alpha\bar{\alpha}}\,e^{-\tilde{E}(u)L_{1}} S(u^{\gamma}, \textbf{u})\frac{i\partial}{\partial u}\log{h(u^{\gamma}, \alpha)}\, .
\eeq
Notice that only the integration over $u$ remains to be done in both (\ref{first}) and (\ref{first}), thanks to the delta function in (\ref{contact}).

\begin{figure}
\begin{center}
\includegraphics[scale=0.45]{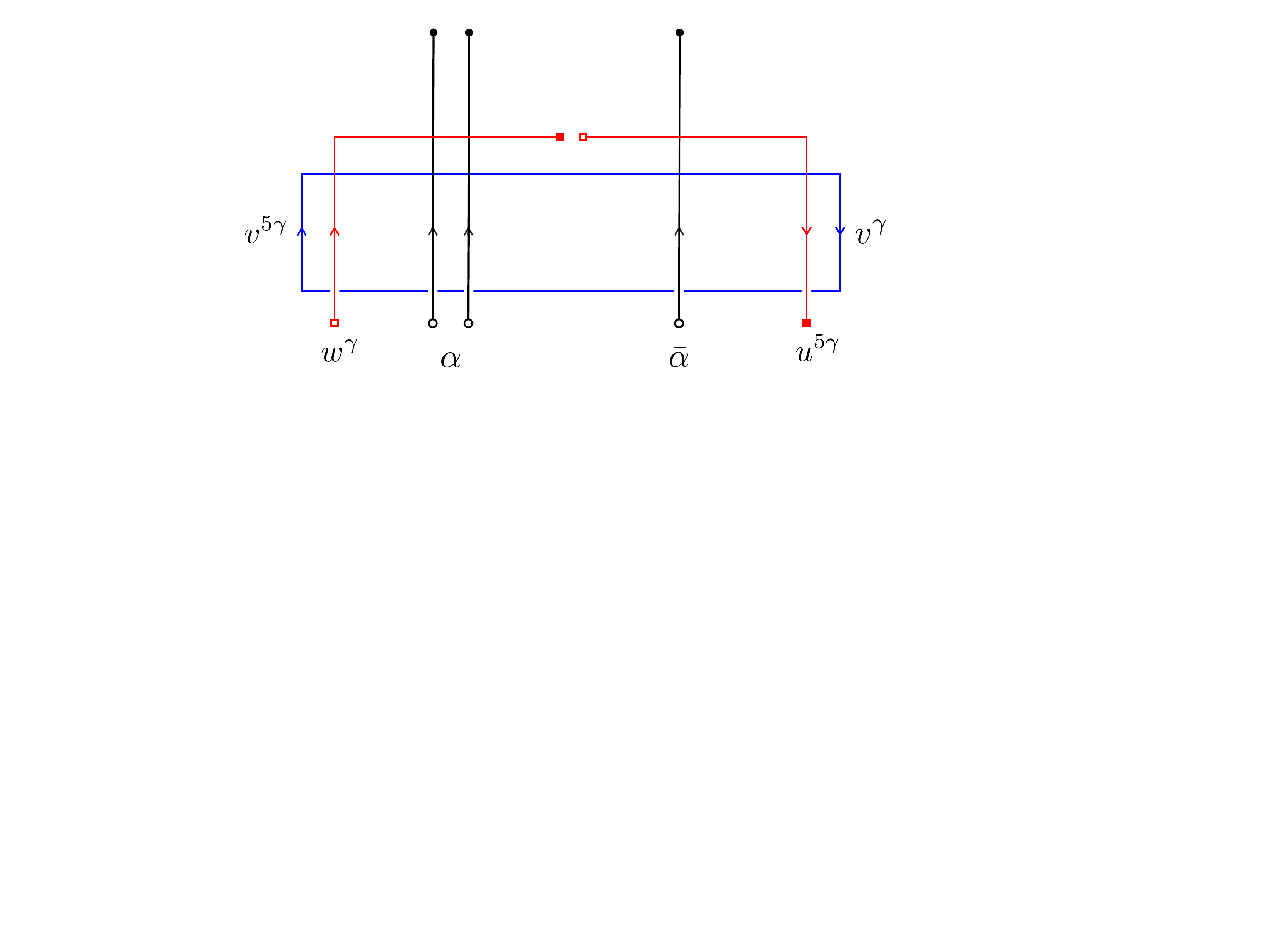}
\end{center}
\vspace{-7.2cm}
\caption{The matrix part of the octagon transition is obtained by contracting two hexagon matrix parts, here standing on the left and on the right. Each open string represents a magnon $\chi_{A\dot{A}}$, with the $SU(2|2)_{L(R)}$ index standing at its left (right) endpoint, and each crossing stands for the action of the $SU(2|2)$ S-matrix $\mathcal{S}$, with the orientation fixing what is incoming and what is outgoing. A loop forms when we identify two magnons and sum over their flavors, as done here for the mirror magnon (closed blue line) shared among the two hexagons. (There is no distinction between one or five mirror rotations at the level of the matrix part, such that the rapidity can be chosen to be the same all along the blue mirror loop.) The wrapped octagon matrix part is obtained by attaching the remaining open mirror lines (in red) together. For the contact term, we must first act with the derivative $\partial_{w}$ and then set the mirror magnons in the wrapping configuration $w = v = u$. It produces two types of contributions, depending on whether we differentiate the mirror-mirror or the mirror-real interaction points. In the latter case, Yang-Baxter and unitarity moves allow us to disentangle the two (red and blue) mirror loops; the blue line factorizing out and giving a transfer matrix. In the former case, the two mirror lines are entangled through the mirror-mirror scattering kernel.}\label{MatG}
\end{figure}

The next step is to include the matrix part. It reduces to two basic objects, which are the matrix counterparts of the above two contact terms. Namely, we can express it in terms of (some differentiated version of) the transfer matrix and of the matrix part of the scattering kernel. We refer the reader to figure \ref{MatG} for a graphical explanation.

{As mentioned in the beginning of this subsection, the analysis so far is for a fundamental mirror magnon. However, it can be generalized simply and naturally to the bound states as shown in Appendix \ref{Ct}.} Combining all the pieces together and decorating the full thing with bound state labels, we arrive at our final expression for the renormalized $(1, 0, 1)$ amplitude,
\beq\label{Aren-final}
\mathcal{A}^{\textrm{ren}}_{(1, 0, 1)} = B + C_{1}\mathcal{A}_{\textrm{asympt}} + C_{2}\, ,
\eeq
with a bulk contribution,
\beq\label{Bulk}
\begin{aligned}
B = \sum_{a,b\geqslant 1} \int \frac{dudv}{(2\pi)^2} \frac{\mu_{a}(u^{\gamma})\mu_{b}(u^{\gamma})T_{a}(u^{\gamma})T_{b}(v^{\gamma})}{h_{ab}(u^{\gamma}+i0, v^{\gamma})h_{ba}(v^{\gamma}, u^{\gamma}+i0)}e^{ip_{a}(u^{\gamma})\ell_{12}+ip_{b}(v^{\gamma})\ell_{31}} \\
\times \sum_{\alpha\cup \bar{\alpha} = \textbf{u}} a_{\alpha\bar{\alpha}}\, \frac{h_{a1}(u^{\gamma}, \alpha)h_{b1}(v^{\gamma}, \bar{\alpha})}{h_{1a}(\alpha, v^{\gamma})h_{1b}(\bar{\alpha}, u^{\gamma})}\, , \\
\end{aligned}
\eeq
which is just the $i0$ shifted version of the bare amplitude (\ref{101}), and with two contact terms,
\beq\label{C12}
\begin{aligned}
&C_{1} = \sum_{a}\int \frac{du}{2\pi} \bigg[\frac{1}{2}Y_{a}(u^{\gamma})K_{aa}(u^{\gamma}, u^{\gamma}) + e^{ip_{a}(u^{\gamma})L_{1}}S_{a1}(u^{\gamma}, \textbf{u}) \mathcal{K}_{aa}(u^{\gamma}, u^{\gamma})\bigg]\, , \\
&C_{2} = \sum_{a, \alpha}a_{\alpha\bar{\alpha}}\int \frac{du}{2\pi}\bigg[Y_{a}(u^{\gamma})i\partial_{u}\log{h_{a1}(u^{\gamma}, \alpha)} + e^{ip_{a}(u^{\gamma})L_{1}}S_{a1}(u^{\gamma}, \textbf{u}) \mathcal{H}_{a1}(u^{\gamma}, \alpha)\bigg]\, .
\end{aligned}
\eeq
Here, $K_{ab}(u, v) = -i\partial_{u}\log{S_{ab}}(u, v)$ and the matrix counterparts of $d\log{h}$ and $d\log{S}$ are defined, respectively, by
\beq
\begin{aligned}
\mathcal{H}_{a1}(u^{\gamma},\alpha) &=  iT_{a}(u^{\gamma})\textrm{tr}_{a} \big\{\mathcal{S}_{a1}(u^{\gamma}, \bar{\alpha})\partial_{u}\mathcal{S}_{a1}(u^{\gamma}, \alpha)\big\} \, ,\\
\mathcal{K}_{ab}(u^{\gamma}, v^{\gamma}) &= -i\textrm{tr}_{a\otimes b} \{\mathcal{S}_{ba}(v^{\gamma}, u^{\gamma})\mathcal{S}_{b1}(v^{\gamma}, \textbf{u})\mathcal{S}_{a1}(u^{\gamma}, \textbf{u})\partial_{u}\mathcal{S}_{ab}(u^{\gamma}, v^{\gamma})\}\, .
\end{aligned}
\eeq
Notice that the $i0$'s in (\ref{Bulk}) are only needed to handle the case where $b=a$, as otherwise the integrand is automatically regular and integrable. Notice also that the contact term $C_{1}$ is partition independent, which is why we could factor out the asymptotic sum in front of it in (\ref{Aren-final}). This is not the case for the contact term $C_{2}$, which explicitly depends on $\alpha$. The former can be interpreted as a wrapping correction to the norm of the state, on top of those already contained in the exact Gaudin norm. The latter can be viewed as a finite size correction to the (asymptotic) sum over the partitions. We refer the reader to equations (\ref{dressed}) and (\ref{dressing}), below, for a precise implementation of this dressing procedure.

\section{Testing the formula}\label{test}

\subsection{Consistency check and thermal dressing}\label{Cc}

The route that we followed to incorporate the wrapping effects in the hexagon formalism does not treat equally the mirror channels $12$ and $31$, adjacent to the excited operator. Indeed, recall that we first sewed the two hexagons into an octagon by gluing along the channel 31 and then wrapped the octagon around the pair of pants along the channel 12. This specific itinerary introduces an asymmetry that is visible in the specific role the roots $\alpha$ play in the second line in (\ref{C12}) and in the specific way one of the two mirror magnons is handled by the $i0$ prescription in (\ref{Bulk}). In the end, it is important to verify that these artifacts do not impinge on the invariance of the three point function under the permutation of the two chiral primaries, or $L_{2}\leftrightarrow L_{3}$. We explain below that the asymmetry present in our L\"uscher formula is actually needed to ensure that the structure constant transforms properly under this permutation at wrapping order. This asymmetry is as it should be to compensate the transformation of the asymptotic amplitude (\ref{asyC}), which becomes anomalous once evaluated on a Bethe state satisfying the \textit{exact} BA equations (\ref{exact}).

Recall first that the asymptotic sum (\ref{Aasympt}) is invariant under $\ell_{12}\leftrightarrow \ell_{31}$ when evaluated on a cyclic Bethe state satisfying the \textit{asymptotic} BA equations (\ref{ABA}). More precisely, the sum maps onto itself,
\beq\label{1231}
\sum_{\alpha} a_{\alpha\bar{\alpha}}(\ell_{31}) = (-1)^M\sum_{\alpha}a_{\alpha\bar{\alpha}}(\ell_{12})\, ,
\eeq
up to a sign.%
\footnote{Only the full three point function, $C_{123}\times \textrm{orbital}_{123}$, ought to be invariant under the permutation of the (bosonic) legs $2$ and $3$. The sign in (\ref{relat}) is a necessary, harmless, ``monodromy" in the transformation property of the structure constant, which cancels the transformation of the orbital part of the three point function: $\textrm{orbital}_{132} = (-1)^M \textrm{orbital}_{123}$, for a spin $M$ conformal primary $\mathcal{O}_{1}$ (with $\mathfrak{su}(4)$ weights $[0, \star, 0]$) in the OPE of two chiral primaries.}
Indeed, the local version of this relation is equivalent to the ABA equations,
\beq\label{relat}
(-1)^{M}a_{\alpha\bar{\alpha}}(\ell_{31})/a_{\bar{\alpha}\alpha}(\ell_{12}) \equiv e^{ip_{\bar{\alpha}}\ell_{13}-ip_{\alpha}\ell_{12}}\frac{h_{\bar{\alpha}\alpha}}{h_{\alpha\bar{\alpha}}} = e^{ip_{\bar{\alpha}}L_{1}}S_{\bar{\alpha}\alpha} = 1 \, ,
\eeq
for a cyclic state, $e^{ip_{\textbf{u}}} = e^{ip_{\alpha}}e^{ip_{\bar{\alpha}}} = 1$, thanks to the Watson relation, $h_{\bar{\alpha}\alpha} = S_{\bar{\alpha}\alpha}h_{\alpha\bar{\alpha}}$. The same transformation property is observed, separately, for each mirror correction in $\mathcal{A}$, if we ignore the divergences.%
\footnote{It is clear from the general formula (\ref{general}) that exchanging $L_{2}$ and $L_{3}$ is the same as exchanging $\alpha$ and $\bar{\alpha}$. In particular, the part of the integrand coming from the channel $23$ is invariant under both operations.} The problem is that the algebra behind (\ref{1231}) no longer works out when we plug exact Bethe roots in the asymptotic amplitude, since the additionnal phase shift in (\ref{exact}) readily spoils the relation (\ref{relat}).

Remarkably enough, our L\"uscher formula introduces the right amount of asymmetry to restore the permutation symmetry of the three point function. To establish this fact, we shall first cast our prediction in a slightly different form, by interpreting the contact terms in our formula as correcting the two main factors entering the asymptotic representation of the structure constant (\ref{asyC}). (This step is not logically needed for the proof, but it leads to a more transparent demonstration.) We start by rewriting the bulk contribution (\ref{bulk}) in a more symmetric fashion, by using principal value integration instead of $i0$'s. The difference between the two prescriptions is a contact term, which can be combined with those in (\ref{C12}) as detailed in Appendix \ref{Ct}. We then absorb the contact terms into a redefinition of the two factors in (\ref{asyC}). The dressed version of the measure in (\ref{asyC}) is written as
\beq\label{dressed}
\sqrt{\frac{\mu(\textbf{u})h(\textbf{u}, \textbf{u})}{G_{1}}}\, \bigg|_{\textrm{dressed}} = \sqrt{\frac{\mu(\textbf{u})h(\textbf{u}, \textbf{u})}{G^{\textrm{exact}}_{1}}}\times(1 + C_{1} + \sum_{a \geqslant 1} \frac{i}{4}\int \frac{d\log{p_{a1}(u^{\gamma}, \textbf{u})}}{2\pi} Y_{a}(u^{\gamma}) + \dots)\, ,
\eeq
where $p_{ab}(u, v) = h_{ab}(u, v)h_{ba}(u, v)$ is the symmetric part of the hexagon form factor, $C_{1}$ is the contact term involving the scattering kernel in (\ref{C12}) and the dots stand for (unknown) higher wrapping contributions. The dressing of the asymptotic amplitude $\mathcal{A}_{(0, 0, 0)}$ absorbs the remaining, partition dependent, contact term,
\beq\label{dressing}
\mathcal{A}_{(0, 0, 0)}|_{\textrm{dressed}} = e^{-i\varphi_{12}-\frac{i}{4}\Phi_{\textbf{u}}}\sum_{\alpha} a_{\alpha\bar{\alpha}}(\ell_{31}) \, e^{\frac{i}{2}\Phi_{\bar{\alpha}}}\, ,
\eeq
where $\varphi_{ij} = \ell_{ij}\Phi_{\textbf{u}}/(2L_{1})$. We stress that these two dressed expressions provide an equivalent representation of the L\"uscher formula, when supplemented with the principal valued bulk integral. Equation (\ref{dressing}) is, in reality, a non-linear extrapolation of our result, based on the ``TBA inspired" exponentiation $1+\tfrac{i}{2}\Phi_{i} \rightarrow \exp{(\tfrac{i}{2}\Phi_{i})}$.%
\footnote{We have much less intuition about what the all order formula for the normalization prefactor should be, though the first term in (\ref{dressed}) is suggestive of the expansion of a Fredholm determinant.}
To prove the permutation symmetry of the three point function at the leading wrapping order, it would be enough to keep only the terms linear in $\Phi_{i} \ll 1$. The complete formula (\ref{dressing}) is a reasonable conjecture for the higher dressing effects that guarantees that the latter property of the three point function holds true in general.

The contribution (\ref{dressed}) is obviously symmetric, regardless of which state $\textbf{u}$ we plug inthere. The same can be said about the symmetrized bulk integral. What is crucial for the proof is that the dressed sum over the partitions (\ref{dressing}) is symmetric for an exact Bethe state. The essential feature is that its summand is controlled by the exact quasi-momenta. This becomes more manifest if we square it,
\beq\label{summand}
(a_{\alpha\bar{\alpha}}(\ell_{31})|_{\textrm{dressed}})^2 = \frac{e^{ip_{\bar{\alpha}}(L_{3}-L_{2})}}{h_{\alpha\bar{\alpha}}h_{\bar{\alpha}\alpha}} \times (-1)^{|\bar{\alpha}|}Y(\bar{\alpha})\, ,
\eeq
and absorb the new dynamical effect into the $Y$-function,
\beq
Y(\bar{\alpha}) = \prod_{i\in \bar{\alpha}}Y(u_{i})\, , \qquad Y(u_{i}) = e^{ip_{i}L_{1}}S_{u_{i}\textbf{u}} T(u_{i})\dot{T}(u_{i}) e^{i\Phi_{i}}\, ,
\eeq
using that the on-shell transfer matrix $T(u_i) = \dot{T}(u_i) = 1$ in the $\mathfrak{sl}(2)$ subsector. When the roots are on-shell w.r.t.~the exact BA equations, we can forget about the last factor in (\ref{summand}), $1+Y(u_{i}) = 0$, and we find the same functional form as in the absence of wrapping corrections (the sole difference being that we must plug exact roots). This is the main reason why the dressed sum (\ref{dressing}) returns a symmetric three point function, at wrapping order.

There is a subtletly in the argument, which is that the cyclic invariance of the state no longer amounts to $p_{\textbf{u}} = 2\pi m$ when wrapping corrections are included. This pitfall is taken care of by the overall phase $\varphi_{12}$ in (\ref{dressing}). For a more careful proof, we can run the same chain of operations as in (\ref{relat}), using the exact BA equations in place of the asymptotic ones. It yields
\beq\label{algebra}
a_{\alpha\bar{\alpha}}(\ell_{31}) \, e^{\frac{i}{2}\Phi_{\bar{\alpha}}} = (-1)^{|\bar{\alpha}|}\frac{e^{-ip_{\bar{\alpha}}\ell_{12}}}{h_{\bar{\alpha}\alpha}} \, e^{-\frac{i}{2}\Phi_{\bar{\alpha}}} = (-1)^{M} e^{i\varphi_{12}-i\varphi_{31}} \times a_{\bar{\alpha}\alpha}(\ell_{12}) e^{\frac{i}{2}\Phi_{\alpha}}\, ,
\eeq
where in the last line we specialized to a cyclic state, $e^{ip_{\textbf{u}}+i\Phi_{\textbf{u}}/L_{1}} = 1$. Hence,
\beq
\mathcal{A}_{(0, 0, 0)}|_{\textrm{dressed}} = (-1)^{M} e^{-i\varphi_{31}-\frac{i}{4}\Phi_{\textbf{u}}}\sum_{\alpha} a_{\alpha\bar{\alpha}}(\ell_{12}) \, e^{\frac{i}{2}\Phi_{\bar{\alpha}}}\, ,
\eeq
which, up to the overall factor $(-1)^{M}$, is the original expression~(\ref{dressing}) with $L_{2}\leftrightarrow L_{3}$. This concludes the proof.

The attentive reader would have noticed that the phase $\Phi_{\textbf{u}}$ in (\ref{dressing}) played no role in the demonstration. This is not surprising since it does not depend on how we partition the spin chain and the Bethe state. {This phase however affects the reality property of the dressed sum over the partitions and the expression \eqref{dressing} is not real as it is.} This problem is inconsequential for the tests carried out in the following subsections, which is why we postpone its discussion to Appendix \ref{Ct}. We also speculate there on its possible resolutions.

\subsection{Diagonal symmetry at wrapping order}\label{SUSY}

In the next subsection, we shall compare our formula with the gauge theory prediction at weak coupling. As a preliminary test, we shall verify that our expression starts at the right loop order for a wrapping correction, that is at four loops for the Konishi super-multiplet. This is not manifestly so if we work with the $\mathfrak{sl}(2)$ representative, which has the minimal length permitted, $L_{1}=2$. In this case, the loop delay must come from a supersymmetric suppression of the matrix components of the various integrands. We would get the right scaling from scrath if we were to work with the $\mathfrak{su}(2)$ representative, which has length $L_{1} = 4$. Showing that our formula does not depend on which one of these two representatives we choose is equivalent to proving that it correctly embodies the (diagonal) supersymmetry preserved by the structure constants. This is what we shall discuss here.

\begin{figure}
\begin{center}
\includegraphics[scale=0.40]{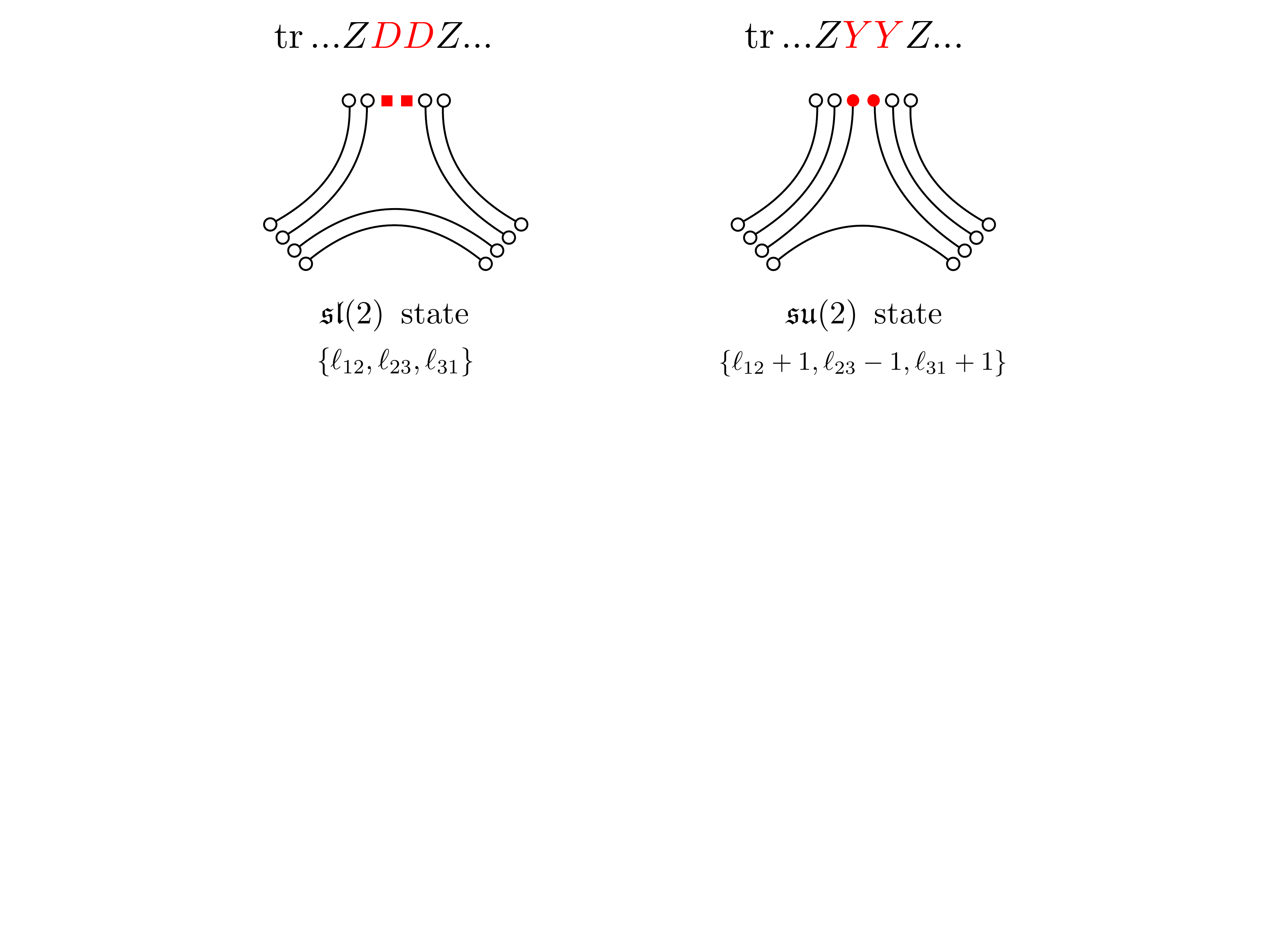}
\end{center}
\vspace{-6.5cm}
\caption{The structure constants are the same for a two-derivative state or a two-scalar state, at the top of the pair of pants. The two states have different lengths and, therefore, their two structure constants exhibit different bridges. Pictorially, converting two derivatives into two scalars forces a link in the bottom bridge in the left panel to break open and reconnect with the two new spin chain sites created at the top in the right panel. It increases by one unit the lengths $\ell_{12}$ and $\ell_{31}$ of the adjacent bridges, at the expense of the bottom bridge that loses one unit of length. The picture explains why the mirror particles propagating close to the top experience a bigger length, and are thus delayed at weak coupling, for states in the $\mathfrak{sl}(2)$ sector. It also predicts a similar effect for mirror particles propagating through the bottom channel when the state at the top lies in the $\mathfrak{su}(2)$ sector. The general lesson is that supersymmetries not only act on the state but also on the lengths of the bridges characterizing the geometry. Besides, as shown here, this action is not restricted to the bridges that are adjacent to the operator which is being transformed. A ``bridges morphism" should also accompany the crossing transformation of excitations on geometries with finite bridges, since this transformation, like the supersymmetric ones, comes at the cost of a length redefinition. These effects would lead to novelties for the bootstrap of geometries with finite bridges, like the octagon for instance. }\label{susy}
\end{figure}

To facilitate the discussion, we first generalize our expression such as to cover at once the two diagonal subsectors of interest, $\mathfrak{su}(2)$ and $\mathfrak{sl}(2)$. The difference entirely resides in the matrix parts of the hexagon form factors and can be encapsulated in the form of an extra weight in the sum over the partitions. (See \cite{Basso:2017khq} for a more general discussion.) Namely, we simply need to switch to
\beq\label{sumT}
\sum_{\alpha\cup \bar{\alpha} = \textbf{u}} a_{\alpha \bar{\alpha}} \, T(\bar{\alpha})\, ,
\eeq
where $T(\bar{\alpha}) = \prod_{i\in \bar{\alpha}}T(u_{i})$ and with $T(u)$ the (left=right) eigenvalue of the fundamental transfer matrix in the (diagonal) state $\textbf{u}$. For the $\mathfrak{sl}(2)$ state, $T(u) = 1$ by convention, and the extra weight trivializes. For the $\mathfrak{su}(2)$ state, we have that $T(\bar{\alpha}) = A_{\bar{\alpha}\alpha} = \prod_{i,j\in \bar{\alpha}, \alpha}A_{ij}$, for an on shell state, in perfect agreement with what the hexagon ansatz prescribes in this subsector \cite{short}. We must also accompany this modification by a change of the Gaudin determinant, which will be based on the $\mathfrak{su}(2)$ Bethe ansatz equations in the latter case. Thanks to this cosmetic rewriting, we can deal with the diagonal supersymmetry transformation of the real and mirror contributions in  pretty much the same way.

At the level of the bare expansion (\ref{bare-exp}), the main mechanism is coming from the transformation property of the transfer matrix. In the spin chain frame, the eigenvalues of the transfer matrix for the two representatives of the same multiplet differ by a simple overall factor
\beq\label{transf}
T_{a}(u)|_{\mathfrak{sl}(2)\, \textrm{state}} = \frac{x^{[+a]}}{x^{[-a]}}\times T_{a}(u)|_{\mathfrak{su}(2)\, \textrm{state}}\, .
\eeq
Its effect in the sum over the partitions (\ref{sumT}), where $a=1$, is readily seen to be equivalent to a redefinition $\ell_{31} \rightarrow \ell_{31}+1$ of the length of the bridge $31$. This is in line with the way the multiplet splitting / joining is achieved in the spin chain description, see e.g.~\cite{BS05}. Namely, the two spin chain primaries belong to two different multiplets in the free theory, with length $L_{1}$ and $L_{1}+2$, respectively, but join into a long multiplet when the theory is interacting. Their structure constants should be the same thanks to the diagonal $PSU(2|2)$ symmetry. This is corroborated asymptotically by (\ref{transf}). 

The same transformation operates at the level of the mirror corrections (\ref{bare-exp}). The only difference is that the transfer matrix is evaluated in the mirror kinematics. In the channels that are adjacent to the operator we are transforming (i.e., channel $12$ or $31$), the argument of the transfer matrix is $u^{\gamma}$ and the prefactor in the RHS of (\ref{transf}) becomes $1/x^{[+a]}x^{[-a]}$. If we are in the mirror channel $23$, we must use $u^{-\gamma}$ and absorb a factor $x^{[+a]}x^{[-a]}$ instead. Hence, the length changing effect reads: $\ell_{12}, \ell_{23}, \ell_{31} \rightarrow \ell_{12}+1, \ell_{23}-1, \ell_{31}+1$. Notice, in particular, that the bridge 23 \textit{loses} one unit of length, when we switch to the $\mathfrak{su}(2)$ representative. This is a bit unusual, but this is needed to compensate the gain in the two adjacent channels. It is in line with what we get by substituting $L_{1} \rightarrow L_{1}+2$ into the bridges' lengths $\ell_{ij} = (L_{i}+L_{j}-L_{k})/2$, at fixed $L_{2}$ and $L_{3}$, or, equivalently, with what is happening at the Feynman diagrammatic level, as shown in figure \ref{susy}. All in all, we can say that the length changing mechanism is realized at the level of the structure constants by transferring one unit of length from the farthest bridge to the closest ones. 

It is slightly harder to verify the diagonal supersymmetry of the wrapping corrections and of the Gaudin norm. At the level of the abelian components in (\ref{C12}) or of the bulk integral (\ref{Bulk}), this is of course the same story since we get the same transfer matrix. For the matrix part of $C_{2}$ in (\ref{C12}) we can use the fact that it can be absorbed in the phase $\Phi_{i}$, like in equation (\ref{dressing}), and as such should be properly supersymmetric (for cyclic states). If that were not true, wrapping effects would lift the energy degeneracy between the $\mathfrak{sl}(2)$ and $\mathfrak{su}(2)$ descendants; a scenario that is clearly excluded (e.g., compare \cite{Bajnok:2009vm} and \cite{Bajnok:2012bz}). The rest of the matrix part in (\ref{C12}) is structurally identical to the kernel that controls the next-to-leading L\"uscher correction to the energy \cite{Bombardelli:2013yka}, if not for the fact that it is evaluated here on a very special kinematical configuration, since the two mirror rapidities coincide. The kernel in the NLO formula also comes with two additional transfer matrices and one extra damping factor $e^{-L_{1} \tilde{E}}$ that delay the full thing all the way to the double wrapping order. It must be clear, however, that if the matrix part in $C_{1}$ were to transform anomalously under supersymmetry, then the same would be true for the NLO energy formula. So, here again, we are confident that the object that we are manipulating in (\ref{C12}) has the right covariance (i.e., transforms as the transfer matrices accompanying the abelian part) under the supersymmetry transformation. The analysis for the Gaudin norm is a bit more delicate, already at the asymptotic level. We refer the reader to Appendix \ref{AsyD} for a thorough discussion. The diagonal symmetry is also checked there for the wrapping corrections to the norm, in the so-called string frame, to leading order at weak coupling. The latter frame, which comes equipped with a non fluctuating length, appears to provide a safer set-up to address this kind of questions at wrapping level.

\subsection{Comparison with the gauge theory}

{Having elucidated all the necessary ingredients, the evaluation of our formula at four loops for the Konishi-BPS-BPS configuration is now a straightforward, but tedius task.}
 The asymptotic part of the structure constant is easily obtained by expanding (\ref{asyC}), for $\mathcal{A}\rightarrow \mathcal{A}_{(0, 0, 0)}$, up to the desired loop order, using the appropriate two-magnon solution to the ABA equations (\ref{ABA}), see Appendix \ref{AsyD}. The computation of the mirror and wrapping corrections is more demanding. We comment below on the main technical novelties of this calculation and refer the reader to Appendix \ref{wrap} for the rest. The final result, for each type of contribution separately, is
\beq\label{final}
\begin{aligned}
&c_{(0, 0, 0)} = 1-6g^2+66g^4-(810-24\zeta_{3})g^6+(\tfrac{20559}{2}-228\zeta_{3}-240\zeta_{5})g^8\, , \\
&c_{(0, 1, 0)} = 36\zeta_{3}g^4-(144\zeta_{3}+360\zeta_{5})g^6+(1224\zeta_{3}+2160\zeta_{5}+432\zeta_{3}^2+4200\zeta_{7})g^8\, , \\
&c_{(1, 0, 0)} = (27+72\zeta_{3}-120\zeta_{5})g^6-(729+468\zeta_{3}+240\zeta_{5}+216\zeta_{3}^2-2100\zeta_{7})g^8\, , \\
&c_{(1, 0, 1)} = -(\tfrac{1215}{2}+324 \zeta_{3}-840\zeta_{5}+216\zeta_{3}^2-420\zeta_{7} )g^8\, ,
\end{aligned}
\eeq
and $c_{(1, 0, 0)} = c_{(0, 0, 1)}$, where $c$ is the ratio%
\footnote{We use here that $(C_{123}^{\bullet\circ\circ}/C_{123}^{\circ\circ\circ})^2|_{\textrm{tree}} = (M!)^2/(2M)! = 1/6$ for spin $M=2$.}
\beq\label{smallc}
c := \frac{C_{123}^{\bullet\circ\circ}}{C_{123}^{\bullet\circ\circ}|_{\textrm{tree}}} = \sqrt{6}\, \frac{C_{123}^{\bullet\circ\circ}}{C_{123}^{\circ\circ\circ}}\, .
\eeq
We stress that these formulae apply to the case where $\ell_{12} = \ell_{23} = \ell_{31} = 1$. They can easily be generalized to the half-split structure constants with $L_{2} = L_{3} > 2$, which correspond to taking $\ell_{23}>1$ while keeping $\ell_{12} = \ell_{31} = 1$, if needed. The only contribution in (\ref{final}) that depends on the value of $\ell_{23}$ is $c_{(0, 1, 0)} \sim g^{2\ell_{23}+2}$. The interested reader can find its four loop expression for $\ell_{23} = 2, 3$ in Appendix \ref{Reg}.%
\footnote{With a bit more work, one could also cover non-symmetric configurations with $\ell_{12}\neq \ell_{31}$. In the $\mathfrak{sl}(2)$ set up, this necessarily implies that one of the two adjacent bridges disappears, since $\ell_{12}+\ell_{31} = L_{1} = 2$ for the Konishi operator. This is an extremal configuration. The situation is not that worrisome, however, since the configuration remains non-extremal in the $\mathfrak{su}(2)$ set-up. As such, the truly adjacent extremal geometries fall in the compact sector and correspond to ``negative bridge length'' in the noncompact set-up.}

The two and three loop results in (\ref{final}) agree with those obtained in \cite{short,Eden:2015ija,3loops}. The four loop corrections to $c_{(1, 0, 0)}$ and $c_{(0, 1, 0)}$ were derived by pushing to one more loop the analysis carried out in \cite{3loops}, as explained in Appendix \ref{Reg}. The coefficient $c_{(1, 0, 1)}$ is the main outcome of our analysis. It takes into account \textit{all} the wrapping effects discussed earlier, including the shift of the roots and the associated modification of the Gaudin norm.
\begin{figure}[t]
\centering
\includegraphics[height=6cm]{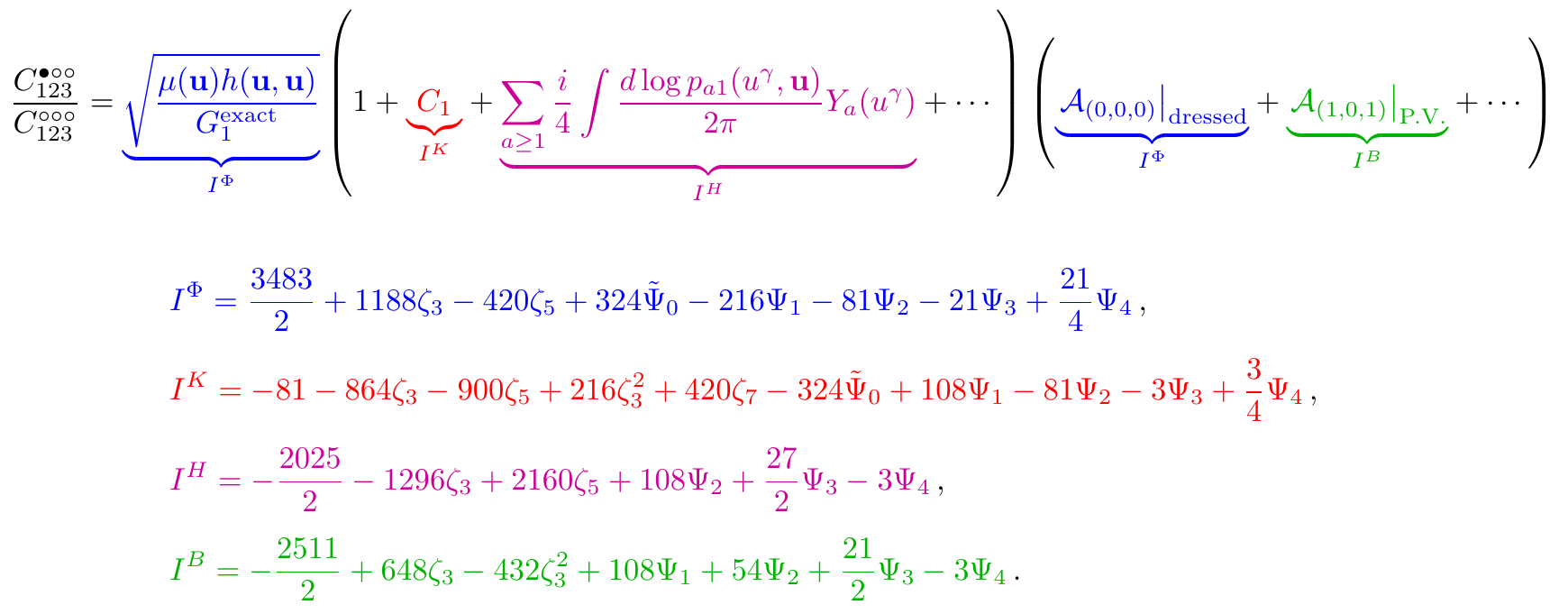}
\vspace{-10pt}\caption{The regularization of $\mathcal{A}_{(1,0,1)}$, together with the wrapping correction to the Gaudin norm, leads to the ``dressed" hexagon series shown at the top. The four loop results for the various contributions are listed above. $I^{\Phi}$ comes from the terms proportional to the finite-size correction to the Bethe roots $\Phi$ whereas $I^{K}$ denotes the contribution involving the scattering kernel $\mathcal{K}_{aa}$. $I^{H}$ is the term containing the symmetric part of the hexagon form factor and $I^{B}$ is the double integral regularized by the principal-value prescription.
 Although each contribution contains non-Riemannian transcendental numbers ($\Psi_n$'s in the formulae above), they cancel out if we sum them all, leaving only standard $\zeta$ functions. The result after the sum is given by the last line in (\ref{final}), $c_{(1,0,1)}$. For details of the calculation of each term, see Appendix \ref{wrap}. }\label{renexpfig}
\end{figure}

{The various contributions to $c_{(1,0,1)}$ and their relation to the renormalized expansion given in section \ref{Cc} are summarized in figure \ref{renexpfig}.
As shown in the figure, each contribution to $c_{(1, 0, 1)}$ looks complicated and appears polluted by non-Riemannian transcendental numbers, similar to those found in the intermediate steps of the five loop computation of the L\"uscher energy formula \cite{Bajnok:2009vm}.} Only after adding everything up do we find the nice number given in the last line of (\ref{final}), expressible in terms of Riemann zeta values $\zeta(z)$ solely. Most of the integrals we had to take are of the type encountered earlier for the wrapping shift of the energy and can be computed using the same techniques, see \cite{Bajnok:2008bm,Bajnok:2008qj,Bajnok:2009vm}. The hardest ones are those akin to the double wrapping corrections for the energy, that is the bulk double integral (\ref{Bulk}) and the single integral $C_{1}$ involving the scattering kernel in (\ref{C12}). {(In figure \ref{renexpfig}, they correspond to $I^{B}$ and $I^{K}$ respectively.)} The former integral is difficult to take because of the many residues and the fact that one needs to double sum them in the end. We sketch the main steps of our calculation in Appendix \ref{wrap}, since each step is too bulky to be shown in full form. The latter integral is demanding for a different algebraic reason, which is that one must first of all obtain the expression for the scattering kernel $\mathcal{K}_{aa}$ of all the bound states ($a = 1 , 2, \ldots\,$). The task is arduous because the bound state S-matrices are bulky and notoriously unhandy. 
\begin{figure}[t]
\centering
\includegraphics[scale=0.55]{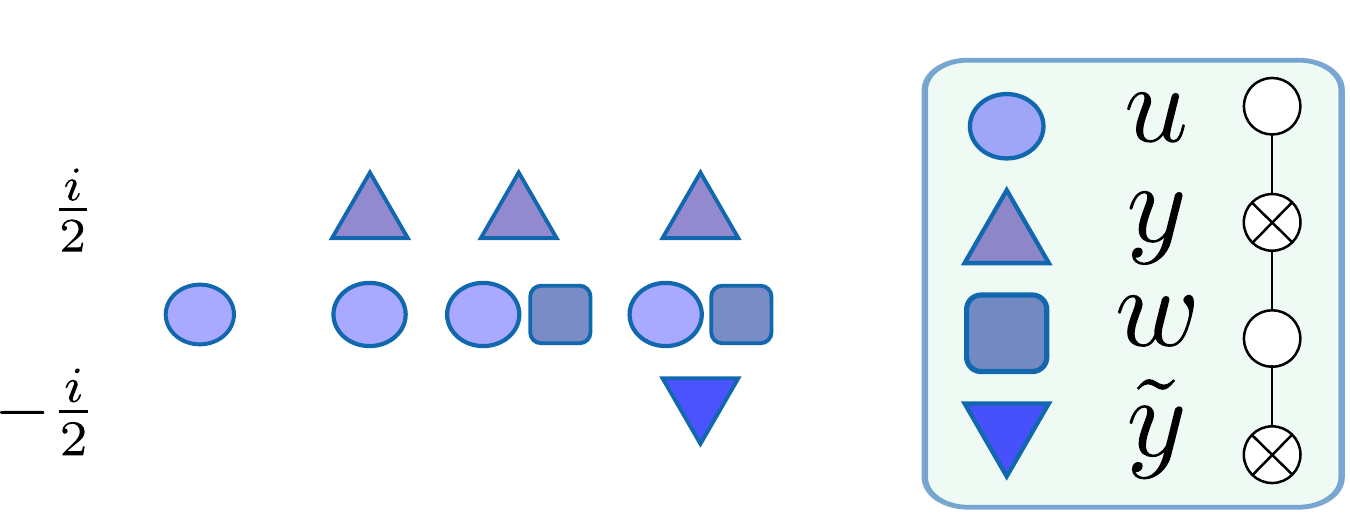}
\vspace{0.5cm}
\caption{The exact Bethe string basis for a fundamental magnon, describing $4$ different states contained in the module. Each string is made out of the momentum-carrying root $u$ and roots of type $y,w,\tilde{y}$, each of which is associated with the Dynkin node of $SU(2|2)$ shown on the right.  They are represented, respectively, as circle, upper triangle, square, and lower triangle. The states are obtained by appropriately attaching the roots $y,w,\tilde{y}$ to the momentum carrying root, as shown in the left figure. These patterns of roots can be read off from the structure of (the generating function of) the transfer matrix \cite{Beisert06}. For details and generalization to the bound states, see Appendix \ref{AFF}.}\label{rootsstring0}
\end{figure}
{To facilitate the analysis, we utilized an alternative basis, akin to the one found in the study of the matrix part of the flux tube OPE integrand \cite{Frank,Mpaper}, which is characterized by ``exact Bethe strings"; namely a set of rapidities each of which is separated by $i$ or $i/2$. See figure \ref{rootsstring0} for a particular example. A gratifying feature of this basis is that it diagonalizes the S-matrix, hence reducing the problem of computing the matrix structure to the multiplication and the summation of the eigenvalues of the S-matrices. (This feature is reminiscent of the standard nested Bethe ansatz, and it is in fact very likely that the basis can be understood as the states satisfying infinitely twisted nested Bethe equations.%
\footnote{This is, indeed, the case for the strings underlying the matrix part of the flux tube integrand \cite{Frank}. In this paper however, we shall not work out such a relation.})
In particular, this specificity allowed us to generate more easily the expressions for the kernels of the bound states, as explained in Appendix \ref{AFF}.}

At the end, after adding all the coefficients in (\ref{final}) together and squaring the result, it is rewarding to find that our expression matches perfectly with all the terms in the field theory answer (\ref{QFT}).

\subsection{Comparison with the string theory}

The check performed at weak coupling is interesting but limited to a special configuration. A more complete series of tests can be made at strong coupling by considering semiclassical configurations for which all quantum numbers ($\ell_{ij}$ and $M$) are of order $\sqrt{\lambda} \gg 1$. Structure constants in this regime should match with their stringy counterparts and fit the formula recently derived for them in \cite{Shota} building on previous analyses \cite{Janik:2011bd,Kazama:2011cp}. The latter expresses the classical area of the splitting string, or, equivalently, the log of the structure constant, directly in terms of the quasi-momenta which characterize the states at the boundary. The string theory formula is a nice laboratory for testing the hexagon recipe, and its amendments, since it resums an infinite number of real and virtual processes. It contains, in particular, the information about \textit{all} the wrapping corrections that survive to the strong coupling limit, in the semiclassical regime. We shall see that the first of these corrections, which is absent in the naive hexagon expansion, is properly captured by our renormalized formula.

The string formula for the structure constant \cite{Shota} splits into two pieces, referred to as the real and the mirror part, respectively. We start with the real component. It is localized on the distribution of Bethe roots and reads
\beq\label{real}
\log{C^{\bullet\circ\circ}_{123}} |_{\textrm{real}} = \oint_{\Gamma}\frac{du}{2\pi}\textrm{Li}_{2}\big(e^{i\hat{\mathfrak{p}}_{31}(x)}\big) -\frac{1}{2}\oint_{\Gamma}\frac{du}{2\pi}\textrm{Li}_{2}\big(e^{2i\hat{\mathfrak{p}}_{1}(x)}\big)\, ,
\eeq
where $\textrm{Li}_2(z)$ is the dilogarithm, $x+1/x = 4\pi u/\sqrt{\lambda} = O(1)$ and the integrals go anticlockwise along a contour $\Gamma$ encircling closely the Bethe roots. The quasi-momentum of the state at the $i$-th puncture is denoted by $\hat{\mathfrak{p}}_{i}(x)$ and we introduced $\hat{\mathfrak{p}}_{ij}(x) := \hat{\mathfrak{p}}_{i}(x)+\hat{\mathfrak{p}}_{j}(x)-\hat{\mathfrak{p}}_{k}(x)$. The string theory prediction (\ref{real}) is structurally identical to the semiclassical spin chain expression of \cite{Gromov:2011jh,Bettelheim:2014gma}. The sole difference resides in the quasi-momenta, which take different forms at weak and strong coupling. The hexagon approach allows us to interpolate between these two semiclassical results, which come, in both cases, entirely from the asymptotic structure constant \cite{short,Jiang:2016ulr}. In particular, one easily proves, using the methods of \cite{Gromov:2011jh,Bettelheim:2014gma} for instance, that the first integral in (\ref{real}) originates from the sum over the partitions $\mathcal{A}_{(0, 0, 0)}$ and the second one from the Gaudin determinant $G_{1}$.

{The real part (\ref{real}) does not contain the information about the wrapping corrections, since the correction to the quasi-momentum due to the finite-size effects, $\hat{\mathfrak{p}}_{1}(x_i) \rightarrow \hat{\mathfrak{p}}_{1}(x_i) + \Phi_{i}$, is one-loop suppressed at strong coupling \cite{Gromov:2009tq} for the class of finite gap solutions we are considering.}\footnote{We are not discussing giant magnons, here; by assumption all our magnons have spin chain momenta of order $1/\sqrt{\lambda}$.}
Schematically, this is so because the phase $\Phi_{i}$ contains a derivative $\partial_{u}$ of the S-matrix, which is small semiclassically $u \sim \sqrt{\lambda}$. (Technically, one needs a large rapidity gap between the two arguments of the S-matrix, to assure that the derivative is suppressed. This is automatically so for a mirror and a real rapidity lying in their respective perturbative domains.)

We move thus to the mirror part of the string theory result. It comprises 5 terms,
\beq\label{mirrorstring}
\log{C^{\bullet\circ\circ}_{123}} |_{\textrm{mirror}} = \mathcal{I}_{(1, 0, 0)} + \mathcal{I}_{(0, 1, 0)} + \mathcal{I}_{(0, 0, 1)} - \mathcal{I}_{(1, 0, 1)} + \mathcal{I}_{(1,1,1)}\, ,
\eeq
where the numbers in brackets indicate which mirror channels are excited. They are all of the same type,
\beq\label{mirror-string}
\mathcal{I}_{\star} = \int_{U^{-}}\frac{du}{2\pi}\left[\textrm{Li}_{2}\big(e^{i\hat{\mathfrak{p}}_{\star}(x)}\big) + \textrm{Li}_{2}\big(e^{-i\hat{\mathfrak{p}}_{\star}(1/x)}\big) - 2\textrm{Li}_{2}\big(e^{i\tilde{\mathfrak{p}}_{\star}(x)}\big)\right]\, ,
\eeq
up to the choice of the quasi-momentum $\mathfrak{p}_{\star}$, which depends on the channel $\star$, e.g.
\beq
\mathfrak{p}_{(1, 0, 0)}(x) = \mathfrak{p}_{12}(x), \qquad \mathfrak{p}_{(1, 0, 1)}(x) = \mathfrak{p}_{12}(x) + \mathfrak{p}_{31}(x) = 2\mathfrak{p}_{1}(x), \qquad \textrm{etc.} \eeq
The integrals are all taken along the mirror contour $U_{-}$ which runs from $x=-1$ to $x=1$ along the lower part or the unit circle $|x|=1$. We also have that $\hat{\mathfrak{p}}_{i}(x) = \tilde{\mathfrak{p}}_{i}(x) = 2\pi x L_{i}/((x^2-1)\sqrt{\lambda})$ for a vacuum state, while for an excited state in the $\mathfrak{sl}(2)$ sector ($i=1$ in our case) one must distinguish between the AdS quasi-momentum $\hat{\mathfrak{p}}_{i}$, which is non trivial, and its sphere counterpart $\tilde{\mathfrak{p}}_{i}$, which takes its vacuum value.

We can isolate the various terms in (\ref{mirrorstring}) by taking appropriate long bridge limits. For instance, were we interested in $\mathcal{I}_{(0, 1, 0)}$, we would take $\ell_{12}, \ell_{31}\rightarrow \infty$ and keep $\ell_{23}$ finite. (Since all the bridges' lengths must be at least $\sim \sqrt{\lambda}\gg 1$ semiclassically, we would actually consider $\ell_{12}, \ell_{31} \gg \sqrt{\lambda} \sim \ell_{23}$.) A systematic analysis of this contribution was carried out recently \cite{Jiang:2016ulr}, showing that it is completely captured by the hexagon series. Here, we want to excite the adjacent channels instead, so we will consider the alternative situation where $\ell_{23}\rightarrow \infty$ and $\ell_{12}, \ell_{31}$ are hold fixed. The term $\mathcal{I}_{(0, 1, 0)}$ decouples in this case and so does the cubic vertex $\mathcal{I}_{(1, 1, 1)}$. {Therefore what remain are the terms $\mathcal{I}_{(1, 0, 0)}$, $\mathcal{I}_{(0, 0, 1)}$ and $\mathcal{I}_{(1, 0, 1)}$. In what follows, we demonstrate that our formula properly reproduces those terms.}

We shall work in the linearized approximation $\textrm{Li}_{2}(z) \rightarrow z$, which amounts to keeping only the lightest string modes. Higher orders correspond to heavier states. Though we do have bound states in our L\"uscher formula, which definitely contribute to the non linearities, we are missing the multiparticle states, which contribute at the same level. (Bound states are at threshold at strong coupling and as such cannot be isolated from the continuum of multiple particle states.) Therefore, we cannot probe the string theory result beyond the term linear in $z$, with our formula.

The main dynamical feature of the hexagon form factor at strong coupling is that their module trivializes,
\beq\label{hh}
h(u, v)h(v, u) = 1+O(1/\lambda)\, ,
\eeq
and so does the measure $\mu(u) = 1+\ldots\,$. The phase dominates and it is governed by the S-matrix, $h(u, v)/h(v, u)  = S(u, v)$. This has two essential consequences: 1) the hexagon series exponentiates, in lack of interactions between the mirror magnons, and 2) all of the interaction with the roots goes into the quasi-momenta. We can illustrate both points on the mirror corrections in the adjacent channels. A particle in these channels interacts with the Bethe state through the combination
\beq
\frac{h(u, \alpha)}{h(\bar{\alpha}, u)} \simeq h(u, \alpha) h(u, \bar{\alpha}) = h(u, \textbf{u}) \simeq \sqrt{S(u, \textbf{u})}\, .
\eeq
It immediately follows that the sum over the partitions factors out,
\beq
\mathcal{A}_{(1, 0, 0)} \simeq  \mathcal{A}_{\textrm{asympt}} \times \mathcal{I}_{(1, 0, 0)}\, ,
\eeq
and, after using some well-known strong coupling identities summarized in Appendix \ref{Strong}, that the remaining factor is controlled by the quasi-momenta,
\beq
e^{ip(u)\ell_{12}} \sqrt{S(u, \textbf{u})} \times T(u) = e^{i\hat{\mathfrak{p}}_{12}(x)} + e^{-i\hat{\mathfrak{p}}_{12}(1/x)} -2e^{i\tilde{\mathfrak{p}}_{12}(x)} \, ,
\eeq
in agreement with the linearized string theory formula (\ref{mirror-string}). The matching is complete after ones rotates the rapidity $u$ to the mirror kinematics, which amounts, in the perturbative strong coupling regime, to continuing it to the lower half unit circle, as in (\ref{mirror-string}). We would of course get the same answer for $\mathcal{A}_{(0, 0, 1)}$, up to the substitution $\mathfrak{p}_{12}\rightarrow \mathfrak{p}_{31}$. It is as straightforward to check the exponentiation at the level of the bulk contribution $B$. In that case, there are two mirror magnons, which only interact through (\ref{hh}) or, in other words, do not interact at all. Their couplings to the Bethe roots are factorized and controlled by the quasi-momenta, resulting in
\beq\label{Bstrong}
B = \mathcal{A}_{\textrm{asympt}} \times \mathcal{I}_{(1, 0, 0)}\times  \mathcal{I}_{(0, 0, 1)} + \ldots \, ,
\eeq
as expected. (We could also average over the two $i0$'s options and work with the principal valued integral. The choice of prescription for $B$ turns out to be subleading at strong coupling.)

In principle one should worry about the dots in (\ref{Bstrong}) as they can contribute to $\mathcal{I}_{(1, 0, 1)}$. E.g., there could be a pair of singularities in the $B$ integrand that pinches the contour of integration at strong coupling \cite{Basso:2013vsa} and provokes the clustering of the two excitations \cite{Jiang:2016ulr}. This scenario is not realized for $B$ if the two excitations are both fundamental magnons. The poles, in this case, stand right on top of each other, at $(u-v+i0)^2 = 0$, and, as such, are totally harmless. In addition, there could be a more regular type of contributions coming directly from the subleading term $\sim 1/\sqrt{\lambda}$ in the link (\ref{hh}) between the two excitations. When such a term exists, it exponentiates and corrects the area $\sim \sqrt{\lambda}$. This is happening e.g.~for the pentagon OPE series at strong coupling \cite{Basso:2013vsa} where the correction generates the piece that depends on the kernel of the TBA equation for the minimal surface \cite{Alday:2009dv,Alday:2010ku}. There is nothing similar for the structure constant $C^{\bullet\circ\circ}$, simply because the first correction in (\ref{hh}) is $O(1/\lambda)$; a regular contribution of the kernel type would actually be incompatible with the structure of the string theory result (\ref{mirror-string}). So, in conclusion, the dots in (\ref{Bstrong}) can be safely discarded; they stand for loop corrections.

It is worth noting that we would reach the same conclusion if we were blindly using the bare hexagon amplitude $\mathcal{A}_{(1, 0, 1)}$ instead of $B$. The divergences affecting the hexagon series $\mathcal{A}$ are superficially subleading at strong coupling, since the double pole in $1/h(u, v)h(v, u)$ is naively suppressed by two powers of $\sqrt{\lambda}$. In this sense, the bare expansion is unable to explain the string wrapping correction $\mathcal{I}_{(1, 0, 1)}$, though its singularities are clearly hinting at the fact that something is missing.

The missing ingredient is coming from the contact term dressing the norm in (\ref{dressed}) and, more precisely, from $C_{1}$. (The $d\log{p}$ component in (\ref{dressed}) is subleading.) The contact term is controlled by the scattering kernel, $K(u, v)\sim \partial_{u}\log{S(u, v)}$, which looks negligible, at first sight, due to the derivative $\partial_{u} \sim 1/\sqrt{\lambda}$. The estimate is naive, however, because there is no large gap between the two rapidities, $u$ and $v$, in the integrand for $C_{1}$. The gap is, in fact, as small as it can be, since the kernel in $C_{1}$ is evaluated at coinciding points, $u=v$. For such nearby excitations, the kernel is $O(1)$ and, hence, $C_{1}$ is $O(\sqrt{\lambda})$. It is then just a matter of algebra to work out its explicit expression, see Appendix \ref{Strong},
\beq
C_{1} = -\int_{U^{-}}\frac{du}{2\pi}\big(e^{2i\hat{\mathfrak{p}}_{1}(x)}+e^{-2i\hat{\mathfrak{p}}_{1}(1/x)}-2e^{2i\tilde{\mathfrak{p}}_{1}(x)}\big) \, ,
\eeq
and conclude that it matches with the linearized version of the string integral $\mathcal{I}_{(1, 0, 1)}$.

{Before closing this section, let us comment on the term $\mathcal{I}_{(1,1,1)}$, which is not captured by our analysis or the one in \cite{Jiang:2016ulr}. As with the term $\mathcal{I}_{(1,0,1)}$, this term cannot be reproduced by the naive hexagon expansion. To compute it, one has to repeat the analysis here by including the mirror magnons in the channel $23$ and carefully deal with all possible singularities that can arise when magnons in different channels have coincident rapidities\footnote{Given the structure of the term, it is tempting to speculate that this term would arise when magnons in all three channels have the same rapidities. If so, it would be a more subtle effect and would require more careful analysis than the one performed in this paper.}. It is an extremely interesting problem to see if such a renormalized expansion could reproduce the term correctly, but we will leave it for a future investigation.}

\section{Conclusion}\label{Sect6}

We studied in this paper the divergences that affect the hexagon gluing procedure at the leading wrapping order. We have seen that they are due to virtual magnons that wrap around a non-BPS operator. Their contributions decouple from the bulk of the geometry and, after regulating them by cutting off the volume of the mirror theory, can be absorbed in the normalization of the operator. What remains, after this renormalization has been performed, is a definite prescription for handling the decoupling singularities and an accompanying series of contact terms that dress the ingredients of the hexagon construction. We have checked their correctness at both weak and strong coupling through comparisons with the gauge and string theory predictions. It would be very interesting to test the method further, by moving to higher loops or to more generic non-BPS operators, not to mention the structure constants for two or three excited operators.

\begin{figure}
\begin{center}
\includegraphics[scale=0.40]{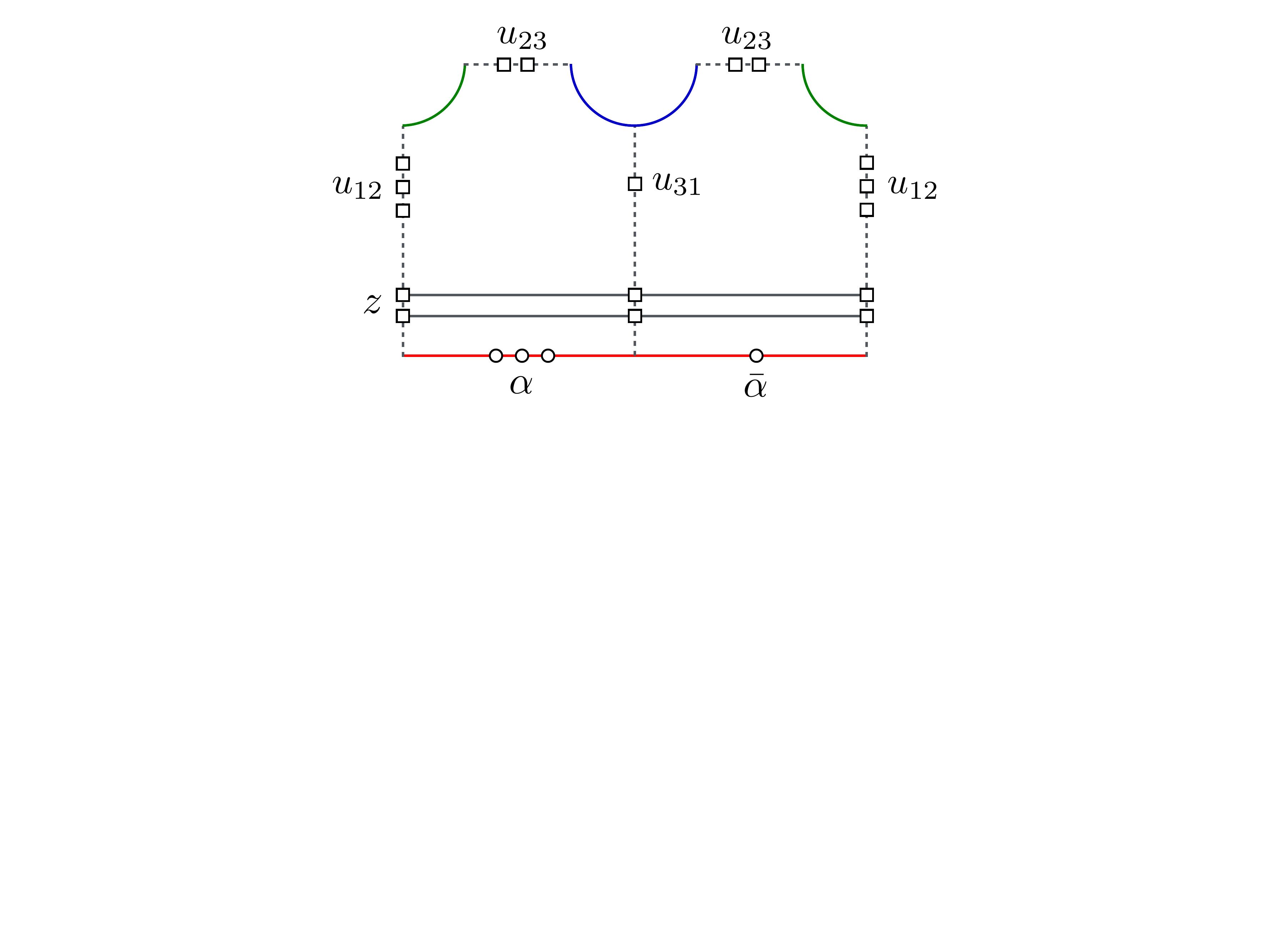}
\end{center}
\vspace{-6.2cm}
\caption{Wrapped hexagons at higher wrapping orders. Our analysis shows that a proper treatment of the divergences plaguing the bare hexagon expansion results in the introduction of ``wrapping magnons", here depicted as straight lines at the bottom of the figure. They live close enough to the edges not to be deflected as they cross the hexagons but can nonetheless influence the physics taking place in the bulk of the pair of pants. They are like thermal loops sourced by the half infinite tube representing the operator. We can have configurations involving any number of bulk and wrapping magnons at the same time. The expectation is that the latter will thermalize the former, resulting in some sort of TBA dressing of the bare hexagon amplitudes.}\label{final-g}
\end{figure}

In the end, our amendment to the hexagon construction boils down to introducing a new object that can ``propagate'' on top of the 3-pt function geometry. It looks like a thermal loop winding around an operator and, as such, is intrinsically tied to the compactification procedure. This magnon does not entirely decouple from the rest of the geometry and sort of lives half way between the bulk and the boundary. We can have arbitrarily many of these ``wrapping magnons'' encircling an operator, besides the more regular mirror magnons that stitch the bulk of the hexagons together. See figure \ref{final-g}.

An important open problem is to understand how these wrapping magnons interact with themselves, with the bulk magnons that live along the bridges and with the real magnons that stand at the boundary. It is tempting to think that their main role is to ``thermalize" the hexagon expansion by injecting some TBA data all over the place. We played with this idea in this paper at the level of the asymptotic structure constant. However, their effects should not be restricted to that specific area and we expect them to affect the mirror corrections as well. At least, this is what is happening for the form factor expansion of thermal correlators of local operators \cite{Leclair:1999ys,Saleur:1999hq,Pozsgay:2007gx,Pozsgay:2009pv,Pozsgay:2010cr,Pozsgay:2014gza}, which shares many similarities with our set-up. More precisely,  in theories with a diagonal scattering at least, the thermal expectation values of local operators are known to be subject to a generalized form factor expansion controlled by the TBA pseudo-energies, which soak up the finite size effects of the cylinder. This is the content of the Le Clair-Mussardo formula \cite{Leclair:1999ys}. Does anything similar exist for the hexagon expansion of the renormalized structure constant? In this respect, it is worthwhile keeping in mind that resumming the thermal loops is a priori independent from our ability to resum the hexagon expansion itself. In fact, it is reasonable to believe that the former resummation can be done at any given configuration of bulk magnons. Still, as thermal as this dressing might be, there is no universal recipe for addressing it and it could also be that resumming the bare expansion would provide a better start for developing the TBA story.

Within our approach, warming up the hexagons entails adding more wrapping magnons. This step is technically challenging. In order to understand the multiparticle divergences and their subtractions, it might be worth exploring the similarities with the treatment of the diagonal structure constants \cite{Bajnok:2014sza,Hollo:2015cda,Jiang:2015bvm,Jiang:2016dsr} (see also \cite{Pozsgay:2007gx,Pozsgay:2009pv} for a discussion of diagonal form factors of local operators in integrable relativistic QFTs). In both cases one faces divergences arising from the decoupling poles and in both cases the divergences are tamed by working in finite volume. One difference is that the cut off is built in for the diagonal structure constants, which deal primarily with magnons in the real channels. The divergences are ultimately resolved because one is considering finite length operators. We were dealing here with magnons in the mirror channels, for which there is no cut off a priori -- the length along the mirror direction being in the dual picture the length of the string ending on the operator, which is infinitely far away. We had therefore to introduce a cut off by hand by slightly moving the operator off the boundary -- the volume of the mirror theory playing the role of a renormalization scale $R \sim \log{\Lambda/\mu}$. The origin of the cut off is optimistically not that relevant and one might hope that the two problems are somehow related through a mirror rotation or, at least, that they can be addressed using similar techniques.

Another source of inspiration and cross checks for our multiparticle problem should come from the study of the finite size corrections to the Neumann coefficients at strong coupling \cite{RZ}. Indeed, both can be seen as resulting from the wrapping of an octagon, albeit the problem of relating precisely these two octagons is not entirely settled.

\section*{Acknowledgments}

We thank Pedro Vieira for collaboration at an early state of this project and invaluable comments on the draft. {We gratefully acknowledge helpful discussions with Frank Coronado on the twisted Bethe basis and with Romuald Janik on the L\"uscher energy formula. We are thankful to Zolt\'an Bajnok, Jo\~ao Caetano, Romuald Janik, Amir-Kian Kashani-Poor, Vladimir Kazakov, Ivan Kostov, Didina Serban and Amit Sever for comments on the manuscript.} Research at the Perimeter Institute is supported in part by the Government of Canada through NSERC and by the Province of Ontario through MRI. V.G. is funded by FAPESP grant 2015/14796-7 and CERN/FIS-NUC/0045/2015. 

\appendix

\section{Distributions, contact terms and a little puzzle}\label{Ct}

In this appendix, we give a detailed derivation of the contact term (\ref{first}). We also convert our formulae from the $i0$ to the principal value prescription and discuss the reality issue of the dressed sum over the partitions (\ref{dressing}). 

\subsection*{Contact terms}

The starting point is the octagon transition depicted in figure \ref{oct1}. It involves two mirror bound states of length $a$, $|w, a\mathcal{i}$ and $|u, a\mathcal{i}$, on the far left edge and on the far right edge of the octagon, respectively, and a third one $|v, b\mathcal{i}$ that mediates the interaction through the middle cut. We specialize to the case where $b=a$, as otherwise there is no decoupling pole to worry about and no contact terms get generated.

We wish to expand this transition in the diagonal limit $w\rightarrow u$. To avoid possible artifacts of our unconventional normalization for the incoming and outgoing states, we shall carry out the analysis in momentum space. The corresponding momentum states are obtained through a simple rescaling of the mirror states,
\beq
|\tilde{p}_{a}(u), a\mathcal{i} = \sqrt{\tilde{\mu}_{a}(u^{\gamma})} |u, a\mathcal{i}\, , \qquad \tilde{\mu}_{a}(u^{\gamma}) = \frac{du}{d\tilde{p}_{a}(u)} \mu_{a}(u^{\gamma})\, ,
\eeq
with $\tilde{p}_{a}(u) = -iE_{a}(u^{\gamma}) = 2u - 2g(1/x^{[+a]}+1/x^{[-a]})$ the momentum of the mirror magnon. This map is designed such as to get the usual (infinite volume) inner product between momentum states,
\beq
\mathcal{h}u, a|v, b\mathcal{i} = \frac{2\pi}{\mu_{a}(u^{\gamma})} \delta_{ab}\delta(u-v) \qquad \Rightarrow \qquad \left<\tilde{p}, a|\tilde{q}, b\right> = 2\pi \delta_{ab}\delta(\tilde{p}-\tilde{q})\, .
\eeq
The hexagon transition between two momentum states is
\beq
\left<\tilde{p}, a\right|\mathcal{H}\left|\tilde{q}, b\right> \equiv \frac{1}{\tilde{h}_{ba}(\tilde{q}, \tilde{p})} = \frac{\sqrt{\tilde{\mu}_{b}(v^{\gamma})}\sqrt{\tilde{\mu}_{a}(u^{\gamma})}}{h_{ba}(v^{\gamma}, u^{\gamma})}\, ,
\eeq
and it has unit residue for $b=a$ in the decoupling limit $\tilde{q} \sim \tilde{p}$\, ,
\beq
\lim\limits_{\tilde{q}\rightarrow \tilde{p}}\, \frac{i(\tilde{q}-\tilde{p})}{\tilde{h}_{ba}(\tilde{q}, \tilde{p})} = \delta_{ab}\, .
\eeq
Equivalently, integration is performed with the measure $d\tilde{p}/(2\pi)$ and nothing else.

In this normalization, the integrand of the octagon transition is
\beq\label{intilde}
\tilde{\textrm{int}}_{a}(w, v, u) = \frac{\tilde{\Psi}_{a}(w, v, u)}{\tilde{h}_{aa}(\tilde{r}+i0, \tilde{q})\tilde{h}_{aa}(\tilde{q}, \tilde{p}-i0)} \, ,
\eeq
where $\tilde{p} = \tilde{p}_{a}(u)$, etc.,
\beq\label{tildePsi}
\tilde{\Psi}_{a}(w, v, u) = e^{-\ell_{31}\tilde{E}_{a}(v)-\ell_{12}\tilde{E}_{a}(u)} \sum_{\alpha\cup\bar{\alpha} = \textbf{u}}a_{\alpha\bar{\alpha}}\frac{h_{a1}(w^{\gamma}, \alpha)h_{a1}(v^{\gamma}, \bar{\alpha})}{h_{1a}(\alpha, v^{\gamma})h_{1a}(\bar{\alpha}, u^{\gamma})}\, \mathcal{N}_{a}(w, v, u)\, ,
\eeq
and with $\mathcal{N}_{a}(w, v, u)$ the matrix part depicted in figure \ref{MatG}. The $\tilde{q}$ integration of (\ref{intilde}) is ill defined if we set $\tilde{r} = \tilde{p}$ because the two $i0$'s are then in conflict with each other (they pinch the contour at $\tilde{q} = \tilde{p} = \tilde{r}$). We can isolate the problem, prior to set $\tilde{r} = \tilde{p}$, by slightly deforming the contour of integration for $\tilde{q}$, such that both the pole at $\tilde{q} = \tilde{p}$ and the one at $\tilde{q} = \tilde{r}$ are integrated from below (i.e., with the same $i0$ prescription). The resulting integral is perfectly well defined for $\tilde{r} = \tilde{p}$ and coincides with the bulk contribution (\ref{Bulk}) for $b=a$. The singularity at $\tilde{r} = \tilde{p}$ now entirely resides in the residue at $\tilde{q} = \tilde{p}-i0$ extracted along the way. Its Taylor expansion around $\tilde{r}=\tilde{p}$ yields both the polar part and the contact term,
\beq\label{res}
\begin{aligned}
\oint\limits_{\tilde{q}=\tilde{p}-i0}\frac{d\tilde{q}}{2\pi} \, \tilde{\textrm{int}}_{a}(w, v, u) &=  -\frac{\tilde{\Psi}_{a}(w, u, u)}{\tilde{h}_{aa}(\tilde{r}+i0, \tilde{p}-i0)} \\
&= \frac{i\tilde{\Psi}_{a}(u, u, u)}{\tilde{r}-\tilde{p}+i0} - \partial_{\tilde{r}}\bigg[\frac{(\tilde{r}-\tilde{p})\tilde{\Psi}_{a}(w, u, u)}{\tilde{h}_{aa}(\tilde{r},\tilde{p})}\bigg]\bigg|_{w=u} + \ldots\, ,
\end{aligned}
\eeq
with a clockwise contour of integration around $\tilde{q} = \tilde{p}-i0$ in the LHS. As explained in Section \ref{Gluing}, see equation (\ref{regularization}) in particular, when the regulator is turned on, the pole $i/(\tilde{r}-\tilde{p}+i0)$ in the transition is replaced by a linear divergence $R/2$ in the mirror volume. Thereby, the first term in the RHS of (\ref{res}) reads, in the regularized theory,
\beq
\frac{R}{2}\times \tilde{\Psi}_{a}(u, u, u) = \frac{R}{2} \mathcal{A}_{(0, 0, 0)}\, e^{-\tilde{E}_{a}(u) L_{1}} S_{a1}(u^{\gamma}, \textbf{u})T_{a}(u^{\gamma})^2 = \frac{R}{2} \mathcal{A}_{(0, 0, 0)} Y_{a}(u^{\gamma})\, ,
\eeq
where we simplified the ratios in (\ref{tildePsi}) using the Watson equation, e.g., $h_{a1}(u^{\gamma}, \alpha)/h_{1a}(\alpha, u^{\gamma})$ $= S_{a1}(u^{\gamma}, \alpha)$, and expressed, using $\mathcal{N}_{a}(u, u, u) = T_{a}(u^{\gamma})^2$, the matrix part in terms of the transfer matrix.

The contact term arises from the derivative in the RHS of (\ref{res}). Dressing the latter with the measure $d\tilde{r}$, for convenience, and relabelling the dummy variables yield
\beq\label{Sec}
d\tilde{p}\, \tilde{\Psi}_{a}(u, u, u)\partial_{\tilde{p}}\bigg[\frac{(\tilde{r}-\tilde{p})}{\tilde{h}_{aa}(\tilde{p},\tilde{r})}\bigg]\bigg|_{w=u}\, ,
\eeq
and
\beq
id\tilde{p}\, \partial_{\tilde{p}}\tilde{\Psi}_{a}(u, w, w)|_{w=u} = idu \, \partial_{u}\tilde{\Psi}_{a}(u, w, w)|_{w=u}\, .
\eeq
The second term was discussed at length in section \ref{Lues}. Here we focus on the first one. To compute the derivative, we use the Watson equation to rewrite
\beq
\frac{(\tilde{r}-\tilde{p})}{\tilde{h}_{aa}(\tilde{p},\tilde{r})} = \frac{if(\tilde{p}, \tilde{r})}{\sqrt{-S_{aa}(u^{\gamma}, w^{\gamma})}}\, ,
\eeq
where $S_{aa}(u^{\gamma}, v^{\gamma})$ is the scalar part of the mirror-mirror S-matrix and where
\beq\label{f-fun}
f(\tilde{p}, \tilde{r}) = f(\tilde{r}, \tilde{p}) = \sqrt{-\frac{(\tilde{p}-\tilde{r})(\tilde{r}-\tilde{q})}{\tilde{h}_{aa}(\tilde{p}, \tilde{r})\tilde{h}_{aa}(\tilde{r}, \tilde{p})}}\, .
\eeq
The symmetric factor $f$ is smooth at $\tilde{p} = \tilde{r}$ and such that $f(\tilde{p}, \tilde{p}) = 1$. Hence, the derivative $\partial_{\tilde{p}}$ acts trivially on it at $\tilde{r} = \tilde{p}$,
\beq\label{nothing}
\partial_{\tilde{p}} f(\tilde{p}, \tilde{r})|_{\tilde{r} = \tilde{p}} = \frac{1}{2}(\partial_{\tilde{p}}+\partial_{\tilde{r}}) f(\tilde{p}, \tilde{r})|_{\tilde{r} = \tilde{p}}  = \frac{1}{2}\frac{d}{d\tilde{p}} f(\tilde{p}, \tilde{p}) = 0\, .
\eeq
Only the S-matrix factor contributes to (\ref{Sec}) allowing us to write the contact term in terms of the scattering kernel,
\beq
\frac{1}{2i}\tilde{\Psi}_{a}(u, u, u)du\, \partial_{u}\log{S_{aa}(u^{\gamma}, w^{\gamma})}|_{w=u} = \frac{1}{2}Y_{a}(u^{\gamma})du\, K_{aa}(u^{\gamma}, u^{\gamma})\, .
\eeq
This is precisely the abelian component of the contact term $C_{1}$ in (\ref{C12}).

\subsection*{Distributions}

Averaging over the $+i0$ and the $-i0$ prescription is, by definition, the same as using a principal value distribution. According to the familiar identity,
\beq
\frac{1}{(u-v+i0)^2} = \frac{1}{2}\bigg[\frac{1}{(u-v+i0)^2} + \frac{1}{(u-v-i0)^2}\bigg] + i\pi \partial_{u}\delta(u-v)\, ,
\eeq
the $i0$ and the principal value prescription only differ in a local  imaginary part. Hence, changing prescription in the integral (\ref{Bulk}) is merely adding a new contact term, obtained by applying the replacement 
\beq\label{rule}
\frac{\mu_{a}(u^{\gamma})\mu_{b}(v^{\gamma})}{h_{ab}(u^{\gamma}+i0, v^{\gamma})h_{ba}(v^{\gamma}, u^{\gamma}+i0)} \rightarrow -i\pi\frac{\mu_{a}(u^{\gamma})(u-v)(v-u)\mu_{a}(v^{\gamma})}{h_{aa}(u^{\gamma}, v^{\gamma})h_{aa}(v^{\gamma}, u^{\gamma})} \delta_{ab}\partial_{u}\delta(u-v)\, ,
\eeq
in the integrand of (\ref{Bulk}). The prefactor of the $\delta$ function is akin to the symmetric function $f$ defined in (\ref{f-fun}) and, as the latter, it has a vanishing derivative on the support of the $\delta$ function, see equation (\ref{nothing}). We can therefore rewrite the RHS in (\ref{rule}) as
\beq
i\pi \delta_{ab}\partial_{u}\delta(u-v)\, .
\eeq
The next step is to plug this distribution in (\ref{Bulk}), integrate the derivative $\partial_{u}$ by part and localize the integral over $v$.

There are as many new contributions as there are factors in (\ref{Bulk}) we can differentiate with $\partial_{u}$. Acting first on the $\ell_{12}$ dependent term we find that the sum over the partitions can be factored out and the remaining factors combine into $Y_{a}(u^{\gamma})$, after setting $v = u$ and using the Watson equation. Hence, up to the overall $\mathcal{A}_{(0, 0, 0)}$, we obtain
\beq
\begin{aligned}
\frac{i}{2}\ell_{12}\sum_{a\geqslant 1}\int \frac{du}{2\pi} \partial_{u}\tilde{E}_{a}(u^{\gamma})\, Y_{a}(u^{\gamma}) = -\frac{i\ell_{12}}{2L_{1}}\Phi_{\textbf{u}}\, .
\end{aligned}
\eeq
This corresponds to the overall phase $\varphi_{12}$ in front the sum over the partitions in (\ref{dressing}). Acting on the partition dependent factors in (\ref{Bulk}) yields
\beq
-\frac{i}{2}Y_{a}(u^{\gamma})\partial_{u}\log{\frac{h_{a1}(u^{\gamma}, \alpha)}{h_{1a}(\bar{\alpha}, u^{\gamma})}}\, ,
\eeq
which must be summed and integrated. This contribution combines naturally with the abelian part of $C_{2}$ in (\ref{C12}) and, after separating the symmetric and the anti-symmetric part of $h$, gives
\beq
\frac{i}{4}Y_{a}(u^{\gamma})\partial_{u}\log{p_{a1}(u^{\gamma}, \textbf{u})} + \frac{i}{4}Y_{a}(u^{\gamma})\partial_{u}\log{\frac{S_{a1}(u^{\gamma}, \alpha)}{S_{a1}(u^{\gamma}, \bar{\alpha})}}\, .
\eeq
The first term is the linearized version of the wrapping correction in (\ref{dressed}). The second one is the abelian component of
\beq\label{phiphi}
\frac{i}{4}\Phi_{\bar{\alpha}}-\frac{i}{4}\Phi_{\alpha} = \frac{i}{2}\Phi_{\bar{\alpha}}-\frac{i}{4}\Phi_{\textbf{u}}\, ,
\eeq
which matches with the remaining dressing factors in (\ref{dressing}). We easily verify that the matrix part of (\ref{phiphi}) is properly reproduced by acting with $\partial_{u}$ on the last factor in (\ref{Bulk}), which is a transfer matrix,
\beq
-\frac{i}{2}T_{a}(u^{\gamma})\partial_{u}T_{a}(u^{\gamma}) = -\frac{1}{2}\mathcal{H}_{a1}(u^{\gamma}, \alpha) -\frac{1}{2}\mathcal{H}_{a1}(u^{\gamma}, \bar{\alpha})\, ,
\eeq
and adding it to the matrix part of $C_{2}$ in (\ref{C12}).

\subsection*{A little puzzle}

As pointed in Section \ref{test}, our result (\ref{dressing}) is a bit at odds with the reality property of the structure constant. Our working assumption is that the dressed sum should have the same reality property as the bare one. The latter is easily seen to transform as
\beq
\mathcal{A}_{(0, 0, 0)}^* = (-1)^{M} \mathcal{A}_{(0, 0, 0)}\, ,
\eeq
under complex conjugation, for a real $\textbf{u}^* = \textbf{u}$ cyclic Bethe state. (This follows directly from $h_{\alpha\bar{\alpha}}^* = h_{\bar{\alpha}\alpha}$.) The former does not follow the same trend however. To make clear where the problem comes from, let us replace the sign in front of $\Phi_{\textbf{u}}$ in (\ref{dressing}) by a free real parameter $\eta$. The value we found in (\ref{dressing}) corresponds to $\eta = -1$. We then complex conjugate the full expression and apply the same algebra as in (\ref{algebra}). It yields
\beq\label{puzzle}
\mathcal{A}_{(0, 0, 0)}|_{\textrm{dressed}}^* = e^{\frac{i}{2}(1-\eta)\Phi_{\textbf{u}}}\times (-1)^{M} \mathcal{A}_{(0, 0, 0)}|_{\textrm{dressed}}\, .
\eeq
So, were $\eta = +1$ we would get what we want and the dressed sum would look even nicer,
\beq\label{wanted}
\mathcal{A}_{(0, 0, 0)}|_{\textrm{dressed}} \xrightarrow{\eta \rightarrow 1} e^{\frac{i}{2}(\varphi_{31}-\varphi_{12})}\sum_{\alpha\cup \bar{\alpha} = \textbf{u}} a_{\alpha\bar{\alpha}}(\ell_{31}) \, e^{\frac{i}{2}\Phi_{\bar{\alpha}}}\, .
\eeq
Unfortunately, according to (\ref{dressing}), we must plug $\eta = -1$ in (\ref{puzzle}). Hence, the dressed sum does not transform in the expected way. The phase of the structure constant is not per se ``observable'' and we should probably not ask our formula to predict more than what can possibly be predicted. We could then conclude here, after adding the  proviso that only the absolute value of our formula should be taken literally. The fact that we got so closed to what we had (perhaps unnecessarily) expected invites us to look for simple mechanisms that could rectify this unpleasant feature of our expression. Here are two plausible options.

The easiest way to fix the problem is to slightly change the rule given in \cite{short} for converting between asymptotic states and finite volume (permutation invariant) states. The prescription used in \cite{short} was adapted from the rule proposed in \cite{Pozsgay:2007kn,Pozsgay:2009pv,Pozsgay:2010cr} to deal with a similar problem in integrable relativistic QFTs (with a diagonal scattering). The recipe is that a finite volume Bethe state, characterized by its set of mode numbers $\textbf{m}$, relates to an ordered asymptotic state $\textbf{u}$ through a Jacobian and a phase
\beq\label{convert}
|\textbf{m}\mathcal{i} = \frac{1}{\sqrt{G}\prod_{i<j}\sqrt{S_{ij}}} |\textbf{u}\mathcal{i}\, .
\eeq
The Jacobian takes care of the conversion between the Dirac and the Kr\"onecker $\delta$ normalization. The phase $\prod_{i<j}S_{ij}$ is the one obtained by reversing the ordering of the state, $12\ldots M \rightarrow M\ldots 21$, and it sort of averages between ingoing and outgoing states. In our formula, this rule was implemented through the overall factor
\beq
\prod_{i<j}\frac{h_{ij}}{\sqrt{S_{ij}}} = \prod_{i< j} \sqrt{h_{ij}h_{ji}}\, ,
\eeq
which is sitting in front of the $\mathcal{A}$ series (\ref{asyC}). The first proposal is to slightly modify the phase in (\ref{convert}) by adopting the prescription that not only the magnons ordering should be reversed but also their position w.r.t.~to the spin chain lattice markers $Z^{L_{1}}$. Namely, we can write our state as $|\chi_{1}\ldots \chi_{M} Z^{L_{1}}\mathcal{i}$ with the spin chain substratum $Z^{L_{1}}$ being exposed at the end in the form of a defect of sort. Reversing the full state then requires scattering the excitations through one another but also through the $Z$'s, such that these ones appear, in the end, at the beginning of the state,
\beq\label{new}
|\chi_{1}\ldots \chi_{M} Z^{L_{1}}\mathcal{i} = e^{ip_{\textbf{u}}L_{1}}\prod_{i<j}S_{ij}\, |Z^{L_{1}}\chi_{M}\ldots \chi_{1}\mathcal{i}\, .
\eeq
The net effect is to multiply by $e^{-\frac{i}{2}p_{\textbf{u}}L_{1}}$ the conversion factor used in (\ref{convert}). It is immaterial asymptotically, for cyclic states. At wrapping level, it produces an extra phase, to be added to the ones already present in (\ref{dressing}), since $e^{-\frac{i}{2}p_{\textbf{u}}L_{1}} = e^{\frac{i}{2}\Phi_{\textbf{u}}}$. This phase is precisely what we need to reverse the sign in front of $\Phi_{\textbf{u}}$ in (\ref{dressing}) and restore the reality property of the dressed sum over the partitions.

The second proposal exploits a loophole in our derivation of the wrapping corrections. There is, indeed, an ambiguity in distributing the effect of $e^{-\tilde{H}L_{1}}$ in the trace (\ref{thermal}). In the derivation of the L\"uscher formula we put all of the weight on the final rapidity $u$, see equation (\ref{psi}). We could also have given it entirely to $w$ and obtain a different result, since acting with $\partial_{w}$ on this extra factor would produce a term proportional to $\Phi_{\textbf{u}}$. If we opt for the most symmetrical choice and give half of it to $u$ and the other half to $w$, then the extra term would be precisely $i\Phi_{\textbf{u}}/2$, which is what we need to obtain (\ref{wanted}).

\section{Asymptotic data and shift of the roots}\label{AsyD}

In this appendix, we derive the four loop expression for the asymptotic part of the structure constant. We also obtain a compact representation for its ``shifted version" which results from combining the shift of the roots, the associated change in the Gaudin norm and the dressing of the asymptotic sum (\ref{dressing}). Finally, we comment on the universal character of these results w.r.t.~to the choice of the representative in a given supermultiplet.

\subsection*{Asymptotic data}

We consider a state with two magnons carrying opposite rapidities $u_{1} = -u_{2} = z \geqslant 0$. The magnitude $z$ is fixed by the asymptotic Bethe ansatz equation \cite{Staudacher:2004tk,BS05}
\beq\label{ABAz}
e^{ipL}S(z, -z)|_{\mathfrak{sl}(2)} = \left(\frac{x^{+}}{x^{-}}\right)^{L+2}\frac{z-\frac{i}{2}}{z+\frac{i}{2}} /\sigma(z, -z)^2 = 1\, ,
\eeq
where $L = L_{1}$ is the length of the excited state, $\sigma = 1+O(g^6)$ is the BES dressing phase \cite{BES}, and $x^{\pm} = x(z\pm \frac{i}{2})$, with $x(z) = (z+\sqrt{z^2-4g^2})/(2g) = z/g+O(g)$. We need the root $z$ up to four loops, for $L=2$. The solution reads
\beq\label{z}
z = \frac{1}{2\sqrt{3}} (1 + 8 g^2-20 g^4+(112+48\zeta_{3}) g^6 - (922+288\zeta_{3}+480\zeta_{5})g^8+ O(g^{10}))\, ,
\eeq
and it gives, for the anomalous dimension,
\beq\label{gamma}
\begin{aligned}
\gamma &= 4 g (\frac{i}{x^+}-\frac{i}{x^-})= 12g^2 -48 g^4+336 g^6- (2820+288\zeta_{3})g^8 + O(g^{10})\, ,
\end{aligned}
\eeq
with all the $\zeta$'s coming from the weak coupling expansion of the dressing phase \cite{BES}. Both (\ref{z}) and (\ref{gamma}) must be corrected at wrapping order \cite{Bajnok:2008bm,Bajnok:2009vm}, that is, at $O(g^8)$. The correction to the dimension will not be needed, here, and the one for the root will be given later on.

Once we have the roots, we can compute the asymptotic amplitude (\ref{Aasympt}). For an half-split configuration, $\ell = \ell_{31} = \ell_{12} = L/2$, only the modulus of the hexagon amplitude matters, because $(e^{ip_{1}\ell}/h_{21})^2 = e^{ip_{1}L}S_{12}/h_{21}h_{12}  = 1/h_{12}h_{21}$ for on-shell rapidities.%
\footnote{More generally, $e^{\frac{i}{2}p_{\bar{\alpha}}L}/h_{\alpha \bar{\alpha}} = (-1)^{m_{\bar{\alpha}}}/\sqrt{h_{\alpha\bar{\alpha}}h_{\bar{\alpha}\alpha}}$ with $m_{\bar{\alpha}} = \sum_{i\in \bar{\alpha}}m_{i}$ and $m_i$ the mode number of the root $u_i$. The minimal energy two-magnon state has $m_{1} = -m_{2} = 1$.}
Hence,
\beq\label{rep-A}
\mathcal{A}_{\textrm{asympt}} = 2+\frac{2}{\sqrt{h(z, -z)h(-z, z)}} = 2 + \frac{2g^2(\sqrt{x^+x^-}+1/\sqrt{x^+x^-})^2}{z\sqrt{z^2+1/4}}\, ,
\eeq
where we used the explicit expression (\ref{huv}) for $h$, and thus
\beq\label{Aapp}
\mathcal{A}_{\textrm{asympt}} = 6 +12 g^2-120 g^4+(1020-144\zeta_{3}) g^6-(8652-432\zeta_{3}-1440\zeta_{5}) g^8 + O(g^{10})\, ,
\eeq
after plugging (\ref{z}) for the root $z$.

The second main ingredient is the determinant $G$, see (\ref{G1}). For a cyclic two-magnon state, it can be brought to the form
\beq\label{Gf}
G = L\frac{dp}{dz} (L\frac{dp}{dz}-i\frac{d}{dz}\log{S(z, -z)}) \, .
\eeq
It results from applying the lower triangulation method to the determinant of the 2-by-2 matrix $\{\partial(Lp_{1}-i\log{S_{12}})/\partial u_{i}, \partial LP/\partial u_{i}\}$, obtained from $G$ after replacing the Bethe ansatz equation for $p_{2}$ by the total momentum condition, $LP = Lp_{1}+Lp_{2} = 2\pi \mathbb{Z}$. Equivalently,
\beq\label{wedge}
G du_{1}\wedge du_{2} = (dLp_{1}-id\log{S_{12}}+\lambda dLP)\wedge dLP\, ,
\eeq
with the Lagrange multiplier $\lambda$ chosen such that the first differential has no component along $du_{2}$. This is readily seen to imply that $G$ is the product of two ``norms'', for the relative and center-of-mass motions, respectively. The same factorization applies to the exact Gaudin norm, after taking into account the finite size modifications of the BA equations.

The last quantity entering the asymptotic structure constant,
\beq\label{Casym}
C_{123}^{\bullet\circ\circ}|_{\textrm{asympt}} = \frac{2\mu\sqrt{L_{2}L_{3}}}{\sqrt{G/L}} (1+\sqrt{h(z, -z)h(-z, z)})\, ,
\eeq
is the magnon measure
\beq
\mu = \frac{(x^{+}x^{-}-1)^2}{((x^{+})^2-1)((x^{-})^2-1)} = 1+O(g^2)\, .
\eeq
Evaluating the full thing for $L = L_{2} = L_{3} =2$ with the help of (\ref{z}) leads to the expression quoted in the first line of (\ref{final}).

\subsection*{Shift of the roots}

At wrapping order, we must switch on the phases $\Phi_{i}(u_1, u_2)$ that correct the quantization conditions, see equation (\ref{exact}). They are such that, for a symmetric state, the shift of $u_1$ is opposite to the shift of $u_2$,
\beq\label{Phi+}
\Phi_{\textbf{u}} = 2\Phi_{+}(z, -z) = 0\, ,
\eeq
with $\Phi_{\pm}(\textbf{u}) = \frac{1}{2}(\Phi_{1}(\textbf{u})\pm \Phi_{2}(\textbf{u}))$. The shift of the absolute value, $z \rightarrow z+\delta z$, is controlled by $\Phi_{-}(\textbf{u})$ and reads, at the linearized level,
\beq\label{deltaz}
\delta z = -\frac{\Phi_{-}(z, -z)}{L\frac{dp}{dz}-i\frac{d}{dz}\log{S(z, -z)}} \simeq \frac{\Phi_{-}(z, -z)}{9}\, ,
\eeq
where in the last equality we specialized to the Konishi multiplet at weak coupling.

In addition to shifting the root, we must correct the Gaudin norm, $G\rightarrow G^{\textrm{exact}}$. This is done by adding $d\Phi_{1}$ to the first differential in (\ref{wedge}) and substituting $LP =L p_{1}+Lp_{2}+\Phi_{+}$ in the second one. It yields
\beq\label{GEf}
G^{\textrm{exact}} = (L\frac{dp}{dz}+ \partial_{+}\Phi_{+}) (L\frac{dp}{dz}-i\frac{d}{dz}\log{S} + \frac{d}{dz}\Phi_{-}) \, ,
\eeq
with $\partial_{\pm} = \partial_{1}\pm \partial_{2}$, after using
\beq
\frac{d}{dz}\Phi_{1}(z, -z) = \frac{d}{dz}\Phi_{-}(z, -z)\, , \qquad \partial_{2}\Phi_{+}(z, -z) = \partial_{+}\Phi_{+} (z, -z)\, ,
\eeq
which both follow from the absence of total momentum shift for a symmetric state, that is from (\ref{Phi+}) and its derivative $\partial_{-}\Phi_{+}(z, -z) = d\Phi_{+}(z, -z)/dz = 0$. Note, however, that $\partial_{+}\Phi_{+}\neq 0$, even upon restriction to a symmetric state, since $\partial_{+}$ is not a derivative along the symmetric subspace, as opposed to $\partial_{-} = d/dz$.

Lastly, we want to take into account the dressing of the asymptotic sum, by switching from (\ref{Aasympt}) to (\ref{dressing}). As noticed earlier, there is not much difference \textit{on shell} between the latter two objects. The overall phase shift in the dressed sum (\ref{dressing}) is irrelevant, thanks to (\ref{Phi+}), and so is the local one in the summand, thanks to (\ref{summand}). Thereby, we can use the same functional expression (\ref{rep-A}) for both the asymptotic sum and its dressed version.

We obtain the shift of the asymptotic structure constant by combining everything together. Replacing $G\rightarrow G^{\textrm{exact}}$ in (\ref{Casym}), plugging $z\rightarrow z+\delta z$ and expanding to first order in $\Phi$, we find, for the case of interest,
\beq\label{Cshift}
\delta C^{\bullet\circ\circ}_{123}/C^{\bullet\circ\circ}_{123}  = \frac{1}{2\sqrt{3}}\Phi_{-} + \frac{1}{18}\frac{d}{dz}\Phi_{-}+\frac{1}{6}\partial_{+}\tilde{\Phi}_{+}\, ,
\eeq
where we defined $\tilde{\Phi}_{+} = \Phi_{+}/L$ for later convenience. As written, the formula (\ref{Cshift}) is only accurate at $O(g^8)$. It could be easily extended all the way up to the onset of the double wrapping corrections, by keeping track of the $g$ dependence of the coefficients in front of the $\Phi$'s. The weak coupling expression for $\Phi_{-} = \Phi_{1}$ and $\partial_{-}\Phi_{-}$ can be found in \cite{Bajnok:2009vm}. We shall recall them in Appendix \ref{wrap} and alongside present the one for $\partial_{+}\tilde{\Phi}_{+}$.

\subsection*{Universality}

We now comment on the independence of our formula w.r.t.~the choice of the (diagonal) representative in a given multiplet. The analysis above is for the $\mathfrak{sl}(2)$ representative. We wish to prove that we would obtain the exact same result if we were to choose the $\mathfrak{su}(2)$ representative instead. To begin with, we recall that the exact same equation for $z$, that is equation (\ref{ABAz}), is found in the $\mathfrak{su}(2)$ subsector, with $L+2$ playing the role of the length.%
\footnote{Equivalently, $S_{\mathfrak{sl}(2)}(z, -z) = (x^+/x^-)^2 S_{\mathfrak{su}(2)}(z, -z)$.}
Hence, the root itself does not depend on the choice of the operator in a given multiplet, as well known.
The hexagon amplitude for the $\mathfrak{su}(2)$ state differs from the one in the $\mathfrak{sl}(2)$ sector by a simple phase, see \cite{short}. However, the representation (\ref{rep-A}) shows that only the module of the amplitude matters, implying that $\mathcal{A}_{(0, 0, 0)}$ is the same for both the $\mathfrak{sl}(2)$ and the $\mathfrak{su}(2)$ representative. The same argument applies to the dressed sum.

We switch to the Gaudin norm. It turns out that the quantity that is invariant is not $G$ itself but the ratio $G/L$. This has to do with cyclicity. Recall indeed that Gaudin's determinant has the meaning of a density of spin chain states (here in rapidity space). However, only the cyclic states are in the physical Hilbert space of the gauge theory. One should therefore exclude the unwanted states when computing the density. Since the ABA equations by themselves only require that $e^{iPL} = 1$, while we want the stronger requirement that $e^{iP} = 1$, their space of solutions contains $L$ times too many states. This gives $G/L$ as a measure of the density of physical states. We can turn the estimate into a rigorous result by enforcing the zero-momentum condition $e^{iP} = 1$ at the level of the determinant. It amounts to substituting the constraint that $P \in 2\pi \mathbb{Z}$ to the equation $LP \in 2\pi \mathbb{Z}$ in (\ref{wedge}). This is equivalent to rescaling $G$ by $L$ and we conclude from it that the density of cyclic states is, indeed, $G/L$ and not $G$.%
\footnote{In the hexagon approach, the quotient by the volume of the group of cyclic permutations $\mathbb{Z}_{L}$ of the spin chain is absorbed in the vacuum structure constant $C_{123}^{\circ\circ\circ} = \sqrt{L_1L_2L_3}/N$ where $N$ is the number of colors.} Finally, since the supersymmetry multiplet joining only works out for cyclic states, it is not surprising to find that only the refined norm $G/L$ has the right property, see \cite{Basso:2017khq} for more details.

We conclude with the more delicate analysis of the invariance of the exact Gaudin norm (\ref{GEf}). Here, it proves useful to switch to the string frame by using that
\beq
\frac{G^{\textrm{exact}}}{J} \bigg|_{\textrm{string}} = \frac{G^{\textrm{exact}}}{L}\bigg|_{\textrm{spin}}\, ,
\eeq
where $J$ is the R-charge of the operator, $L$ the length of the spin chain and where the state is assumed to obey the zero momentum condition. For an $\mathfrak{sl}(2)$ state, we have $J = L$ and the string equations are identical to the spin chain ones (even before imposing the zero momentum condition). For an $\mathfrak{su}(2)$ state, the two sets of equations agree only on the subspace of cyclic states, and $J = L-M$ where $M$ is the number of scalars. There are several reasons to prefer to work in the string frame rather than in the spin chain one at the level of the finite size corrections, one of them being that the R-charge does not fluctuate, contrarily to the length $L$.

It is clear that on the symmetric subspace $\Phi_{-} = \Phi_{1} = -\Phi_{2}$ must be a supersymmetric invariant, as otherwise the wrapping corrections to the root $z$ (and then to the energy) would be different for the $\mathfrak{sl}(2)$ and the $\mathfrak{su}(2)$ state in the same supermultiplet. The density of cyclic states $G^{\textrm{exact}}/J$ will be similarly independent of the choice of a representative if $\partial_{+}\tilde{\Phi}_{+} = \partial_{+}\Phi_{+}/J$ is a supersymmetric invariant. It amounts to proving that
\beq\label{Phi+}
\partial_{+}\tilde{\Phi}_{+} = -\frac{1}{2} \sum_{a\geqslant 1} \int\frac{d\tilde{E}_{a}(u)}{2\pi} \partial_{+}\log{(1+Y_{a}(u^{\gamma}))}
\eeq
does not depend in which sector we choose the state, if the state is cyclic, $u_{1}+u_{2} = 0$. We do not know how to prove this property in general. We did, however, check it to leading order at weak coupling, which is at $O(g^8)$. It follows from the fact that the term linear in $u_{1}+u_{2}$ in the difference of the two string-frame transfer matrices,
\beq
T_{a}(u^{\gamma})|_{\mathfrak{sl}(2)}-T_{a}(u^{\gamma})|_{\mathfrak{su}(2)}\, ,
\eeq
is even in $u$ and thus integrates to zero. This is easily verified using the formulae (\ref{Tsu2}) and (\ref{Tsl2}). It would be very interesting to find a more general argument.

\section{Amplitudes and form factors}\label{AFF}

When computing the structure constant, we must deal with complicated quantities that involve summing over the $SU(2|2)$ flavors of the magnons and their bound states. We introduce below a convenient algorithm for doing that. It builds on an observation of \cite{Beisert06} regarding the structure of the fundamental transfer matrix and draws inspiration from the way the sums over $SU(4)$ or $SL(2|4)$ descendants are implemented in the pentagon OPE series \cite{Frank,Mpaper,Cordova:2016woh}. It uses a particular basis of states for the $SU(2|2)$ module we want to trace over. These states are exact Bethe strings made out of the auxiliary roots of the $SU(2|2)$ Bethe equations. Their scattering is abelian, reflectionless and entirely controlled by the S-matrix eigenvalues entering the latter equations. These features greatly simplify the calculations of the expectation values of matrix observables, like the scattering kernel for instance. We start with the presentation of the method. We then report the expression for the scattering kernel at weak coupling in the mirror kinematics. Finally, we summarize the information for the dynamical parts of the various S-matrices and hexagon form factors.

\subsection{T-matrix and kernel}

The simplest example of flavour average is found for the transfer matrix. As already mentioned, we work with the forward transfer matrix, defined as
\beq\label{fT}
T_{a}(u) = \textrm{tr}_{a}\, \mathcal{S}_{a1}(u, \textbf{u}) =\textrm{tr}_{a}\,  \prod_{j} \mathcal{S}_{a1}(u, u_j)\, ,
\eeq
where the sum is taken over the $4a$-dimensional bound state module $V_{a}$ and where $\mathcal{S}_{a1}(u, v)$ is the $SU(2|2)$ S-matrix between the bound state $|u, a\mathcal{i}$ and the fundamental magnon $|v\mathcal{i}$. Our normalization is such that $\mathcal{S}_{a1}(u, v) = 1$ for the scattering between the top components of each module, in the $\mathfrak{sl}(2)$ grading. (More precisely, we set $D_{12} = -1$ in the S-matrix of \cite{Beisert06} and apply a graded permutation. We also introduce a grading into the trace in (\ref{fT}), such that the contributions of the fundamental bosons come with an overall minus sign. This effectively corresponds to the normalization used in \cite{3loops}.)

\begin{figure}
\begin{center}
\includegraphics[scale=0.40]{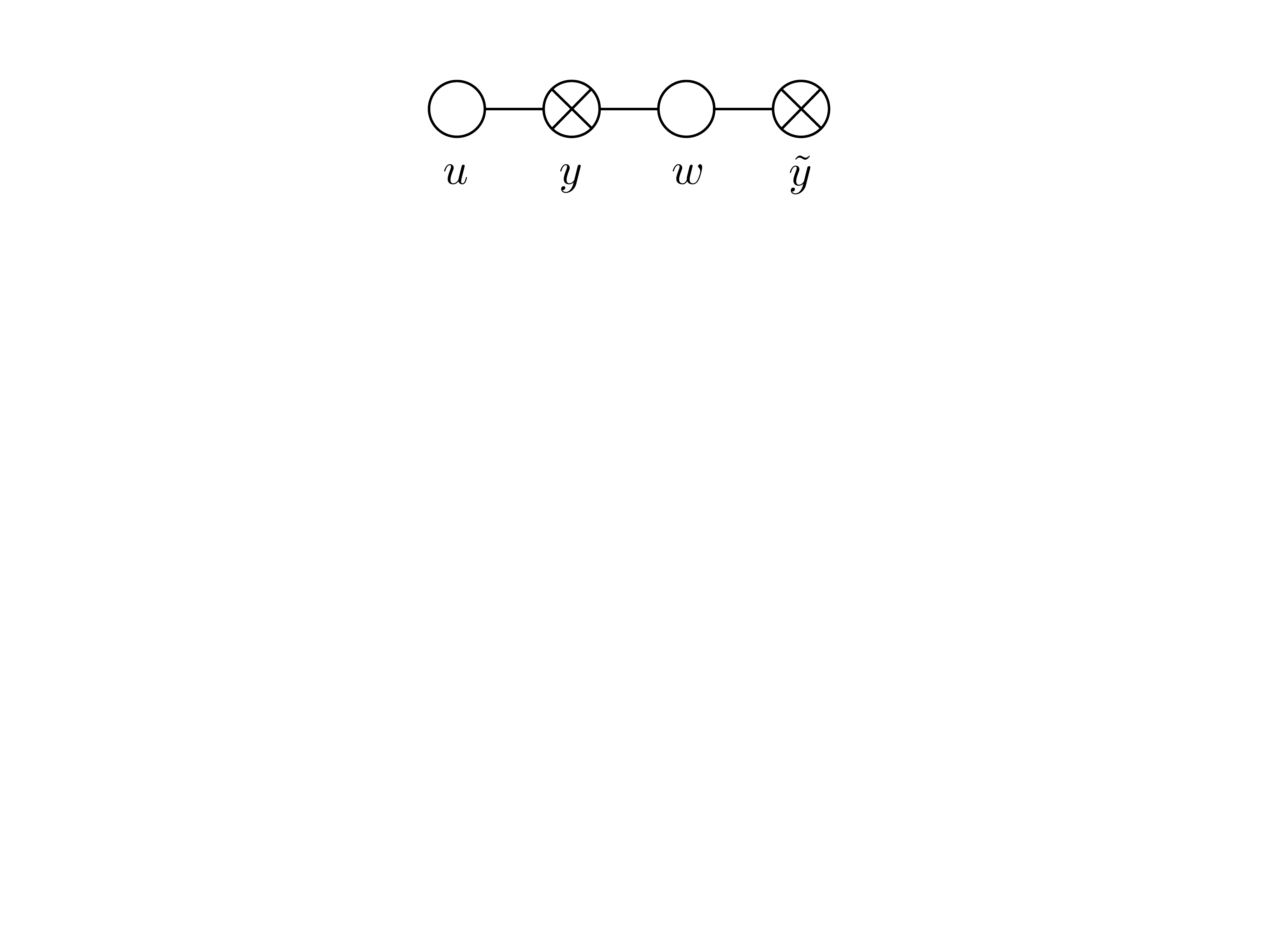}
\end{center}
\vspace{-8.5cm}
\caption{There is one type of root for each node of the $\mathfrak{psu}(2|2)$ Dynkin diagram. The momentum carrying root $u$ corresponds to the top component in the multiplet, which in our case is a fermion $\psi_{1}$. The fermionic root $y$ is associated to a SUSY generator that changes the fermion $\psi_{1}$ into a boson $\phi_{1}$. The $\mathfrak{su}(2)_{b}$ root $w$ flips the spin of the bosonic component: $\phi_{1}\rightarrow \phi_{2}$. Finally, the fermionic root $\tilde{y}$ maps back to a fermion: $\phi_{2}\rightarrow \psi_{2}$, which is the bottom component of the $\mathfrak{su}(2)_{f}$ multiplet the two fermions belong to, as well as the bottom component of the fundamental super-multiplet.}\label{dynkin}
\end{figure}

The expression for the transfer matrix $T_{a}$ can be obtained via the fusion relations. It yields, for the $a$-th anti-symmetric product \cite{Beisert06},
\beq\label{fusion}
T_{a}(u) = \sum_{i_{k}\textrm{'s}}\prod_{k=-(a-1)/2}^{(a-1)/2}t_{i_{k}}(u+ik)\, ,
\eeq
where we sum over all $a$-uplet $\{i_{k}\}$ with $i_{k} = 1|2,3|4$, $i_{k}\geqslant i_{k+1}$ and with all terms with an $i$-sequence containing consecutive $2$'s or $3$'s being removed. The seeds of the algebra are the four components $t_{1,2,3,4}$ of the fundamental transfer matrix $T_{a=1}$, which correspond to the contributions of the four states $\psi_{1}|\phi_{1}, \phi_{2}|\psi_{2}$ in the fundamental module. For illustration, they read
\beq\label{ts}
t_{1} = 1\, , \qquad t_{2} = t_{3} = -\prod_{j}\frac{1}{\xi_{j}}\frac{x^{+}-x^{+}_{j}}{x^{+}-x^{-}_{j}}\, , \qquad t_{4} = \prod_{j}\frac{x^{+}-x^{+}_{j}}{x^{+}-x^{-}_{j}}\frac{1-1/x^{-}x^{+}_{j}}{1-1/x^{-}x^{-}_{j}}\, ,
\eeq
for an $\mathfrak{sl}(2)$ state $\textbf{u} = \{u_{j}, j=1, \ldots , M\}$, with the minus signs in front of the bosonic contributions $t_{2}, t_{3}$ resulting from the grading factors, which we absorbed in the definition of the $t$'s. The general expressions for an arbitrary Bethe state can be found in \cite{Beisert06}.

The important point is that we can view each contribution in (\ref{ts}) as resulting from the factorized scattering between a particular string of Bethe rapidities and the roots in the Bethe state $\textbf{u}$. The set of all these strings form a particular basis for the ``probe'' module we trace over, to which we refer as the string basis. It allows us to treat in the same way the probe state and the Bethe state, since they are both made out of the same auxiliary roots that diagonalize and abelianize the scattering matrix.

The string basis comes with the obvious drawback that the symmetries are no longer manifest. This disadvantage is, however, largely compensated by the benefit of dealing directly with the eigenvalues of the scattering matrix and their corresponding diagonalized excitations. All the scattering data that is needed can be immediately found in the nested levels of the Bethe ansatz equations, without recourse to any explicit construction of the scattering matrix.

In the case at hand, we have four types of ``elementary" roots, $\{u, y, w, \tilde{y}\}$, one for each node of the $\mathfrak{su}(2|2)$ Dynkin diagram in figure \ref{dynkin}. Their scattering amplitudes, read off from the Bethe ansatz equations of \cite{Beisert06}, are given by
\beq\label{stringSm}
\begin{aligned}
&S_{uy} =\xi\,  \frac{x^{-}-y}{x^{+}-y}\, , &&S_{yw} = \frac{v-w+\frac{i}{2}}{v-w-\frac{i}{2}}\, , \,\,\,\,\, &&&S_{\tilde{y}w}= \frac{\tilde{v}-w+\frac{i}{2}}{\tilde{v}-w-\frac{i}{2}}\, , \\
&S_{ww} = \frac{w_{1}-w_{2}-i}{w_{1}-w_{2}+i}\, , \,\,\,\,\, &&S_{u\tilde{y}} =\frac{1}{\xi}\, \frac{1-1/x^{-}\tilde{y}}{1-1/x^{+}\tilde{y}}\, ,
\end{aligned}
\eeq
with $u^{[\pm a]} = u \pm \frac{ia}{2} = g(x^{[\pm a]}+1/x^{[\pm a]})$, $y = x(v), \tilde{y} = x(\tilde{v})$ and $\xi = \sqrt{x^+/x^{-}}$ (see comment below). The remaining amplitudes are either $1$ or related to the ones in (\ref{stringSm}) via unitarity $S_{AB}S_{BA} = 1$. (Note, in particular, that $S_{uu} = 1$ in our normalization.) The content in roots of type $\{u, y, w, \tilde{y}\}$ of each state $A$ in the fundamental multiplet is found by matching the component $t_{A}$ of the eigenvalue of the transfer matrix with a product of elementary amplitudes in (\ref{stringSm}). One must scatter $A$ with a generic Bethe state to see all of its content. Using the general expression for the eigenvalue of the transfer matrix \cite{Beisert06}, one immediately obtains the map between states in the $\textbf{2}|\textbf{2}$ multiplet and Bethe strings, 
\beq\label{f-string}
\begin{aligned}
&\psi_{1}(u) = \{u, \varnothing, \varnothing, \varnothing\}\, , && \phi_{1}(u) = \{u, \,x^{+}, \varnothing, \varnothing\}\, ,\\
&\phi_{2}(u) = \{u, \, x^{+}, u, \varnothing\}\, , && \psi_{2}(u) = \{u, \, x^{+}, u, \, x^{-}\}\, ,
\end{aligned}
\eeq
as depicted in figure \ref{rootsstring0}. One can check this dictionary by comparing the eigenvalues (\ref{ts}) of the transfer matrix in the $\mathfrak{sl}(2)$ sector with the amplitudes (\ref{stringSm}), fused according to the root content of each Bethe string in (\ref{f-string}).

\begin{figure}
\begin{center}
\includegraphics[scale=0.60]{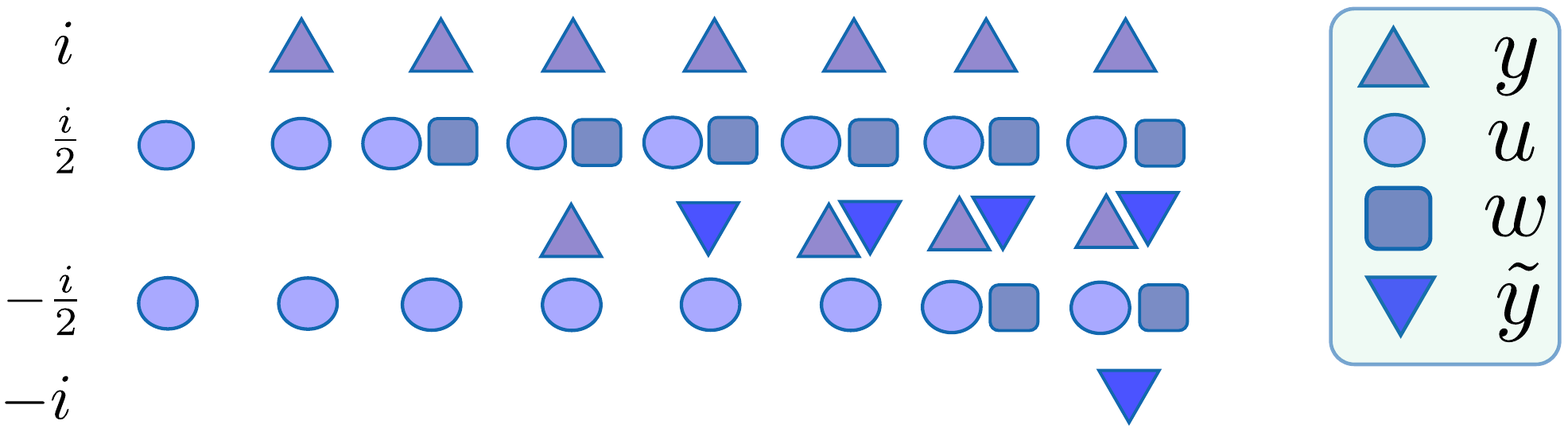}
\end{center}
\vspace{-13.5cm}
\caption{The bound state module $a=2$ counts $8$ strings. Each string is made out of roots of type $y,u,w,\tilde{y}$, represented as upper triangle, circle, square and lower triangle, respectively. The roots are placed between $i$ and $-i$ in the Bethe rapidity plane. The generalization to the higher bound states is easily inferred.}\label{rootsstring}
\end{figure}

We can then span any bound state module with similar strings, by using the fusion relation (\ref{fusion}) and the knowledge of the fundamental strings. For the first bound state, for instance, we have 4 bosonic components, 
\begin{align}
\hat{t}_1= t_1(u^{+})t_1(u^{-}),\,\ \ \ \hat{t}_4=t_3(u^{+})t_2(u^{-}),\, \ \ \ \hat{t}_5=t_4(u^{+})t_1(u^{-}),\, \ \ \hat{t}_8=t_4(u^{+})t_4(u^{-})\, ,
\end{align}
and four fermionic ones,
\begin{align}
\hat{t}_2= t_2(u^{+})t_1(u^{-}),\, \ \ \hat{t}_3=t_3(u^{+})t_1(u^{-}),\, \ \ \hat{t}_6=t_4(u^{+})t_2(u^{-}),\, \ \ \ \hat{t}_7= t_4(u^{+})t_3(u^{-})\, . 
\end{align}
They correspond to the $8$ strings shown in figure \ref{rootsstring}. The simplest one contains two roots only, both of type $u$, located at $u^+$ and $u^-$ in the Bethe rapidity plane. It stands as the top component of the $\textbf{4}|\textbf{4}$ module in our grading and corresponds to the $T$ eigenvalue $\hat{t}_1$. The other strings have more decorations. The second string has a $y$ attached at $y=x^{[+2]}$ and the next one a $w$, right below it, at $w=u^+$. The fifth string is obtained from the latter one by adding a root $\tilde{y}$ at $\tilde{y}=x$. The procedure, up to this point, is the same as for the fundamental module, with all the strings shifted by $i/2$ in the upper half plane. The essential difference is that there is a spectator root at $u^-$, which allows us to continue the dressing procedure. For instance, the component $\hat{t}_4$ is obtained from $\hat{t}_5$ by replacing the root $\tilde{y}=x$ with its antipodal partner $y=x$. The component $\hat{t}_6$ has both roots,  $y = \tilde{y} = x$, and this is the maximal amount one can have in a given slot. So, to continue, one should move further down and add decorations in the bottom slots. The next to last string is obtained by attaching a root $w$ at $w=u^-$ and the last one ends with a root $\tilde{y}$ at the very bottom, $\tilde{y}=x^{[-2]}$.

The higher bound state module follows a similar pattern, with a string of length $a$, $\{u^{[+a-2k-1]}, k=0, \ldots, a-1\}$, at the top of the multiplet.

\subsubsection*{Side comments}

Let us add a few more words about the transfer matrix, prior to start discussing the computation of the scattering kernel.

First, the $\xi$ factor in (\ref{stringSm}) distinguishes between the spin chain frame, for which $\xi = 1$, and the string frame, for which $\xi = \sqrt{x^{[+a]}/x^{[-a]}}$ for a bound state $a$. It originates from two different choices of lattice: spin chain lattice versus R-charge lattice. When converting between the two pictures, the rule is to dress the bosonic excitations with some length changing factors prior to move them around, see \cite{short},
\beq
\phi_{1,2}(u)|_{\textrm{string}} = Z^{\frac{1}{4}}\, \phi_{1,2}(u)|_{\textrm{spin}}Z^{\frac{1}{4}}\, ,
\eeq
while for the fermions nothing has to be done, $\psi_{1,2}|_{\textrm{string}} = \psi_{1,2}|_{\textrm{spin}}$. In a scattering experiment, one must carry the $Z$-factors alongside the excitation, using \cite{Beisert06} $\chi Z^{n} = e^{ipn}Z^{n}\chi$, resulting in the $\xi$ twisting of the amplitudes (\ref{stringSm}).

In most circumstances, the $\xi$ twisting is immaterial, since $\prod_{j}\xi_{j} = 1$ for cyclic states. There are situations where it is nonetheless preferable to use the string frame. The main reason is that the spin chain length can fluctuate, while the R-charge does not. Some of the manipulations used in this paper (which are at best formal in the dynamic spin chain set up), like extracting a derivative of the S-matrix at given length, for instance, are then put on a firmer ground in the string frame. The string frame twisting is also relevant for the mirror and crossing transformations. (It is important to get an S-matrix satisfying the generalized unitarity condition $\mathcal{S}(u^{\gamma}, v^{\gamma})^* = \mathcal{S}(v^{\gamma}, u^{\gamma})$ in the mirror channel, for instance; see \cite{Arutyunov:2009ga} for a review of the properties of the string frame S-matrix.)

We always assume, in the appendices of this paper, that quantities are defined in the string frame. We can always go back to the spin chain frame in the end, if desired. When computing an amplitude involving a probe and a Bethe state, for example, one can switch on the $\xi$ twisting in the intermediate steps and turn to the spin chain frame eventually. The latter operation amounts to an overall $\xi$ rescaling, which comes from the scattering between the roots of type $u$ in the probe and the roots of type $y$ and $\tilde{y}$ in the Bethe state, see (\ref{stringSm}). The precise factor that we should strip off is
\beq\label{xi-map}
\prod_{i\in \textrm{probe}}\xi_{i}^{N}\, ,
\eeq
with the $\xi$ factors being evaluated in the kinematics of the probe and where $N=N_{y}-N_{\tilde{y}}$ with $N_{y}$ and $N_{\tilde{y}}$ the number of roots of type $y$ and $\tilde{y}$ in the Bethe state, respectively. Notice in particular that the factor is absent for states in the $\mathfrak{sl}(2)$ subsector.

As an illustration, consider the transfer matrices for the two magnon $\mathfrak{su}(2)$ and $\mathfrak{sl}(2)$ states. In the $\mathfrak{sl}(2)$ grading, the former state differs from the latter one by the presence of two auxiliary roots, at $y=0$ and $y=\infty$. (This is for cyclic states, $u_{1}+u_{2} = 0$.) This combination of auxiliary roots decouples in the scattering with the probe, resulting in the equality of the two transfer matrices in the string frame, see equation (\ref{su2=sl2}) below. According to (\ref{xi-map}), we must strip out the overall factor $x^{[+a]}/x^{[-a]}$, for a bound state of length $a$, to get the eigenvalue of the transfer matrix in the $\mathfrak{su}(2)$ state in the spin chain frame. It results in the relation (\ref{transf}) used in Section \ref{SUSY} between the two spin chain frame transfer matrices.

We also mentioned that another frequently used definition of the transfer is found by reverting the scattering between $u$ and $\textbf{u}$,
\beq
\overleftarrow{T_{a}}(u) = \textrm{tr}_{a}\, \mathcal{S}_{1a}(\textbf{u}, u)\, .
\eeq
This is the backward transfer matrix that was used in \cite{3loops} and simply denoted $T_{a}(u)$ there. The forward and backward transfer matrices are of course not independent \cite{Beisert06}. They are related by complex conjugation
\beq\label{ccr}
\overleftarrow{T_{a}}(u^{\mp n\gamma}) =  (T_{a}(u^{\pm n\gamma}))^*\, ,
\eeq
where $n=0, 1, 2, \ldots\,$, and also by crossing symmetry. Namely, the backward transfer is, up to an overall factor, the crossed version of the forward one (\ref{fT}),
\beq\label{crr}
T_{a}(u^{2\gamma}) = f_{a}(u) \overleftarrow{T_{a}}(u)\, ,
\eeq
where
\beq\label{fa}
f_{a}(u) = \prod_{j}f_{a1}(u, u_{j}) = \prod_{j}\frac{x^{[-a]}-x_{j}^{+}}{x^{[+a]}-x_{j}^{+}}\frac{1-1/x^{[-a]}x^{-}_{j}}{1-1/x^{[+a]}x^{-}_{j}}\, .
\eeq
The latter factor compensates the crossing transformation of the scalar part of the scattering matrix,
\beq
S_{a1}(u^{2\gamma}, \textbf{u}) T_{a}(u^{2\gamma})\dot{T}_{a}(u^{2\gamma}) = S_{1a}(\textbf{u}, u) \overleftarrow{T_{a}}(u)\dot{\overleftarrow{T_{a}}}(u)\, ,
\eeq
such that scattering a mirror excitation with the state $\textbf{u}$ from the left or from the right amount to the same thing (after tracing over the flavors of the mirror magnon). Relations (\ref{ccr}) or (\ref{crr}) are useful to convert between the formulae used in this paper and the ones given in the appendices of \cite{3loops}.

It is also possible to get closed expressions for all the transfer matrices of interest. The simplest one is found for an $\mathfrak{su}(2)$ state. It has a simple 4-term structure for any $a$ and reads \cite{Beisert06,Arutyunov:2011mk} (in the string frame)
\beq\label{Tsu2}
\begin{aligned}
&T_{a}(u)|_{\mathfrak{su}(2)} = \bigg(\frac{x^{[+a]}}{x^{[-a]}}\bigg)^{\frac{M}{2}}\bigg[(a+1)\prod_{j}\frac{x^{[-a]}-x^{-}_{j}}{x^{[+a]}-x_{j}^{-}}-a\prod_{j}\frac{1}{\xi_{j}}\frac{x^{[-a]}-x^{+}_{j}}{x^{[+a]}-x_{j}^{-}}\\
&\,\,\,\,\, -a\prod_{j}\frac{1}{\xi_{j}}\frac{x^{[-a]}-x^{-}_{j}}{x^{[+a]}-x_{j}^{-}}\frac{1-1/x^{[+a]}x^{-}_{j}}{1-1/x^{[+a]}x^{+}_{j}}+(a-1)\prod_{j}\frac{1}{\xi^2_{j}}\frac{x^{[-a]}-x^{+}_{j}}{x^{[+a]}-x_{j}^{-}}\frac{1-1/x^{[+a]}x^{-}_{j}}{1-1/x^{[+a]}x^{+}_{j}}\bigg]\, ,
\end{aligned}
\eeq
where $M$ is the number of magnons in the Bethe state. (Note that the overall factor would be absent in the spin chain frame; it is absorbed in the conversion from the R-charge to the spin length.) In comparison, the transfer matrix for an $\mathfrak{sl}(2)$ state appears much more involved. It can be written as
\beq\label{Tsl2}
\begin{aligned}
T_{a}(u)|_{\mathfrak{sl}(2)} &= \sum_{n=0, \pm 1}(3n^2-2)\prod_{m=0}^{n}\frac{R^{(-)}(u^{[a-2m]})}{R^{(+)}(u^{[a-2m]})}\\
&\qquad\qquad  \times \sum_{j=0}^{a-n-1}\prod_{k=1}^{j}\frac{R^{(-)}(u^{[a-2k-2n]})B^{(-)}(u^{[a-2k]})}{R^{(+)}(u^{[a-2k-2n]})B^{(+)}(u^{[a-2k]})}\, ,
\end{aligned}
\eeq
in terms of
\beq
R^{(\pm)}(u) = \prod_{j}\frac{x(u)-x_{j}^{\mp}}{(\xi_{j})^{\mp \frac{1}{2}}}\, , \qquad B^{(\pm)}(u) = \prod_{j}\frac{\frac{1}{x(u)}-x_{j}^{\mp}}{(\xi_{j})^{\mp \frac{1}{2}}}\, .
\eeq
As mentioned before, we can set $\prod_{j}\xi_{j} = 1$ in most of our computations. However, we need to keep track of the $\xi$ factors when we compute the integral for $\partial_{+}\tilde{\Phi}_{+}$, since this one is sensitive to the first derivative of $T_{a}$ around the symmetric point $u_{1}+u_{2} = 0$.

For a two-magnon multiplet, we can choose either the $\mathfrak{su}(2)$ or the $\mathfrak{sl}(2)$ representative. The agreement between the two string frame transfer matrices is recovered,
\beq\label{su2=sl2}
T_{a}(u)|_{\mathfrak{sl}(2)} = T_{a}(u)|_{\mathfrak{su}(2)}\, ,
\eeq
as soon as we impose that the state is cyclic $u_{1}+u_{2} = 0$. (Without the latter condition this is of course not true.)

\subsubsection*{Kernel}

\begin{figure}
\begin{center}
\includegraphics[scale=0.45]{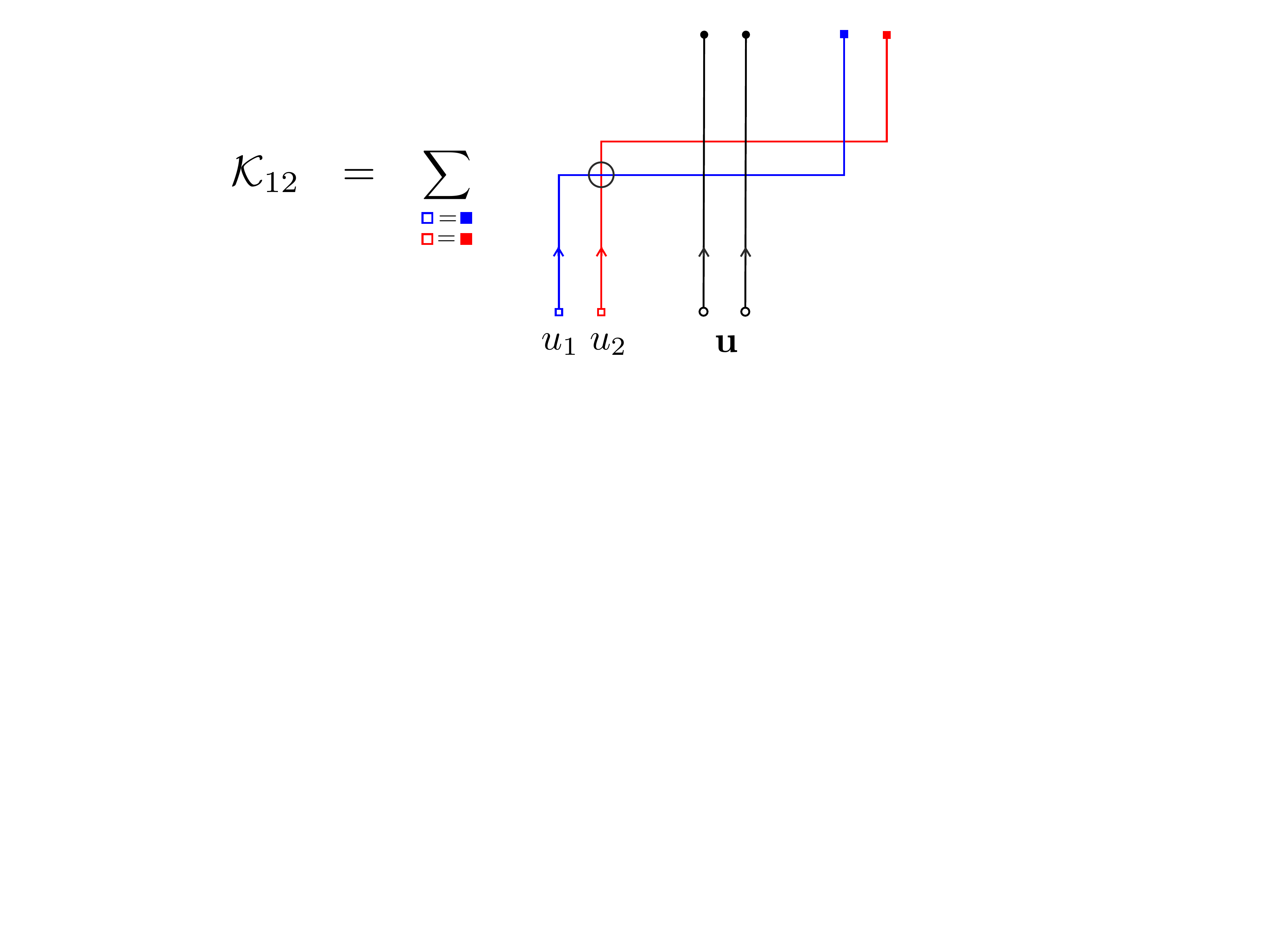}
\end{center}
\vspace{-8cm}
\caption{For the scattering kernel, two probe rapidities $u_{1}, u_{2}$ are scattered with one another, carried through the Bethe state $\textbf{u}$ and scattered back to their original positions. We identify the incoming and outgoing probe states and sum over all the flavors. The circle around the first scattering event indicates that a derivative w.r.t.~$u_{1}$ is taken at this interaction point.}\label{kernel}
\end{figure}

We now consider the $SU(2|2)$ matrix part of the scattering kernel. We recall that it is defined by 
\beq\label{calK}
\mathcal{K}_{12} = \textrm{tr}_{a_1\otimes a_2}\{ \mathcal{S}_{a_{2}a_{1}}(u_{2}, u_{1})\mathcal{S}_{a_{2}1}(u_{2}, \textbf{u})\mathcal{S}_{a_{1}1}(u_{1}, \textbf{u})\frac{\partial}{i\partial u_{1}}\mathcal{S}_{a_{1}a_{2}}(u_{1}, u_{2})\}\, ,
\eeq
where the trace is over the $SU(2|2)$ module $V_{a_{1}}\otimes V_{a_{2}}$ of the probe excitations $1$ and $2$. Pictorially, it looks as in figure \ref{kernel}.

Computing the kernel (\ref{calK}) for generic $a_{1,2}$ using the (rather complicated) matrix expressions for the various bound-state S-matrices in the component basis is challenging. To our knowledge, it could only be done for the case where the Bethe state $\textbf{u}$ is replaced by a twisted vacuum \cite{Ahn:2011xq}. As alluded to before, we shall follow a different route here and evaluate (\ref{calK}) by using the Bethe string basis. (More precisely, we shall focus on the diagonal configuration $a_{1} = a_{2} = a$ and $u_{1} = u_{2} = u$, which the case of interest here.) Since in this basis the scattering is abelian and reflectionless, we can carry out the trace in (\ref{calK}) by simply adding up the individual kernels, defined independently for each pair of strings,
\beq\label{Kabel}
\mathcal{K}_{12} = \sum_{A,B} K_{AB}(u_{1}, u_{2})t_{A}(u_{1})t_{B}(u_{2})\, ,
\eeq
where
\beq\label{KAB}
K_{AB}(u_{1}, u_{2}) = S_{BA}(u_{2}, u_{1})\frac{\partial}{i\partial u_{1}}S_{AB}(u_{1}, u_{2}) = \frac{\partial}{i\partial u_{1}}\log{S_{AB}(u_{1}, u_{2})}\, ,
\eeq
and with $t_{A}(u)  =S_{A1}(u, \textbf{u})$ the string $A$ component of the eigenvalue of the transfer matrix in the Bethe state $\textbf{u}$. We can then compute the kernel $K_{AB}$ as easily as we computed $t_{A}$. Take for example the case of two fundamental magnons $a_{1} = a_{2}=1$. Then $A$ and $B$ in (\ref{Kabel}) can be any of the 4 fundamental strings (\ref{f-string}) and each of the $16$ components of $S_{AB}$ in (\ref{KAB}) can be obtained by fusing the eigenvalues of the S-matrix (\ref{stringSm}). The scattering between two $\phi_{1}$'s, for instance, is given by the product of the $u$-$u$ and $u$-$v$ scatterings. Since $S_{uu} = 1$, we are left with
\begin{align}
S_{\phi_1\phi_1}(u_{1}, u_{2})= \xi_1\frac{x_1^{-}-y_2}{x_1^+-y_2}\times \frac{1}{\xi_2}\frac{y_1-x_2^{+}}{y_1-x_2^-}\bigg|_{y_{1,2}=x_{1,2}^+}=\frac{\xi_1}{\xi_2}\frac{x_1^--x_2^+}{x_1^+-x_2^-}\,.
\end{align}
The other components are obtained similarly : $S_{\psi_1\psi_1}=S_{\psi_2\psi_2}=1$, $S_{\phi_{1}\phi_1}=S_{\phi_2\phi_2}$, and
\beq\label{scatt}
\begin{aligned}
&S_{\psi_1\phi_1}=S_{\psi_1\phi_2}=\xi_1\frac{x_1^{-}-x_2^+}{x_1^+-x_2^+}\,,&&S_{\phi_1\psi_1}=S_{\phi_2\psi_1}=\frac{1}{\xi_2}\frac{x_1^{+}-x_2^+}{x_1^+-x_2^-}\,, \\
&S_{\phi_1\psi_2}=S_{\phi_2\psi_2}=\frac{1}{\xi_2}\frac{x_1^--x_2^+}{x_1^--x_2^-}\,, &&S_{\psi_2\phi_1} = S_{\psi_2\phi_2} =\xi_1\frac{x_1^--x_2^-}{x_1^+-x_2^-}\,, \\
&S_{\psi_1\psi_2}=\frac{x_1^--x_2^+}{x_1^+-x_2^+}\frac{1-1/x_1^-x_2^-}{1-1/x_1^+x_2^-}\,, &&S_{\psi_2\psi_1}=\frac{x_1^+-x_2^+}{x_1^+-x_2^-}\frac{1-1/x_{1}^-x_2^+}{1-1/x_1^-x_2^-}\, ,\\
&S_{\phi_1\phi_2}=\frac{\xi_1}{\xi_2}\frac{x_1^--x_2^+}{x_1^+-x_2^-}\frac{u_1-u_2+i}{u_1-u_2}\,, \,\,\,\,\, &&S_{\phi_2\phi_1}=\frac{\xi_1}{\xi_2}\frac{x_1^--x_2^+}{x_1^+-x_2^-}\frac{u_1-u_2}{u_1-u_2-i}\,.
\end{aligned}
\eeq
They are seen to fulfill the unitarity relation $S_{AB}(u, v)S_{BA}(v, u) = 1$.

We immediately remark that all the amplitudes in (\ref{scatt}) either have a pole or a zero at coinciding rapidities $u_1=u_2$. The general property, that lifts to the bound state amplitudes for the case where $a_{1} = a_{2} = a$, is that $S_{AB}(u_1, u_2)$ has a pole or a zero whenever $A\neq B$ (see the comment below regarding the degree of the pole or zero). The diagonal component $S_{AA}(u_1, u_2)$ and the symmetric product $S_{AB}(u_1,u_2)S_{BA}(u_1,u_2)$ are pure phases, however, and as such are regular at $u_{1} = u_{2}$. Following these observations, it proves useful to decompose the kernel in (\ref{KAB}) into its symmetric and anti-symmetric components,
\beq
K_{AB} = \frac{\partial}{2i\partial u_{1}}\ln \{S_{AB}(u_1,u_2)S_{BA}(u_1,u_2)\}+\frac{\partial}{2i\partial u_{1}}\ln \frac{S_{AB}(u_1,u_2)}{S_{BA}(u_1,u_2)}\, .
\eeq
The anti-symmetric part should naively drop out from the full sum in (\ref{Kabel}) when $u_{2}\rightarrow u_{1}$. This is ignoring that poles and zeros give rise to divergent contributions in this limit. Though the divergences cancel out on average in the full kernel, which must be regular at $u_{2}=u_{1}$, a finite part is left behind. Indeed, averaging over the pole and zero yields
\beq
\begin{aligned}
&\lim_{u_{i}\rightarrow u}\frac{1}{2}(t_{A}(u_1)t_{B}(u_2) -t_{B}(u_{1})t_{A}(u_{2}))\frac{\partial}{2i\partial u_{1}}\ln \frac{S_{AB}(u_1,u_2)}{S_{BA}(u_1,u_2)}\\
&=\lim_{u_i\rightarrow u} \frac{d}{2i(u_1-u_2)}(t_{A}(u_1)t_{B}(u_2)-t_{B}(u_1)t_{A}(u_2))= \frac{d}{2i}(t'_{A}(u) t_{B}(u)-t'_{B}(u)t_{A}(u))\, ,
\end{aligned}
\eeq
for each pair $(A, B)$, with $t'(u) = \partial_{u}t(u)$. It is non zero for $d\neq 0$, which stands for the degree of the zero of $S_{AB}$, if $d$ is positive, or of its pole, if $d$ is negative. Accordingly, the summand for the kernel $\mathcal{K}_{aa}(u, u)$ at coinciding rapidities (and for $a_{1} = a_{2} = a$) can be written as
\begin{align}\label{symanti}
t_{A}(u)t_{B}(u)\times \bigg[\frac{\partial}{2i\partial u_{1}}\ln \{S_{AB}(u_1,u_2)S_{BA}(u_1,u_2)\}\big|_{u_{i}\rightarrow u}+c_{AB}\frac{\partial}{2i\partial u}\ln \frac{t_A(u)}{t_B(u)}\bigg]\, .
\end{align}
(It is manifestly finite in the limit $u_{i}\rightarrow u$ since the symmetric product is a pure phase.) It features the ``anomalous'' contribution coming from the zeros/poles of the scattering amplitudes, which is controlled by the (integer valued) anti-symmetric matrix $c_{AB} = -c_{BA}$. Upon inspection of the scattering amplitudes for a generic bound state, one easily infers that its entries can be either $0,\pm1$ or $\pm2$. Furthermore, the matrix element $c_{AB}$ is positive \textit{iff} the string $A$ is longer than the string $B$ (that is, if the string $A$ contains more auxiliary roots than the string $B$). The fundamental amplitudes (\ref{scatt}) have at most simple poles or zeros. Double poles or zeros appear for the first time for the first bound state. For example, the amplitude $S_{36}$ between the third and sixth strings in figure \ref{rootsstring} has a double pole,
\begin{align}
S_{36}=\frac{\xi_1}{\xi_2}\frac{x_1^{[-2]}-x_2^{[+2]}}{x_1^{[+2]}-x_2^{[-2]}}\frac{(u_1-u_2)^2+1}{(u_1-u_2)^2}\, .
\end{align}
More generally, double poles occur whenever the bottom auxiliary roots in the string $A$ and $B$ are at a distance $i$ from each other and consist in a root of type $w$ (for $A$) and a stack $(y, \tilde{y})$ of antipodal fermionic roots (for $B$). This is so if we choose $A = 3$ and $B = 6$ in figure \ref{rootsstring} and this is the sole configuration of this type in the first bound state module. The number of such configurations grows linearly with the bound state label $a$ and we have $a-1$ of them in general. This configuration is impossible for the fundamental strings, which never host pairs of antipodal fermionic roots.

Lastly, we must sum over $A$ and $B$ in (\ref{Kabel}). For the fundamental magnons, we easily obtain, after adding everything up, 
\beq\label{Ksym}
\begin{aligned}
\mathcal{K}_{11}(u, u)|_{\textrm{sym}} &= \frac{(x^{+}+x^{-})^2(1-\frac{1}{x^{+}x^{-}})^2}{2(x^{+}-\frac{1}{x^{+}})(x^{-}-\frac{1}{x^{-}})}(t_{2}^2 +t_{2}t_{3}+t_{3}^2 +(t_{2}+t_{3})(t_{1}+t_{4}) + t_{1}t_{4})\\
&\,\,\, +\frac{(x^{+}-x^{-})^2(1+\frac{1}{x^{+}x^{-}})^2}{2(x^{+}-\frac{1}{x^{+}})(x^{-}-\frac{1}{x^{-}})}(t_{2}t_{3}-t_{1}t_{4})\, ,
\end{aligned}
\eeq
for the symmetric part of the kernel, while the anti-symmetric part is obtained by plugging $c_{AB} = +1$ in (\ref{symanti}) for $A>B$. We have checked the resulting formula against a direct matrix computation using the S-matrix of \cite{Beisert06}, for states in the $\mathfrak{sl}(2)$ subsector. (In this way of doing, one must use the Yang-Baxter equation to relate the various contributions and bring them to the ``$t$ form'' obtained here.) In the bound state case, though each coefficient is straightforwardly computed, there are many of them and it is not easy to cast their sum in a closed form. Our ambition is more modest for the kernels of the bound states and we do not really need concise formulae for them. We only want their leading weak coupling expressions for a two magnon $\mathfrak{sl}(2)$ state, with the probe rapidity $u$ in the mirror kinematics. With that in mind, we generated the kernels for the first few bound states with the help of Mathematica and, based on the expressions so obtained, we identified the pattern for all $a$'s. We found, to leading order at weak coupling,
\beq\label{matK}
\frac{\mathcal{K}_{aa} (u^{\gamma}, u^{\gamma})}{2a^2 g^4} = \frac{(a^2-1)(u^2+\frac{a^2}{4})^2-(u^2-z^2+\frac{a^2-1}{4})^2(\frac{2(1+a)}{a}u^2-z^2+\frac{2a^2-2a-1}{4})}{(u^2+\frac{a^2}{4})^3 Q(u^{[a-1]})^2(z^2+\frac{1}{4})^2} \, ,
\eeq
for a state $\textbf{u} = \{z, -z\}$ and with $Q(u) = (u-\textbf{u})$ the Baxter polynomial. We verify that it is suppressed by two powers of the coupling, in agreement with the supersymmetry delay discussed in Section \ref{SUSY}.

It would be interesting to study the algebraic properties of these kernels and perhaps derive a set of fusion relations for them. It would allow one to generate the kernles more systematically, with no commitment to any specific basis.

\subsubsection*{Abelian factor}

The abelian part of the kernel is easily obtained. To the leading order at weak coupling the mirror-mirror $SL(2)$ S-matrix is given by
\beq
S_{ab}(u^{\gamma}, v^{\gamma}) = s_{ab}(u, v) \frac{\Gamma(1+\frac{b}{2}+iv)^2\Gamma(1+\frac{a+b}{2}+iu-iv)^2\Gamma(1+\frac{a}{2}-iu)^2}{\Gamma(1+\frac{b}{2}-iv)^2\Gamma(1+\frac{a+b}{2}-iu+iv)^2\Gamma(1+\frac{a}{2}+iu)^2}\, ,
\eeq
where $s_{ab}$ is the $SU(2)$ string factor for XXX magnon scattering,
\beq
s_{ab}(u, v) = \frac{\Gamma(1+\frac{a+b}{2}-iu+iv)\Gamma(1+\frac{a-b}{2}+iu-iv) \Gamma(\frac{a+b}{2}-iu+iv)\Gamma(\frac{a-b}{2}+iu-iv)}{\Gamma(1+\frac{a+b}{2}+iu-iv)\Gamma(1+\frac{a-b}{2}-iu+iv)\Gamma(\frac{a+b}{2}+iu-iv)\Gamma(\frac{a-b}{2}-iu+iv)}\, .
\eeq
It is intriguingly similar to the S-matrix for gluonic bound states on the GKP string~\cite{Basso:2014nra}. Here we are primarily interested in the log derivative for $b=a$ and $v=u$. It reads
\beq\label{abelK}
\begin{aligned}
K_{aa}(u^{\gamma}, u^{\gamma}) &= \frac{\partial}{i\partial u}\log{(-S_{aa}(u^{\gamma}, v^{\gamma}))}\big|_{v = u} \\
&= -2\left[\psi(1+\frac{a}{2}+iu)+\psi(1+\frac{a}{2}-iu)-\frac{1}{a}-2\psi(1)\right] + O(g^2)\, ,
\end{aligned}
\eeq
where $\psi(x) = \partial_{x}\log{\Gamma(x)}$ is the Euler $\psi$ function.

\subsection{S-matrix and hexagon}

We collect here the expressions for the $SL(2)$ S-matrix and hexagon form factor in the mixed kinematics. The most complicated component comes in both cases from the dressing phase. We write it as a product of three terms,
\beq\label{dphmr}
\sigma_{a1}(u^{\gamma}, v) = \textrm{prefactor} \times F(u, v)\times e^{\tilde{\theta}(u, v)}\, ,
\eeq
with
\beq\label{prefactor}
\textrm{prefactor} = \prod_{k=-\frac{a}{2}}^{\frac{a}{2}-1} \frac{1-1/x^{[+2k]}y^{+}}{1-1/x^{[+2k]}y^{-}}\, ,
\eeq
and
\beq\label{tildetheta}
\tilde{\theta}(u, v) = \sum_{n, m\geqslant 1} c_{2n, 2m+1} (\tilde{q}^{u}_{2n}q^{v}_{2m+1}- q^{v}_{2n}\tilde{q}^{u}_{2m+1}) \, .
\eeq
The latter is the mirror-real counterpart of the BES dressing \cite{BES}, with the usual coefficient,
\beq
c_{2n, 2m+1} = 2(-1)^{n+m}(2n-1)(2m)\int_0^{\infty} \frac{dt}{t}\frac{J_{2n-1}(2gt)J_{2m}(2gt)}{e^{t}-1}\, ,
\eeq
and with
\beq
\tilde{q}^{u}_{r} = \frac{1}{r-1}((x^{[+a]})^{1-r} + (x^{[-a]})^{1-r})\, , \qquad q^{v}_{r} = \frac{i}{r-1}((y^{+})^{1-r} - (y^{-})^{1-r} )\, ,
\eeq
the mirror and the real higher conserved charges, respectively. Finally, we have
\beq\label{logF-G}
\begin{aligned}
\log{F} &= \sum_{m\geqslant 1}2(2m)(-1)^{m}\int_0^{\infty}\frac{dt}{t}\frac{\sin{(ut)}e^{-at/2}J_{2m}(2gt)}{e^{t}-1} q^{v}_{2m+1} \\
& + \sum_{m\geqslant 1}2(2m-1)(-1)^{m}\int_0^{\infty}\frac{dt}{t}\frac{(\cos{(ut)}e^{-at/2}-J_{0}(2gt))J_{2m-1}(2gt)}{e^{t}-1} q^{v}_{2m}\, .
\end{aligned}
\eeq
Later on, we will need to expand $F$ at weak coupling, for a symmetric state. We will then forget about the first line above, since there will be no odd charges, $q_{2m+1} = 0$. Furthermore, up to $O(g^4)$ included, only the $m=1$ contribution in the second line, that is the one proportional to the anomalous dimension $2gq_{2}$, is needed. After expanding the Bessel functions and integrating over $t$, this contribution gives
\beq\label{logF}
\log{F} = gq_{2}^{v}\bigg[H_{1}(\tfrac{a}{2}+iu) -g^2(H_{3}(\tfrac{a}{2}+iu)+2\zeta_{3}) + \textrm{c.c.} \bigg] + O(g^6)\, ,
\eeq
where $H_{1}(z) = \psi(1+z)-\psi(1)$ and $H_{3}(z) = \frac{1}{2}(\psi_{2}(1+z)-\psi_{2}(1))$ are harmonic sums. The $\tilde{\theta}$ part (\ref{tildetheta}) starts at $O(g^6)$, as well known, and will not be needed.

The dressing phase (\ref{dphmr}) is not a nice quantity on its own, because of the prefactor (\ref{prefactor}) which contains ``bad" cuts $x^{[\pm k]}$ for $k< a$. These cuts cancel out in physical quantities. For instance, the abelian part of the S-matrix does not have this problem, 
\beq\label{Sa1}
S_{a1}(u^{\gamma}, v) = \frac{(gy^{-})^2 (u-v+i\frac{a-1}{2})}{(u-v-i\frac{a-1}{2})(u-v+i\frac{a+1}{2})(u-v-i\frac{a+1}{2})}\frac{(1-1/x^{[+a]}y^{-})^2(1-1/x^{[-a]}y^{-})^2}{F(u, v)^2e^{2\tilde{\theta}(u, v)}}\, .
\eeq
Similarly, the abelian part of the hexagon amplitude is much nicer,
\beq\label{ha1}
h_{a1}(u^{\gamma}, v) = \frac{-gy^{+}}{u-v-i\frac{a+1}{2}}\frac{(1-1/x^{[+a]}y^{+})(1-1/x^{[-a]}y^{+})}{F(u, v)e^{\tilde{\theta}(u, v)}}\, .
\eeq
By using the crossing relation,
\beq
h_{a1}(u^{2\gamma}, v)h_{a1}(u, v) = \frac{1}{f_{a1}(u, v)}\, ,
\eeq
and similarly for $S_{a1}$ with $1/f_{a1}^2$ in the RHS, where $f_{a1}$ is the function appearing in the product (\ref{fa}), we can access to the hexagon and the scattering amplitude for other mirror rotated configurations. Further fundamental and very useful relations for moving excitations around are \cite{short,3loops}
\beq
 h_{ab}(u^{4\gamma}, v)h_{ba}(v, u) = 1\, ,  \qquad h_{ab}(u^{2\gamma}, v^{2\gamma}) = h_{ab}(u, v) \, ,
\eeq
and
\beq
p_{ab}(u^{2\gamma}, v)p_{ab}(u, v) = 1\, ,
\eeq
where $p_{ab}(u, v)$ is the symmetric part of the hexagon amplitude
\beq\label{pab}
p_{ab}(u, v) = h_{ab}(u, v)h_{ba}(v, u) = \frac{(u-v)^2+\frac{(a-b)^2}{4}}{(u-v)^2+\frac{(a+b)^2}{4}}\bigg(\frac{1-\frac{1}{x^{[+a]}y^{[-b]}}}{1-\frac{1}{x^{[+a]}y^{[+b]}}}\frac{1-\frac{1}{x^{[-a]}y^{[+b]}}}{1-\frac{1}{x^{[-a]}y^{[-b]}}}\bigg)^2\, .
\eeq
The latter has a double zero at $u=v$ for $a=b$ with a coefficient that is the square of the bound state measure. After rotating to the mirror kinematics, it reads
\beq\label{mua}
\mu_{a}(u^{\gamma}) = \frac{a}{g^2 (x^{[+a]}x^{[-a]})^2(1-1/x^{[+a]}x^{[-a]})^2 (1-1/(x^{[+a]})^2)(1-1/(x^{[-a]})^2)}\, .
\eeq
Using these relations, one can, for instance, re-express the hexagon amplitude in the mirror channel $23$, see equation (\ref{general}), as
\beq
\frac{1}{h_{1a}(\textbf{u}, u^{-\gamma})} = h_{a1}(u^{3\gamma}, \textbf{u}) = \frac{1}{f_{a1}(u^{\gamma}, \textbf{u})h_{a1}(u^{\gamma}, \textbf{u})}\, .
\eeq
This specific identity shows the equivalence between the formula used here and the one in \cite{3loops} for this channel,
\beq
\frac{T_{a}(u^{-\gamma})}{h_{1a}(\textbf{u}, u^{-\gamma})} = \frac{\overleftarrow{T}_{a}(u^{\gamma})}{h_{a1}(u^{\gamma}, \textbf{u})}\, ,
\eeq
given (\ref{crr}) and the $4\gamma$ invariance of the transfer matrix, $T_{a}(u^{4\gamma}) = T_{a}(u)$. It is as straightforward, using the above web of relations, to check the equivalence in the other two mirror channels.

We can also apply all that to the computation of the $Y$ function at weak coupling, which will be needed later on. Using (\ref{Sa1}) and (\ref{Tsu2}) immediately give \cite{Bajnok:2008bm}
\beq\label{Yapp}
Y_{a}(u^{\gamma}) = \frac{4a^2(u^2-z^2+\frac{a^2-1}{4})^2g^8}{(u^2+\frac{a^2}{4})^4Q(u^{[a+1]})Q(u^{[a-1]})Q(u^{[-a+1]})Q(u^{[-a-1]})} + O(g^{10})\, ,
\eeq
where we specialized to the Konishi state, $Q(u) = u^2-z^2$ and $J=L=2$.

\section{Regular mirror corrections}\label{Reg}
%%%%%%%%%%%%%%%%%%%%%%%%%%%%%%%%%%%%%%%%%%%%%%%%%%%

In this appendix, we recall how to compute the regular mirror corrections $\mathcal{A}_{(1, 0, 0)}, \mathcal{A}_{(0, 0, 1)}$ and $\mathcal{A}_{(0, 1, 0)}$.

\subsection*{Integral in the adjacent channel}

The integral for the $(1, 0, 0) = (0, 0, 1)$ amplitude is given by
\beq
\begin{aligned}
&\mathcal{A}_{(1, 0, 0)} = \sum_{a\geqslant 1}\int \frac{du}{2\pi} \mu_{a}(u^{\gamma})e^{ip_{a}(u^{\gamma})\ell}T_{a}(u^{\gamma}) \sum_{\alpha \cup \bar{\alpha} = \textbf{u}} (-1)^{|\bar{\alpha}|}e^{ip_{\bar{\alpha}}\ell_{31}}\frac{h_{a1}(u^{\gamma}, \alpha)}{h_{\alpha\bar{\alpha}}h_{1a}(\bar{\alpha}, u^{\gamma})}\,\label{adjacent} ,
\end{aligned}
\eeq
where $\ell = \ell_{12} (= \ell_{31} = \ell_{23}) = 1$ in the case of interest. It starts at 3 loops for $\ell =1$, see \cite{short,Eden:2015ija,3loops}. Its evaluation at four loops presents no major difficulty. The integrand has a fast decay at large $u$ and is meromorphic in $u$, at any given loop order, permitting the use of the method of residues. At four loops, one first generates the integrand using the expressions given in the appendix \ref{AFF} and then compute its residues. These ones can be split, following \cite{Bajnok:2008bm}, into a dynamical and a kinematical part, depending on whether the pole relates or not to the Bethe roots $\pm z$. We have checked numerically that the contributions to the dynamical part sum up to zero, in agreement with the proposal of \cite{Bajnok:2008bm}. Hence the only poles that matter are the kinematical ones, at $v=ia/2$ and at $v=ia/2+im$, with $m=1, 2, \ldots\,$. The latter half infinite tail arises from the $\psi$ functions in the dressing phase, see equation (\ref{logF}). These $\psi$ terms are the trickiest ones to integrate. Fortunately, they appear in the same combination as for the computation of the 5 loop anomalous dimension \cite{Bajnok:2009vm}. The reason is that the sum over the partitions in the above integral gives a transfer matrix at leading order at weak coupling, see \cite{3loops} for the precise formula, reproducing therefore the integrand (\ref{Yapp}) of the leading L\"uscher formula. The dressing phase dresses it with the $\psi$ functions, as it does for the five loop anomalous dimension. For short, one can reuse the results of \cite{Bajnok:2009vm} for that part of the integral. The rest of the integrand is rational and easily integrated. Putting everything together leads to
\beq
\mathcal{A}_{(1, 0, 0)} = (162+432\zeta_{3}-720\zeta_{5})g^6 -(3078-648\zeta_{3}+1296\zeta_{3}^2+7200\zeta_{5}-12600\zeta_{7})g^8\, .
\eeq
Its ratio with $\mathcal{A}_{(0, 0, 0)}$, see equation (\ref{Aapp}), yields the result quoted in the third line of (\ref{final}).

\subsection*{Integral in the bottom channel}
%%%%%%%%%%%%%%%%%%%%%%%%%%%%%%%%%%%%%%%%%%%%%%%%%%%
%%%
The integral in the $(0, 1, 0)$, so-called bottom, channel is given by
\beq
\begin{aligned}
&\mathcal{A}_{(0, 1, 0)} = \mathcal{A}_{\textrm{asympt}}\sum_{a\geqslant 1}\int \frac{du}{2\pi} \mu_{a}(u^{\gamma})e^{ip_{a}(u^{\gamma})\ell} \frac{T_{a}(u^{-\gamma})}{h_{1a}(\textbf{u}, u^{-\gamma})}\label{eq:finiteSizeCorrectionbottom}\,,
\end{aligned}
\eeq
where $\ell = \ell_{23}$. A specificity of that integral is that its integrand is not bounded at large $u$. For instance, at weak coupling and for a generic $\mathfrak{sl}(2)$ state, the integrand reads \cite{short}
\beq\label{010w}
\frac{a}{(u^2+\frac{a^2}{4})^{2+\ell}}\frac{Q(u^{[a+1]})+Q(u^{[-a-1]})-Q(u^{[a-1]})-Q(u^{[-a+1]})}{Q(i/2)}\, .
\eeq
It is clearly not integrable at large $u$, for $Q(u)\sim u^{M}$, if the spin $M$ is much greater than the bridge length $\ell$. The cure comes from the summation over the bound states in (\ref{eq:finiteSizeCorrectionbottom}). This sum leads to the telescoping of the various terms in the integrand (\ref{010w}) and makes the integral converge for any spin \cite{short} as long as $\ell >0$.

More precisely, the integral converges as long as $\ell  > \frac{1}{2}\gamma$, where $\gamma$ is the anomalous dimension of the excited operator, $\gamma = \Delta-L-M$. This is so because the \textit{effective} length of the bridge in the background of the excited operator is $\ell_{\textrm{eff}} = \ell - \frac{1}{2}\gamma$, with the shift following asymptotically from the large $u$ scaling of the dressing factor (\ref{logF-G}), $\log{F} \sim \gamma \log{u}$. Thereby, the integral (\ref{eq:finiteSizeCorrectionbottom}) blows up when $\ell_{\textrm{eff}} = 0$, which happens when the twist $\tau$ of the excited operator matches with the total twist $\tau_{2}+\tau_{3}$ of the two chiral primaries, $\tau = \Delta-M = \tau_{2}+\tau_{3} = 2\ell +L$. There is no fix for this divergence, if one sticks to the planar theory. Indeed, based on the general discussion of \cite{Korchemsky:2015cyx,Penedones} and the findings of \cite{Alday:2013cwa}, one expects to find poles in the planar structure constant at $\gamma = 2\ell +2n$, for $n\in \mathbb{N}$, as a result of the mixing with double trace operators. In the hexagon approach, these poles are coming from the integral (\ref{eq:finiteSizeCorrectionbottom}) and its higher particle siblings. This is, of course, a finite coupling issue, as long as $\ell>0$.

At weak coupling, we can efficiently deal with (\ref{eq:finiteSizeCorrectionbottom}) by using the method introduced in  \cite{3loops}. The basic idea is to replace the Baxter polynomial  in (\ref{010w}) by a plane wave, $Q(u)\rightarrow e^{iut}$, such as to make the integral converge at fixed $a$. Each integral can then be done directly by taking residues and the sum over $a$ expressed in terms of harmonic polylogarithms in the variable $e^{t}$. One finally generates the result for an arbitrary state $Q(u)$ by acting with the differential operator $Q(-i\partial_{t})$ and setting $t=0$ at the end. The fact that the original integral is convergent, after summation over $a$, is reflected here in the regularity of the expansion of the plane wave integral around $t=0$.

There is nothing technically different in applying the method at four loops as compared to what was done in \cite{3loops} at three loops. One simply needs to expand one loop further the various ingredients in (\ref{eq:finiteSizeCorrectionbottom}). One must in particular use the NNLO expression (\ref{logF}) for the dressing phase in the mirror kinematics.
\begin{figure}[t]
\begin{center}
%\vspace{-1cm}
\includegraphics[scale=.32]{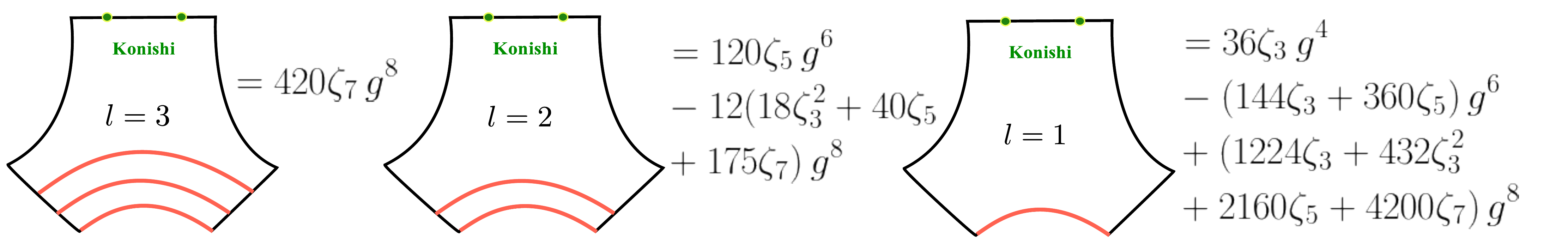}
\end{center}
\caption{Four-loop result for $c_{(0, 1, 0)}$, see definition (\ref{smallc}), when $\ell = 1, 2, 3$.}
\label{Bottomdifferentlengths} \vspace{-.2cm} 
\end{figure}
All terms in the integrand of (\ref{eq:finiteSizeCorrectionbottom}) can be computed in the same manner for any length $\ell$. The final results for $\ell=1,2,3$ are summarized in figure \ref{Bottomdifferentlengths} for the Konishi operator.

To conclude, let us mention that the Konishi operator is a bit special in that the original integral turns out to be convergent at weak coupling even \textit{before} summing over $a$, already for the smallest bridge $\ell = 1$. As such, for spin $2$, one can deal with (\ref{eq:finiteSizeCorrectionbottom}) without invoking any fancy technique. The harmonic polylogs method remains advantageous if one wants to lift the result to arbitrarily high spins. It is also relatively easy to systematize the evaluation of the higher loop plane wave integral in terms of harmonic polylogarithms.

\section{Wrapping corrections}\label{wrap}

In this appendix we evaluate the wrapping contributions to the structure constant $C^{\bullet\circ\circ}$. They comprise the various integrals for the shift of the roots and its derivatives, the integrals for the two contacts terms, and the double integral for the bulk contribution. The main results are the single integrals (\ref{one}), (\ref{two}), (\ref{three}), (\ref{intH}) and (\ref{intK}). After plugging (\ref{Yapp}) inthere, they can be taken by conventional methods (Cauchy residue theorem, etc.) and expressed in terms of Riemann $\zeta$- and Euler $\psi$-values. For the more complicated bulk double integral (\ref{Bulk}), we highlight the main steps of the analysis. The final individual expressions are summarized in equation (\ref{last}).

\subsection{Shift integrals}

The shift of the roots, for a symmetric two magnon state $\textbf{u} = \{u_{1}, u_{2}\} = \{z, -z\}$, is controlled by the integral \cite{Bajnok:2009vm}
\beq\label{phi1}
\Phi_{1} = -\sum_{a\geqslant 1}\int \frac{du}{2\pi} \frac{1}{(x^{[+a]}x^{[-a]})^{J}\, }\textrm{tr}_{a}\, \{\mathbb{S}_{a1}(u^{\gamma}, -z)\partial_{u}\mathbb{S}_{a1}(u^{\gamma}, z)\}\, ,
\eeq
where $J$ is the R-charge of the state ($J=2$ for the Konishi multiplet) and $\mathbb{S}_{a1} = S_{a1}  \mathcal{S}_{a1} \dot{\mathcal{S}}_{a1}$ is the full S-matrix. For the diagonal states of interest, with left = right, we can of course use that $\mathcal{S}_{a1} = \dot{\mathcal{S}}_{a1}$. To evaluate the integrand (\ref{phi1}) we use the expressions for the transfer matrix and diagonal factor given in appendix \ref{AFF}. For the diagonal part, we use (\ref{Sa1}) at weak coupling and immediately obtain
\beq
\begin{aligned}
&\partial_{u}S_{a1}(u^{\gamma}, z)= \bigg[\frac{1}{u-z+i\frac{a-1}{2}}-\frac{1}{u-z-i\frac{a-1}{2}}-\frac{1}{u-z+i\frac{a+1}{2}}-\frac{1}{u-z-i\frac{a+1}{2}}\bigg] S_{a1}(u^{\gamma}, z)\, .
\end{aligned}
\eeq
For the $SU(2|2)$ matrix part, we use the transfer matrix for the $\mathfrak{sl}(2)$ representative (\ref{Tsl2}), mirror rotate the $u$ rapidity, $x^{[\pm a]}\rightarrow (x^{[\pm a]})^{\mp 1}$, and differentiate w.r.t.~$u$ the factors in $B$ and $R$ that contain the root $u_{1} = z$. To the leading order at weak coupling, this yields
\beq\label{dSa1}
\textrm{tr}_{a}\,\{\mathcal{S}_{a1}(u^{\gamma}, -z)\partial_{u}\mathcal{S}_{a1}(u^{\gamma}, z)\} = \bigg[-\frac{u}{u^2+\frac{a^2}{4}}-\frac{1}{u-z+i\frac{a-1}{2}}+\frac{u+z}{u^2-z^2+\frac{a^2-1}{4}}\bigg] T_{a}(u^{\gamma})\, .
\eeq
Combining both expressions (with a factor of $2$ for the matrix part, since left and right give the same) we arrive at
\beq
\begin{aligned}
&\textrm{tr}_{a}\{\mathbb{S}_{a1}(u^{\gamma}, -z)\partial_{u}\mathbb{S}_{a1}(u^{\gamma}, z)\} = \bigg[-\frac{2u}{u^2+\frac{a^2}{4}}-\frac{1}{u-z+i\frac{a-1}{2}}-\frac{1}{u-z-i\frac{a-1}{2}}\\
&\qquad -\frac{1}{u-z+i\frac{a+1}{2}}-\frac{1}{u-z-i\frac{a+1}{2}}+\frac{2(u+z)}{u^2-z^2+\frac{a^2-1}{4}}\bigg] S_{a1}(u^{\gamma}, z)S_{a1}(u^{\gamma}, -z) T_{a}(u^{\gamma})^2\, .
\end{aligned}
\eeq
The integrand for $\Phi_{2}$ is obtained by swapping the roles of the two roots, $z\rightarrow -z$. The sum and the difference $\Phi_{\pm} = \frac{1}{2}(\Phi_{1}\pm \Phi_{2})(z, -z)$ are then found to match, respectively, with the odd and even part in $u$ of the integrand for $\Phi_{1}$. The integrand for $\Phi_{+}(z, -z)$ integrates to zero, while the one for $\Phi_{-}(z, -z)$ coincides with the $d/dz$ derivative of $Y_{a}(u^{\gamma})$. Hence, to leading order at weak coupling,
\beq\label{one}
\Phi_{-}(z, -z) = \frac{1}{2}\sum_{a\geqslant 1}\int \frac{du}{2\pi}\, \frac{d}{dz}Y_{a}(u^{\gamma})\, ,
\eeq
and therefore
\beq\label{two}
\frac{d}{dz}\Phi_{-}(z, -z)  = \frac{1}{2}\sum_{a\geqslant 1} \int \frac{du}{2\pi}\, \frac{d^2}{dz^2}Y_{a}(u^{\gamma})\, ,
\eeq
in agreement with the findings of \cite{Bajnok:2009vm} and \cite{Balog:2010vf}.

The last third of the shift of the Gaudin determinant (\ref{Cshift}) is controlled by the first term in the expansion of $\tilde{\Phi}_{+} = \Phi_{+}/J$ around the symmetric point, $\tilde{\Phi}_{+} = \frac{(u_{1}+u_{2})}{2}\partial_{+}\tilde{\Phi}_{+}+\ldots\,$. Using (\ref{Phi+}) and expanding at weak coupling, we get
\beq
\tilde{\Phi}_{+} = -\sum_{a\geqslant 1}\int \frac{du}{\pi} \frac{u}{u^2+\frac{a^2}{4}}Y_{a}(u^{\gamma}; \{u_{1}, u_{2}\}) \, .
\eeq
We must act on it with $\partial_{+} = \partial_{1}+\partial_{2}$ and then take the limit $u_{1} = -u_{2} = z$. Clearly, only the odd part in $u$ of $\partial_{+}Y_{a}$ matters here. We evaluated it using both the $\mathfrak{sl}(2)$ and the $\mathfrak{su}(2)$ expression for $Y_{a}(u^{\gamma}; \{u_{1}, u_{2}\})$. We found that they both give the same result, which we can write as
\beq\label{three}
\begin{aligned}
\partial_{+}\tilde{\Phi}_{+}(z, -z) &= -\int \frac{du}{2\pi} \frac{4u^2}{u^2+\frac{a^2}{4}}\bigg[\frac{u^2-z^2 +\frac{(a+1)^2}{4}}{Q(u^{[a+1]})Q(u^{[-a-1]})}+\frac{u^2-z^2 +\frac{(a-1)^2}{4}}{Q(u^{[a-1]})Q(u^{[-a+1]})}\\
&\qquad \qquad \qquad  -\frac{1}{z^2+\frac{1}{4}} -\frac{1}{u^2-z^2+\frac{a^2-1}{4}}\bigg]Y_{a}(u^{\gamma})\, .
\end{aligned}
\eeq
Note that it was essential to keep the $\xi$ factors in the transfer matrix to obtain this result and to get the agreement in the two sectors.

\subsection{Contact terms}

In this appendix, we shall use the principal value prescription for the bulk integral. Hence, we should combine it with the contact terms given in Section \ref{Cc}. The contact terms captured by the dressed sum (\ref{dressing}) were treated in Appendix \ref{Ct}. The remaining ones are given in (\ref{dressed}). The one containing $d\log{p}$ reads
\beq
C_{H} = \sum_{a=1}^{\infty}\int \frac{du}{2\pi} \, \textrm{int}^{H}_{a}(u)\, ,
\eeq
where
\beq\label{intH}
\textrm{int}^{H}_{a}(u) = 2\bigg[\frac{(a+1)(u^2+z^2+\frac{(a+1)^2}{4})}{Q(u^{[a+1]})Q(u^{[-a-1]})} -\frac{(a-1)(u^2+z^2+\frac{(a-1)^2}{4})}{Q(u^{[a-1]})Q(u^{[-a+1]})}\bigg]Y_{a}(u^{\gamma}) + O(g^{10})\, ,
\eeq
It follows from mirror rotating (\ref{pab}) to evaluate $\frac{i}{4}\partial_{u}\log{p_{a1}(u^{\gamma}, \textbf{u})}$, expanding at weak coupling and specializing to a two magnon state $\textbf{u} = \{z, -z\}$. The other one is the contact term $C_{1}$ involving the scattering kernel, see (\ref{C12}). Here, we simply put together the abelian and the matrix part, found in (\ref{abelK}) and (\ref{matK}), and get
\beq
C_{1} = \sum_{a=1}^{\infty}\int \frac{du}{2\pi}\, \textrm{int}^{K}_{a}(u) \, ,
\eeq
with
\beq\label{intK}
\begin{aligned}
\textrm{int}^{K}_{a}(u) &= Y_{a}(u^{\gamma})\bigg[\frac{1}{a}-H(\tfrac{a}{2}+iu)-H(\tfrac{a}{2}-iu) \\
&\,\, + \frac{(a^2-1)(u^2+\frac{a^2}{4})^2-(u^2-z^2+\frac{a^2-1}{4})^2(\frac{2(1+a)}{a}u^2-z^2+\frac{2a^2-2a-1}{4})}{2(u^2+\frac{a^2}{4})(u^2-z^2+\frac{a^2-1}{4})^2} \bigg] + O(g^{10})\, ,
\end{aligned}
\eeq
and with $H(x) = \psi(1+x)-\psi(1)$ the harmonic sum.

\subsection{Bulk integral}

The double integral (\ref{Bulk}) is by far the most bulky of all the integrals. Its integrand can nonetheless be obtained straightforwardly, at weak coupling, by expanding the various ingredients with the help of the formulae in Appendix \ref{AFF}. We get, after specializing to $\ell_{12} = \ell_{31} = 1$,
\beq\label{intB}
\textrm{int}^{B}_{ab}(u, v) = \frac{ab g^4}{(u^{2}+\frac{a^2}{4})(v^{2}+\frac{b^2}{4})} \frac{T_{a}(u^{\gamma})T_{b}(v^{\gamma})}{((u-v)^2+\frac{(a-b)^2}{4})((u-v)^2+\frac{(a+b)^2}{4})} \times \textrm{sum}_{ab}(u, v)\, ,
\eeq
with the sum
\beq
\begin{aligned}
\textrm{sum}_{ab}(u, v) =\,\, &\frac{Q(\frac{i}{2})^2 Q(v^{[b-1]})}{Q(u^{[-a-1]})Q(v^{[-b+1]})Q(v^{[b+1]})}\sum_{\bar{\alpha}}(-1)^{|\bar{\alpha}|}\bigg(\frac{\bar{\alpha}+\frac{i}{2}}{\bar{\alpha}-\frac{i}{2}}\bigg)\frac{\alpha-\bar{\alpha}-i}{\alpha-\bar{\alpha}} \\
&\qquad \qquad \times \frac{u-\bar{\alpha}-i\frac{a+1}{2}}{u-\bar{\alpha}+i\frac{a+1}{2}}\frac{u-\bar{\alpha}+i\frac{a-1}{2}}{u-\bar{\alpha}-i\frac{a-1}{2}}\frac{v-\bar{\alpha}+i\frac{b+1}{2}}{v-\bar{\alpha}-i\frac{b+1}{2}}\frac{v-\bar{\alpha}-i\frac{b-1}{2}}{v-\bar{\alpha}+i\frac{b-1}{2}}\, .
\end{aligned}
\eeq
It has a double pole at $u=v$ for $b=a$, $\textrm{int}^{B}_{aa}\sim \mathcal{A}_{\textrm{asympt}} Y_{a}(u^{\gamma})/(u-v)^2$. We integrate it with the principal value prescription,
\beq\label{pvp}
\frac{1}{(u-v)^2} \rightarrow \frac{1}{2}\bigg[\frac{1}{(u-v+i0)^2}+\frac{1}{(u-v-i0)^2}\bigg]\, ,
\eeq
which treats symmetrically the $u$ and $v$ integrals. It combines nicely with the fact that the integrand is symmetric under the exchange $(u, a)\leftrightarrow (v, b)$, since we are dealing with a symmetric splitting $\ell_{12} = \ell_{31} = 1$.

To calculate the integral, we found it useful to split it into two by applying
\beq\label{split}
\frac{1}{((u-v)^2+\frac{(a-b)^2}{4})((u-v)^2+\frac{(a+b)^2}{4})} = \frac{1}{ab}\bigg[\frac{1}{(u-v)^2+\frac{(a-b)^2}{4}}-\frac{1}{(u-v)^2+\frac{(a+b)^2}{4}}\bigg]
\eeq
to (\ref{intB}), and integrate each term separately. The first term in brackets captures the singularity at $(u, a) = (v, b)$ and must be handled with (\ref{pvp}) when $b=a$. The second term has no such issue. At the end, we must also sum over $a, b = 1, 2, \ldots\,$. We can take advantage of the fact that each integral is symmetric w.r.t.~permutation of $a$ and $b$ to write this sum as
\beq\label{applying}
\sum_{a, b} = 2\sum_{a<b} + \sum_{a = b}\, .
\eeq
This turns out to be useful when dealing with the first term in (\ref{split}). The reason is that we eventually compute the integral by summing over the residues in a given half plane. The pole we will have to pick in the first term (\ref{split}) will then depend on whether $a$ is greater or smaller than $b$. It is then convenient to stick to one case, say $a<b$, once and for all. There is no real gain in applying (\ref{applying}) to the second term in (\ref{split}), however, since the pole in this case depend on the sum $a+b$ and not the difference. On the contrary, the original sum is complete and easier to take.

Our next step was to integrate out one of the two mirror magnons, say $(v, b)$. It results in a sort of effective integrand for the remaining magnon. We evaluated the integral over $v$ by the method of residues, picking those in the upper half plane, i.e.,
\beq
v = i\tfrac{b}{2}\, , \qquad v = \pm iz+i\tfrac{b+1}{2}\, , \qquad v = \pm iz+i\tfrac{b-1}{2}\, , \qquad v = u+i\tfrac{b+\eta a}{2}\, ,
\eeq
where $\eta = -/+$ corresponds to the first/second term in (\ref{split}) and where $z = 1/\sqrt{12}$ is the weak coupling Bethe root of the Konishi multiplet. (For the boundary case $b=a$ one must only take half of the residue at $v = u$ because of (\ref{pvp}).) We then carried out the summation over $b$, using (\ref{applying}) when $\eta = -$. The resulting effective integrands are rational functions of $u$ and $a$ dressed with $\psi$ function and its derivative $\psi_{1}(x) = \partial_{x}\psi(x)$. An example of contribution that is found is
\beq\label{piece}
Y_{a}(u^{\gamma})\bigg[\psi(1+a)+\psi(1)-\psi(1+\tfrac{a}{2}+iu)-\psi(1+\tfrac{a}{2}-iu)\bigg]\, .
\eeq
The other non-rational ``words" are
\beq
\psi_{1}(1+a)\, , \qquad \psi_{1}(1+\tfrac{a}{2}\pm iu)\, , \qquad \psi_{0, 1}(\tfrac{1}{2}+a\pm \tfrac{i}{2\sqrt{3}})\, , \qquad \psi_{0}(a\pm \tfrac{i}{\sqrt{3}})\, ,
\eeq
plus those obtained by plugging $a=0$ upthere. The corresponding integrals are similar to those found before, as one can see by comparing for instance (\ref{piece}) with (\ref{intK}). They can be taken by residues and summed explicitly over $a$, after a lot of work using Mathematica. The final expression is given below, together with the other integrals.

\subsection{Summary}

We summarize here the expressions found for the individual contributions. We first introduce some notations,
\beq
\begin{aligned}
&\tilde{\Psi}_{0} = \psi(\tfrac{1}{2}+\tfrac{i}{2\sqrt{3}})-\tfrac{1}{2}\psi(1+\tfrac{i}{\sqrt{3}})-\tfrac{1}{2}\psi(1) + \textrm{c.c.}\, , \\
&\Psi_{2n} = \psi_{2n}(\tfrac{1}{2}+\tfrac{i}{2\sqrt{3}})+\psi_{2n}(\tfrac{1}{2}-\tfrac{i}{2\sqrt{3}})\, , \\
&\Psi_{2n-1} = i\sqrt{3}(\psi_{2n-1}(\tfrac{1}{2}+\tfrac{i}{2\sqrt{3}})-\psi_{2n-1}(\tfrac{1}{2}-\tfrac{i}{2\sqrt{3}}))\, ,
\end{aligned}
\eeq
where $\psi_{m}$ is the $m$-th derivative of the $\psi$ function. Then we have the integrals
\beq
\begin{aligned}
&I^{\Phi}_{1} = \frac{1}{2\sqrt{3}}\Phi_{-} \,, \qquad I^{\Phi}_{2} = \frac{1}{18}\frac{d}{dz}\Phi_{-}\, , \qquad I^{\Phi}_{3} = \frac{1}{6} \partial_{+}\tilde{\Phi}_{+}\, , \\
&I^{K} = \sum_{a}\int \frac{du}{2\pi} \, \textrm{int}^{K}_{a}(u)\, , \qquad I^{H} = \sum_{a}\int \frac{du}{2\pi} \, \textrm{int}^{H}_{a}(u) \, , \\
&I^{B} = \frac{1}{6}\sum_{a, b}\, PV \int \frac{du dv}{(2\pi)^2} \, \textrm{int}^{B}_{ab}(u, v)\, ,
\end{aligned}
\eeq
which we evaluate at weak coupling. (The overall factors in the $I^{\Phi}$'s come from (\ref{Cshift}), while the one in $I^{B}$ is because we stripped out $\mathcal{A}_{(0, 0, 0)} = 6$.) We found, up to an overall common factor $g^8$,
\beq\label{last}
\begin{aligned}
&I_{1}^{\Phi} = -\frac{405}{2} -324\zeta_{3}-360\zeta_{5}+ 162\Psi_{1}-81\Psi_{2}-\frac{9}{2}\Psi_{3}\, , \\
&I_{2}^{\Phi} = 1782+1080\zeta_{3}+120\zeta_{5}+972\tilde{\Psi}_{0}-216\Psi_{1}+54\Psi_{2}-6\Psi_{3}+3\Psi_{4}\, , \\
&I_{3}^{\Phi} =  162+432\zeta_{3}-180\zeta_{5}-648\tilde{\Psi}_{0}-162\Psi_{1}-54\Psi_{2}-\frac{21}{2}\Psi_{3}+\frac{9}{4}\Psi_{4}\, , \\
&I^{K} = -81 -864\zeta_{3}-900\zeta_{5}+ 216\zeta_{3}^2+420\zeta_{7}-324\tilde{\Psi}_{0}+108\Psi_{1}-81\Psi_{2}-3\Psi_{3}+\frac{3}{4}\Psi_{4}\, ,\\
&I^{H} = -\frac{2025}{2}-1296\zeta_{3}+2160\zeta_{5}+108\Psi_{2}+\frac{27}{2}\Psi_{3}-3\Psi_{4}\, , \\
&I^{B} = -\frac{2511}{2}+648\zeta_{3}-432\zeta_{3}^2+108\Psi_{1}+54\Psi_{2}+\frac{21}{2}\Psi_{3}-3\Psi_{4}\, .
\end{aligned}
\eeq
The sum of the first three lines, $I^{\Phi} = I^{\Phi}_{1}+I^{\Phi}_{2}+I^{\Phi}_{3}$, reproduces the expression reported in figure \ref{renexpfig}. The complete sum gives the last line in (\ref{final}), which is free of $\Psi$'s.

\section{Strong coupling}\label{Strong}

The leading contribution at strong coupling comes from the fundamental magnons, that stand for the lightest excitations. In order to evaluate our integrals in this limit we must use the semiclassical expressions for the fundamental transfer matrix and the kernel. We record both of them below.

The fundamental transfer matrix contains 4 terms. For a semiclassical state in the $\mathfrak{sl}(2)$ subsector, they are given by
\beq
t_{1} = 1\, , \qquad t_{2} = t_{3} = -e^{-i\hat{\mathfrak{p}}(x)+i\tilde{\mathfrak{p}}(x)}\, , \qquad t_{4} = e^{-i\hat{\mathfrak{p}}(x)-i\hat{\mathfrak{p}}(1/x)}\, ,
\eeq
where we recall that in our convention the grading factor is part of the definition of the $t$-components. Here, $\hat{\mathfrak{p}}(x)$ and $\tilde{\mathfrak{p}}(x)$ are the $\mathfrak{sl}(2)$ and the $\mathfrak{su}(2)$ quasi-momentum, respectively. The latter is trivial and takes its vacuum value, for an $\mathfrak{sl}(2)$ state on a spin chain of length $L$,
\beq
\tilde{\mathfrak{p}}(x) = \frac{1}{2} p(u)L = \frac{2\pi x L}{\sqrt{\lambda}(x^2-1)}\, ,
\eeq
while the former includes all of the information about the smooth distribution of roots,
\beq
\hat{\mathfrak{p}}(x)  = \frac{1}{2} p(u)L -\frac{i}{2}\log{S(u, \textbf{u})}\, .
\eeq
Equivalently, we can write
\beq
T(u) = e^{-i\hat{\mathfrak{p}}(x)} (e^{i\hat{\mathfrak{p}}(x)}+e^{-i\hat{\mathfrak{p}}(1/x)}-2e^{i\tilde{\mathfrak{p}}(x)})\, ,
\eeq
and in our integrals it always appears dressed with the dynamical part
\beq
h(u, \textbf{u}) = e^{i\hat{\mathfrak{p}}(x)-i\tilde{\mathfrak{p}}(x)}\, .
\eeq

The other ingredient is the kernel $\sim S(v, u)\partial_{u} S(u, v)$. It is subleading $\sim 1/\sqrt{\lambda}$ in the semiclassical regime, when the two rapidities $u, v$ are of order $\sqrt{\lambda}$ as well as their difference. This is no longer so when the two rapidities are close to each other, i.e., when $u-v = O(1)$. In our case, the two rapidities are as close as they can be, since we need $v = u$. To compute the expression of the kernel in this set up it is sufficient to keep track of the rational part of the S-matrix; the S-matrix becomes then a function of the difference of rapidities. For illustration, the scalar factor of the S-matrix can be replaced by the usual XXX expression it reduces to in the rational limit,
\beq
S(u, v) \rightarrow \frac{u-v+i}{u-v-i}\, ,
\eeq
yielding for the kernel at coinciding rapidities
\beq
\frac{1}{2}K(u, u) = -\frac{i}{2}\partial_{u}\log{S(u, v)}\big|_{v=u} = -1.
\eeq
This factor is multiplying the product of transfer matrices in our expression. Hence, the diagonal part evaluates to
\beq
-T(u)^2\, .
\eeq
Evaluating the matrix part of the kernel at strong coupling shows that only the first line in (\ref{Ksym}) survives, that is, 
\beq
2(t_{2}^2 +t_{2}t_{3}+t_{3}^2 +(t_{2}+t_{3})(t_{1}+t_{4}) + t_{1}t_{4})\, .
\eeq
It is identical, up to an overall factor, to the eigenvalue of the transfer matrix in the symmetric product representation (in a regime where the shifts of the arguments are irrelevant). Putting everything together yields the full kernel as a graded sum of squares
\beq\label{squares}
-(t_{1}^2-t_{2}^2-t_{3}^2+t_{4}^2)\, .
\eeq
It comes multiplied with
\beq
S(u, \textbf{u}) = \frac{h(u, \textbf{u})}{h(\textbf{u}, u)} = e^{2i\hat{\mathfrak{p}}(x)-2i\tilde{\mathfrak{p}}(x)}\, .
\eeq
This is precisely what is needed to connect with the string theory prediction for the $(1, 0, 1)$ integral at strong coupling (\ref{mirrorstring}).

\bibliography{biblio}
\bibliographystyle{JHEP}

\end{document}